\documentclass{aa}

\usepackage[utf8]{inputenc}
\usepackage{color}
\usepackage{amsmath}
\usepackage{multirow}
\usepackage{xspace}
\usepackage{xstring}
\usepackage{supertabular}
\usepackage{tabularx}
\usepackage{hyperref}
\hypersetup{
    colorlinks=true,
    linkcolor=blue,
    filecolor=magenta,      
    urlcolor=cyan,
    citecolor=blue,
}
\usepackage{placeins}

\newcommand\who{WhoSGlAd\xspace}
\newcommand\AMLT{\alpha_{\textrm{MLT}}}
\newcommand{\mkcolor}[1]{%
    \IfEqCase{#1}{%
        {AGSS09}{dark blue}%
        {GN93}{light blue}%
        {OP}{light brown}
        {OPLIB}{beige}
        {CEFF}{dark brown}
        {OPAL05}{grey}
        {NoDiff}{light pink}
        {Dturb}{purple}
        {Alpha}{yellow}
        {Vernazza}{orange}
        {VernazzaAMLT}{dark pink}
        {NoCorr}{dark green}
        {Aov}{red}
        {Aun}{khaki}
        {TR}{light green}
        {FeH}{grey brown}
        {OPLIBDturb}{cyan}
        {Obs}{black}
    }[\PackageError{mkcolor}{Undefined option to color: #1}{}]%
}

\newcommand{\mkshape}[1]{%
    \IfEqCase{#1}{%
        {AGSS09}{diamond}%
        {GN93}{small circle}%
        {OP}{large circle}
        {OPLIB}{downward triangle}
        {CEFF}{upward triangle}
        {OPAL05}{left triangle}
        {NoDiff}{right triangle}
        {Dturb}{square}
        {Alpha}{pentagon}
        {Vernazza}{diamond}
        {VernazzaAMLT}{small circle}
        {NoCorr}{large circle}
        {Aov}{downward triangle}
        {Aun}{upward triangle}
        {TR}{facing left triangle}
        {FeH}{facing right triangle}
        {OPLIBDturb}{square}
        {Obs}{box}
    }[\PackageError{mkshape}{Undefined option to color: #1}{}]%
}

\newcommand{\cshape}[1]{\mkcolor{#1} \mkshape{#1}}

\newcolumntype{L}[1]{>{\raggedright\arraybackslash}p{#1}}
\newcolumntype{C}[1]{>{\centering\arraybackslash}p{#1}}
\newcolumntype{R}[1]{>{\raggedleft\arraybackslash}p{#1}}

\title{Thorough characterisation of the 16 Cygni system}
\subtitle{Part I: Forward seismic modelling with \who}
\titlerunning{Forward seismic modelling of 16 Cygni with \who}
\author{M.~Farnir\inst{\ref{ULg}}
\and M-A.~Dupret\inst{\ref{ULg}}
\and G. Buldgen\inst{\ref{Gen}}
\and S.J.A.J.~Salmon\inst{\ref{ULg}}
\and A. Noels\inst{\ref{ULg}}
\and C. Pin\c{c}on\inst{\ref{ULg},\ref{Par}}
\and C. Pezzotti\inst{\ref{Gen}}
\and P. Eggenberger\inst{\ref{Gen}}}
\institute{Institut d’Astrophysique et G\'eophysique de l’Universit\'e de Li\`ege, All\'ee du 6 ao\^ut 17, 4000 Li\`ege, Belgium \\ \email{martin.farnir@uliege.be}\label{ULg}
\and Observatoire de Gen\`eve, Universit\'e de Gen\`eve, 51 Ch. Des Maillettes, CH$-$1290 Sauverny, Suisse
\label{Gen}
\and LESIA, Observatoire de Paris, PSL Research University, CNRS, Universit\'e Pierre et Marie Curie, Universit\'e Paris Diderot, 92195 Meudon, France \label{Par}}
\date{Received 28/05/2020 /
Accepted 07/10/2020}

\abstract {Being part of the brightest solar-like stars, and close solar analogues, the 16 Cygni system is of great interest to the scientific community and may provide insight into the past and future evolution of our Sun. It has been observed thoroughly by the \emph{Kepler} satellite, which provided us with data of an unprecedented quality.} 
{This paper is the first of a series aiming to extensively characterise the system. We test several choices of micro- and macro-physics to highlight their effects on optimal stellar parameters and provide realistic stellar parameter ranges.}{We used a recently developed method, \protect\who, that takes the utmost advantage of the whole oscillation spectrum of solar-like stars by simultaneously adjusting the acoustic glitches and the smoothly varying trend. For each choice of input physics, we computed models which account, at best, for a set of seismic indicators that are representative of the stellar structure and are as uncorrelated as possible. The search for optimal models was carried out through a Levenberg-Marquardt minimisation. First, we found individual optimal models for both stars. We then selected the best candidates to fit both stars while imposing a common age and composition.} 
{We computed realistic ranges of stellar parameters for individual stars. We also provide two models of the system regarded as a whole. We were not able to build binary models with the whole set of choices of input physics considered for individual stars as our constraints seem too stringent. We may need to include additional parameters to the optimal model search or invoke non-standard physical processes.} 
{}
\keywords{asteroseismology -- stars:oscillations -- stars:solar-type -- stars:abundances}
\begin{document}
\maketitle

\section{Introduction}
In the past decade, the CoRoT \citep{2009IAUS..253...71B} and \emph{Kepler} \citep{2010AAS...21510101B} space missions have provided the stellar physics community with a wealth of data of unprecedented quality for solar-like stars. Such data allow stellar scientists, through the use of asteroseismology, to put their models to the test and to provide stringent constraints on the physical processes at hand, therefore highlighting the current shortcomings in the modelling (e.g. mixing processes, angular momentum transport, star-planet interaction). In addition, studying solar-like stars enables us to gather invaluable insight into the past and future of our Sun.

The 16 Cygni system is of great interest as it consists of binary solar twins which have been observed continuously for $928$ days. Both stars are therefore among the solar-like pulsators with the best data available for seismic studies. Moreover, a great amount of information has yet to be accounted for. For example, differences in superficial lithium abundances remain unexplained (\citet{1993A&A...274..825F} and \citet{1997AJ....113.1871K} observed that the B component is at least four times more Li depleted than its twin). The presence of a jovian companion to 16 Cygni B \citep{1997ApJ...483..457C} has been argued by \citet{2015A&A...584A.105D} to be the possible cause. This specific example illustrates that the system is an ideal test-bench to constrain stellar models as well as to test non-standard physical processes while taking advantage of asteroseismic techniques. 

Solar-like oscillations, as both stars display, are stochastically excited by the outer convective layer. Such oscillation spectra may present the following two main features: a regular pattern, referred to as the smooth part of the spectrum, and an oscillating pattern of low amplitude, the glitch. An acoustic glitch is the oscillating signal observed in frequencies, which is caused by a sharp variation -- compared to the wavelength of the oscillating mode -- variation in the stellar structure. The first mention of the possible use of such signatures was by \citet{1988IAUS..123..151V} and \citet{1990LNP...367..283G} who theoretically demonstrated the effect of a sharp feature in the stellar structure on oscillation frequencies, either directly or in the second frequency differences. For example, in solar-like stars, we have the helium glitch, caused by a depression in the first adiabatic index\footnote{ We recall the definition of the first adiabatic index: $\Gamma_1 = \frac{d ln P}{d ln \rho}\vert_S$, where $\rho$ is the density, $P$is  the pressure, and $S$ is the entropy.} in the second helium ionisation zone, and the convection zone glitch, due to the variation of the temperature gradient at the base of the convective envelope zone. They may constrain the surface helium content, inaccessible by other means for solar-like stars, (e.g. \citealt{2004MNRAS.350..277B,2014ApJ...790..138V}) and the total extent of the envelope convective zone, as well as the mixing processes at hand that might explain such an extent (e.g. \citealt{2000MNRAS.316..165M}).

In a previous paper, \citet{2019A&A...622A..98F} described a new method to provide a robust analysis of the solar-like oscillation frequencies simultaneously accounting for the smooth and glitch parts, the \who (\textbf{Who}le \textbf{S}pectrum and \textbf{Gl}itches \textbf{Ad}justment) method. This method relies on the \emph{Gram-Schmidt} orthonormalisation process to define seismic indicators as uncorrelated as possible. It shows a great potential to provide precise, accurate and statistically relevant constraints on stellar physics. Compared to other seismic methods accounting for the glitches signature, the \who method has the advantage of decorrelating the information contained in both components of the oscillation spectrum (smooth and glitch parts) while accounting for them simultaneously. This leads to constraints which are, in turn, the least correlated possible and more stringent. Moreover, the measured frequencies are not fitted individually as it introduces large correlations. Rather, we use seismic indicators defined to be representative of the stellar structure and as little correlated as possible. This enables us to compute optimal models as accurate as possible.

This paper is part of a series of publications dedicated at providing the most accurate and complete picture of the 16Cyg binary system. In this first study, our goal is to establish a large sample of reliable structural models analysing the degeneracies stemming from variations in the micro- and macro-physical prescriptions using our new consistent seismic modelling technique, \who. We model the system using asteroseismic, spectroscopic and interferometric constraints considering both 16CygA (\object{KIC12069424}) and 16CygB (\object{KIC12069449}) independently and as a joint system. We provide a suitable set of models for structural inversions to be studied in a second paper. Our thorough analysis also paves the way for an in-depth description of potential traces of non-standard processes acting (or having acted) during the history of the system. These include the effects of angular momentum transport processes \citep{2010A&A...509A..72E,2019A&A...626L...1E} as well as the effects of planetary formation and accretion on the lithium abundances of both stars \citep{2015A&A...584A.105D,2017A&A...598A..64T}.

The paper is structured as follows. First, we present in Sect. \ref{Sec:Met} the general methodology and recall the basics of the \who method. We then model the system. This is done in two steps. To take advantage of the great precision of the data for each star, we first provide, in Sect. \ref{Sec:ABSep}, separate adjustments while testing different choices of micro- and macro-physics. This allows to provide robust stellar parameter ranges accounting for the modelling uncertainties as well as to show discrepancies in the modelling for some cases. We select in Sect. \ref{Sec:AB} the models having consistent ages and initial compositions as initial guesses to compute models imposing a common age and initial composition, as those stars should have formed from a single molecular cloud. Even though no specific interaction between both stars is taken into account during their evolution, we refer to those models as binary models. We discuss the results in Sect. \ref{Sec:Dis}. Finally, we conclude our paper in Sect. \ref{Sec:Con}.

\section{Methodology}\label{Sec:Met}
In the current section, we describe the optimisation scheme and the seismic and non-seismic constraints used. We then present the basic principle of the \who method. Finally, we describe the physics included in the models as well as the considered variations.

\subsection{Best-fit model search}\label{Sec:BestFit}
The search for best-fit models is undertaken by a \emph{Levenberg-Marquardt} (L-M) algorithm. In doing so, we compare observed values of a set of constraints with model values, computed on the fly, through a $\chi^2$ function, to be minimised, defined as:
\begin{equation}
\chi^2 = \sum\limits^N_{i=1} \frac{\left( C_{\textrm{obs},i}-C_{\textrm{mod},i} \right)^2}{\sigma_i^2},
\end{equation}
where $C$ represents the $N$ constraints, the $\textrm{obs}$ (resp. $\textrm{mod}$) subscript the observed (resp. model) values and $\sigma$ their associated standard deviations.

Except when mentioned otherwise, the set of constraints consists of the $\Delta$, $\hat{r}_{01}$, $\hat{r}_{02}$, and $A_{\textrm{He}}$ seismic indicators \citep{2019A&A...622A..98F} presented in Sect.\ref{Sec:SeisInd} and the free parameters adopted in the modelling procedure are the mass ($M$), age ($t$), initial hydrogen abundance ($X_{0}$) and, metallicity $(Z/X)_{0}$ of the considered star. Other non-seismic data, such as the effective temperature ($T_{\textrm{eff}}$), interferometric radius ($R$), or the metallicity ($\left[ \textrm{Fe/H} \right]$), are used to discriminate between the several choices of input physics. In some cases, and when so stated, non-seismic data may be used as constraints to the model search while relaxing the mixing length parameter or including turbulent diffusive mixing with a free coefficient (see Sect. \ref{Sec:Mod} for a description of the physics included in the models). Those data are gathered in Table \ref{Tab:Const}. Finally, we do not use as a constraint the luminosity ($L$) from \citet{2012ApJ...748L..10M} as it results from asteroseismic modelling and would not be independent of our study. Instead, we compute it from the observed interferometric radius \citep{2013MNRAS.433.1262W} and the definition of the effective temperature: $s T_{\textrm{eff}}^4 = \frac{L}{4\pi R^2}$, where $s$ is the Stefan-Boltzmann constant.

\begin{table}
\centering
\small
\caption{Set of non-seismic data used througout this paper. Ref.: a: \citet{2013MNRAS.433.1262W}, b: \citet{2012ApJ...748L..10M}, c: {\citet{2009A&A...508L..17R}}, d: \citet{2014ApJ...790..138V}.}\label{Tab:Const}
\begin{tabular}{cccc}
\hline
Quantity & 16CygA & 16CygB & Ref.\\ 
\hline
\hline\\[-0.8em]
$R (R_{\odot})$ & $1.22 \pm 0.02$ & $1.12 \pm 0.02$ & a   \\
$T_{\textrm{eff}} (K)$ & $5839 \pm 42$  & $5809 \pm 39$ & a \\
$L (L_{\odot})$ & $1.56 \pm 0.05$ & $1.27 \pm 0.04$ & b \\
$\left[ \textrm{Fe/H} \right] (\textrm{dex})$ & $0.096 \pm 0.026$ & $0.052 \pm 0.021$ & c \\
$Y_f$ & $\left[ 0.23,0.25 \right]$ & $\left[ 0.218,0.260 \right]$ & d \\
\hline 
\end{tabular}
\end{table}

\subsection{\who principle and seismic indicators}\label{Sec:SeisInd}
We recall here the set of \who seismic indicators used in the fitting procedure as well as the basics of the method. For a more detailed description, see \citet{2019A&A...622A..98F}. 

\paragraph{Principle:}
The \who method relies on \emph{Gram-Schmidt}'s orthogonalisation. To represent the observed frequencies, we define a Euclidean vector space of functions of the spherical degree, $l$, and radial order, $n$ (only $m=0$ modes are considered). The N observed frequencies at a given value of $l$ are regarded as unknown vector functions of $n$ and $l$ which we write $\boldsymbol{\nu_{l}}=(\nu_{l,n_1},...,\nu_{l,n_N})$. Two notable functions are the identity, $\boldsymbol{1}$, and linear function of the radial order, $\boldsymbol{n_l}=(n_{l,1},...,n_{l,N})$. 

Given two vector quantities, say the observed and theoretical vectors of frequencies, $\boldsymbol{\nu_{\textrm{obs}}}$ and $\boldsymbol{\nu_{\textrm{t}}}$, we may define their scalar product as:
\begin{equation}
\left\langle \boldsymbol{\boldsymbol{\nu_{\textrm{obs}}}} \vert \boldsymbol{\nu_{\textrm{t}}} \right\rangle=\sum\limits_{i=1}^N\frac{\nu_{\textrm{obs},n_i} \nu_{\textrm{t},n_i}}{\sigma_i^2},
\end{equation}
with $\sigma_i$ the uncertainties associated with each component.
From this scalar product is defined the norm of a vector $\boldsymbol{\nu_{\textrm{obs}}}$:
\begin{equation}
\Vert \boldsymbol{\nu_{\textrm{obs}}} \Vert=\sqrt{\left\langle \boldsymbol{\nu_{\textrm{obs}}} \vert \boldsymbol{\nu_{\textrm{obs}}} \right\rangle}.
\end{equation}
We may also define the weighted mean of a quantity, according to our scalar product and normalisation, as:
\begin{equation}
\overline{\boldsymbol{\nu_{\textrm{obs}}}}=\frac{\left\langle \boldsymbol{\nu_{\textrm{obs}}} \vert \boldsymbol{1} \right\rangle}{\Vert\boldsymbol{1}\Vert^2}=\frac{\sum\limits^N_{i=1} \nu_{\textrm{obs},n_i}/\sigma^2_i}{\sum\limits^N_{i=1}1/\sigma^2_i}.
\end{equation}

The functions used to describe the frequencies are separated into two contributions: a smooth part, represented by second-order polynomials of $n$, and a glitch part, represented by oscillating functions of the frequency. The form of those functions is given in App. \ref{Sec:WhoDec}. Using \emph{Gram-Schmidt}'s orthogonalisation process, we build a basis of functions over that vector space. To provide an adjustment of the observed frequencies, those are projected over the basis elements. This provides completely independent coefficients which are combined into seismic indicators as little correlated as possible. One of the main advantages of this approach is that the glitch part of the adjustment is completely independent of the smooth part, even though both adjustments are carried out simultaneously. We define the seismic indicators as follows:

\paragraph{The large separation for modes of degree $l$:}
Corresponds to the value of the slope of the frequencies decorrelated from the contribution of the acoustic glitches and expressed as a linear function of the radial order $n$ for each spherical degree $l$:
\begin{equation}\label{Eq:Dnul}
\Delta_l~=~\frac{\left\langle \boldsymbol{\nu_l} \vert \boldsymbol{n_l} \right\rangle/\Vert \boldsymbol{1} \Vert^2-\overline{\boldsymbol{n_l}}~\overline{\boldsymbol{\nu_l}}}{\Vert \boldsymbol{n_l} \Vert^2/\Vert \boldsymbol{1} \Vert^2-\overline{\boldsymbol{n_l}}^2}.
\end{equation}
This is equivalent to its standard definition obtained through a linear regression (see e.g. \citealt{2012A&A...539A..63R}).

\paragraph{The large separation:} 
Corresponds to the weighted mean value, $\overline{\boldsymbol{\Delta_l}}$, of the individual large separations for each spherical degree $l$.
\begin{equation}\label{Eq:Dnu}
\Delta~=~\overline{\boldsymbol{\Delta_l}}~=~\frac{\sum\limits_l \Delta_l/\sigma^2\left(\Delta_l\right)}{\sum\limits_l 1/\sigma^2\left(\Delta_l\right)},
\end{equation}
with $\sigma\left(\Delta_l\right)$ the uncertainty on the large separation of degree $l$.

\paragraph{The normalised small separations between degrees $0$ and $l$:}
\begin{equation}
\hat{r}_{0l}~=~\frac{\overline{\boldsymbol{\nu_0}}-\overline{\boldsymbol{\nu_l}}}{\Delta_0}+\overline{\boldsymbol{n_l}}-\overline{\boldsymbol{n_0}}+\frac{l}{2}.\label{Eq:r0l}
\end{equation}
Those indicators are analogous to the mean value of the local small separation ratios defined by \citet{2003A&A...411..215R} but are again completely independent from the contribution of acoustic glitches.

\paragraph{The helium glitch amplitude:}
\begin{equation}
A_\textrm{He}~=~\Vert \boldsymbol{\delta\nu_\textrm{He}} \Vert,
\end{equation}
where $\boldsymbol{\delta\nu_\textrm{He}}$ is the helium glitch component.

Beside those indicators, we may define complementary seismic indicators which are presented in App. \ref{Sec:AddSeiInd}. Those were not part of the constraints.

Table \ref{Tab:SeiInd} gathers the values of the considered seismic indicators computed using the modes defined over the full length of the \emph{Kepler} mission determined by \citet{2015MNRAS.446.2959D}. We take out from those modes the ones with uncertainties above $1.5 \mu Hz$. This mostly corresponds to high frequency modes. A brief discussion of this choice is given in App. \ref{Sec:Freq}. We have corrected the frequencies for the surface effects by using the power law prescribed by \citet{2008ApJ...683L.175K} and the $a$ and $b$ coefficients fitted by \citet{2015A&A...583A.112S} as a function of $T_\textrm{eff}$ and $g$. The authors have undertaken this coefficient adjustment by comparing the adiabatic frequencies of patched models based on 3D simulations and that of unpatched standard 1D models.

\begin{table}
\centering
\small
\caption{Observed seismic indicators. The standard deviations result from the propagation of the uncertainties on the observed frequencies.}\label{Tab:SeiInd}
\begin{tabular}{ccc}
\hline
Indicator & 16CygA & 16CygB \\ 
\hline
\hline\\[-0.8em]
$\Delta (\mu Hz)$ & $104.024 \pm 0.005$ & $117.911 \pm 0.004$  \\
$A_\textrm{He}$ & $30 \pm 1$  & $36 \pm 1$ \\
$\hat{r}_{01}$ & $\left(3.62 \pm 0.02\right)\cdot 10^{-2}$ & $\left(2.52 \pm 0.02\right)\cdot 10^{-2}$ \\
$\hat{r}_{02}$ & $\left(5.75 \pm 0.03\right)\cdot 10^{-2}$ & $\left(5.53 \pm 0.03\right)\cdot 10^{-2}$ \\
\hline 
\end{tabular}
\end{table}

\subsection{Models}\label{Sec:Mod}
Unless specified otherwise all the models are computed using the CLES stellar evolution code \citep{2008Ap&SS.316...83S} with the AGSS09 solar chemical mixture \citep{2009ARA&A..47..481A}, the OPAL opacity table \citep{1996ApJ...464..943I} combined with that of \citet{2005ApJ...623..585F} at low temperatures, the FreeEOS software to generate the equation of state table \citep{2003ApJ...588..862C}, and the nuclear reactions rates prescribed by \citet{2011RvMP...83..195A}. We also use the mixing length theory \citep{1968pss..book.....C}, with the solar calibrated value of $\alpha_{\textrm{MLT}} = l/H_{p} = 1.82$ (where $l$ is the mixing length and $H_p$ the pressure scale height), to parametrise the mixing inside convective regions. This value is the result of a solar calibration that we carried out using the same set of input physics as described in the present section. The microscopic diffusion of elements is included and treated as in \citet{1994ApJ...421..828T}. The models do not include rotation and, therefore, rotation-induced mixing. Unless specified otherwise, models do not include overshooting at the boundary of convective layers. The temperature conditions above the photosphere are determined using an Eddington $T(\tau)$ relation, $\tau$ being the optical depth. We choose such a relation to remain consistent with \citet{2015A&A...583A.112S} whose fitted coefficients are used to correct surface effects on the observed frequencies. From now on, to distinguish from the several physical variations, we refer to the models with this specific set of input physics as the reference models. Finally, we compute theoretical adiabatic oscillation frequencies for each model via the LOSC oscillation code \citep{2008Ap&SS.316..149S}.

\subsection{Variations in the input physics}\label{Sec:PhysVar}
As mentioned earlier, to provide the most reliable set of stellar parameter ranges while accounting for the choice of micro- and macro-physics, we test choices by changing one ingredient at the time from the reference models. Those variations are:
\begin{itemize}
\item The GN93 solar reference mixture \citep{1993oee..conf...15G}, in \mkcolor{GN93} in the figures. (See Sec. \ref{Sec:InpPhy});
\item The opacities from the opacity project \citep{2005MNRAS.360..458B}, denoted OP in \mkcolor{OP}, the Los Alamos opacities \citep{2016ApJ...817..116C}, written OPLIB in \mkcolor{OPLIB} (See Sec. \ref{Sec:InpPhy});
\item The CEFF equation of state \citep{1992A&ARv...4..267C}, in \mkcolor{CEFF}, and the revised OPAL equation of state \citep{2002ApJ...576.1064R}, written OPAL05 in \mkcolor{OPAL05}. (See Sec. \ref{Sec:InpPhy}); 
\item A different choice of mixing length coefficient ($\alpha_{\textrm{MLT}}=1.7$), in \mkcolor{Alpha}. (See Secs. \ref{Sec:InpPhy} and \ref{Sec:NonSei});
\item The inclusion (or not) of microscopic diffusion, in \mkcolor{NoDiff}. (See Sec. \ref{Sec:Dif});
\item The inclusion of turbulent mixing of chemical elements following the relation for the turbulent mixing coefficient $DDT = D_{\textrm{turb}}\left(\frac{\rho}{\rho_0}\right)^n + D_{ct}$ (in $cm^2 s^{-1}$), where $\rho$ is the density, $\rho_0$ the density at the bottom of the convective envelope and $D_{\textrm{turb}}$, $n$ and $D_{ct}$ are fixed at $7500$, $-3$ and $0$ respectively \citep{1991ApJ...380..238P}, shown in \mkcolor{Dturb}. (See Secs. \ref{Sec:Dif} and \ref{Sec:NonSei});
\item The inclusion of overshooting extending outside convective regions over a distance $d=\alpha_{\textrm{ov}} \textrm{min}\left(H_p,h\right)$ where $\alpha_{\textrm{ov}}$ is the overshooting parameter, $H_p$ the local pressure scale height and $h$ the thickness of the convective region. The temperature gradient in the overshooting region is set to the radiative one and the mixing is instantaneous. We either include overshooting above the convective core, denoted $\alpha_{\textrm{ov}}$ and shown in \mkcolor{Aov}, or below the convective envelope, written $\alpha_{\textrm{un}}$ in \mkcolor{Aun} and referred to as `undershoot'. Both values are set to $0.1$. (See Sec. \ref{Sec:Ove});
\item The effect of a different choice of temperature profile above the stellar photosphere, in \mkcolor{Vernazza}. We use the model temperature profile of the quiet sun by \citet{1981ApJS...45..635V} for which an analytical formulation may be found in \citet{2013ApJS..208....4P}. (See Sec. \ref{Sec:Atm}); 
\item The impact of the surface effects, computing a model fitting seismic indicators defined with stellar frequencies which are not corrected for surface effects in \mkcolor{NoCorr}. Their values are shown in Table \ref{Tab:SeiIndNoCorr}. See also Sec. \ref{Sec:SurEff} and App. \ref{Sec:NoCorr}.
\end{itemize} 

\section{16 Cygni A and B seen as separate stars}\label{Sec:ABSep}
In the present section, we look for individual models of each star representative of the observed data and accounting at best for the modelling uncertainties. The stellar parameters for every best model estimates are displayed in App. \ref{Sec:IndMod}. To find individual models, we test several choices of input physics without any specific hypothesis about the binarity of the stars. This allows to take advantage of the unprecedented quality of the data. The first part of this study is subdivided in two steps. We start by only considering seismic constraints. Then, we add non-seismic constraints, in Sect. \ref{Sec:NonSei}, to further improve the models.
The advantage of first computing individual models for each star is that it allows to have the same amount of constraints as free parameters and to obtain an exact solution, from a statistical point of view, to the minimisation process.

\subsection{Fitting seismic constraints only}
In the present section, we present the results of the modelling considering only seismic indicators. This allows to show the impact of the seismic indicators alone on optimal results as well as the possible limitations of such an exclusive approach. Furthermore, we test several choices of micro- and macro-physics. This enables us to highlight their influence on the set of optimal parameters we retrieve. The models are computed as described in Sect. \ref{Sec:Mod} while changing only one physical ingredient at a time.

The individual models for both components are shown in Figs. \ref{Fig:ACal} and \ref{Fig:BCal}. (See Table \ref{Tab:StePar} for individual parameter values.) We show, in the upper panel of the figures the age versus the mass of the optimal model for each variation in the input physics along with the associated uncertainties. The middle panel displays the position of those models in a Hertzsprung-Russell (HR) diagram. We also represent, as a \mkcolor{Obs} box, the observed effective temperature and luminosity computed from the interferometric radius \citep{2013MNRAS.433.1262W}.  Finally, the lower panels represent the initial hydrogen versus initial metallicity and their uncertainties for each model. 

In both figures, the reference model is represented in \mkcolor{AGSS09} and denoted AGSS09. For 16 Cyg A, it actually corresponds to the model presented in Sect. 5 of \citet{2019A&A...622A..98F}. Their Fig. 14 illustrates that the use of the \who seismic indicators as constraints allows to provide a proper agreement between observed and model frequencies.

\subsubsection{Influence of seismic constraints on stellar parameters}\label{Sec:ConStePar}
\citet{2019A&A...622A..98F} showed that the \who helium glitch amplitude is a good proxy of the surface helium abundance. This means that, when requiring our models to reproduce the observed helium glitch amplitude, we require a specific helium abundance. To illustrate this statement, we plot in Figs. \ref{Fig:AFeH} and \ref{Fig:BFeH} the surface helium content as a function of the surface metallicity of each model. We also represent, as a \cshape{Obs}, the spectroscopic metallicity from \citet{2009A&A...508L..17R} and the asteroseismic $Y_s$ range computed by \citet{2014ApJ...790..138V}, taking advantage of the information contained in the helium glitch, along with their associated uncertainties. We indeed observe that the surface helium abundance is well constrained and, in most cases, in agreement with the study of \citet{2014ApJ...790..138V}. They computed ranges of $\left[0.231,0.251\right]$ and $\left[0.218,0.266\right]$ for the A and B components respectively, which encapsulate most of our values. However, we also note a small scatter in the values. Again, \citet{2019A&A...622A..98F} showed that the helium glitch amplitude is both correlated to the surface helium and metal abundances, with the helium abundance being the dominant factor. They indeed observe that, at constant surface helium abundance, a lower surface metal abundance or, at constant surface metal abundance, a higher surface helium abundance may both lead to a greater glitch amplitude. This is in fact due to a shift in the position of the adiabat in the second helium ionisation zone, where the first adiabatic index, $\Gamma_1 = \frac{d ln P}{d ln \rho}\vert_S$, presents a large depression. This allows us to account for the small scatter observed.
A direct consequence of the helium glitch amplitude constraint is the anti-correlation between the initial metallicity and helium abundance that we observe in the lower panels of Figs. \ref{Fig:ACal} and \ref{Fig:BCal}.

Moreover, we point out that most models do not account for the spectroscopic metallicity constraint. This is clearly visible in Figs. \ref{Fig:AFeH} and \ref{Fig:BFeH}. This does not come as a surprise as the presented models do not yet include the metallicity constraint in the fitting procedure. This also shows that the information contained in the helium glitch amplitude and the surface metallicity are complementary and it comes as a necessity to take advantage of both to provide the most accurate model possible.

Now looking at the middle panels of Figs. \ref{Fig:ACal} and \ref{Fig:BCal}, the first striking feature is the fact that most models lie on a straight line. Such line corresponds to the locus of models of constant radius. This almost constance of the radii stems from the $\Delta$ indicator which provides a constraint on the mean stellar density \citep{2019A&A...622A..98F,1986ApJ...306L..37U}. Thus, the models of constant mean density have almost constant stellar radii, as long as the mass remains close to constant. Actually, this is what we observe as the mean radius values of our models are $1.22 R_{\odot}$ and $1.11 R_{\odot}$ for 16CygA and B respectively (Typical uncertainties are of $0.02 R_{\odot}$ and $0.01 R_{\odot}$ respectively. Individual values are shown in Table \ref{Tab:StePar}). This is in good agreement with the values in Table \ref{Tab:Const}. We note that some models do not lie on the straight line with the other models. Such models are those with masses values that differ significantly from the mean value of other models. Finally, we observe that many models do not fall in the effective temperature and luminosity observational boxes. This comes from the fact that those constraints are not yet part of the modelling procedure and shows that their inclusion is necessary to provide the most accurate picture of the system.

\subsubsection{Effect of the metallicity reference, opacity table, and equation of state}\label{Sec:InpPhy}
As is clearly visible in Figs. \ref{Fig:ACal} and \ref{Fig:BCal} using either the GN93 solar reference mixture or a greater value of the mixing length parameter produces models for both stellar components which are more luminous and have a greater effective temperature than the reference models. Furthermore, looking at Figs. \ref{Fig:AFeH} and \ref{Fig:BFeH}, we observe that the GN93 solar reference tends to produce more metallic models, directly stemming from the fact that this solar reference is indeed more metallic than the AGSS09 one. However, in terms of metallicity, the mixing length parameter has opposite impacts on the two stellar components. A decrease of its value leads, in the case of 16 Cyg A, to models which become less metallic, while it produces more metallic models for 16 Cyg B.

Now looking at the influence of the opacity tables, we note that the OPLIB table leads to a model for 16 Cyg A which has a greater effective temperature while the effect is barely visible for 16 Cyg B. Conversely, the OP opacity table leads to models which are cooler for both stars but the effect is not as pronounced. The effect of the opacity tables is not clear on the surface composition as both models react in different ways. 

Finally, the use of a different equation of state table also produces differential effects on both stars. On the first hand, in the case of 16 Cygni A, using either the CEFF or OPAL05 tables lead to hotter, more luminous stars and with decreased surface helium and metal abundances. On the other hand, for 16 Cygni B, both tables have very little influence on the position of the star in the HR diagram (see Fig. \ref{Fig:BCal}). However, the use of the OPAL05 table leads to a model of the B component which is both richer in helium and metals at its surface. The impact of the CEFF equation of state is barely visible.

\subsubsection{Impact of diffusion}\label{Sec:Dif}
We note that the models we compute without diffusion of chemical elements are both older, heavier and richer in hydrogen than the reference, as represented in \mkcolor{NoDiff} in Figs. \ref{Fig:ACal} and \ref{Fig:BCal}.  As showed by \citet{2019A&A...622A..98F}, at a specific composition, more massive models present a stronger helium glitch signature. Therefore, to reproduce the observed signature of both stars, the models need to be poorer in helium as more massive models are favoured. This is even reinforced by the fact that no diffusion is included and the initial helium abundance has to match that of the surface. Moreover, the difference in surface helium abundance between models with and without diffusion is systematically of about $0.02$ as was noted by \citet{2019MNRAS.483.4678V}. This confirms their observation of the importance of diffusion in low mass stars of solar metallicity.

We observe that non-seismic data, that will be considered in Sect. \ref{Sec:NonSei}, are not accounted for in most cases. We note that one way to account for them is to reduce the impact of microscopic diffusion, either partially for 16 Cygni A -- by involving additional mixing processes as turbulent mixing -- or completely for 16 Cygni B. This contradicts the conclusions of \citet{2016A&A...585A.109B} who determined that models with increased diffusion efficiency (with diffusion velocities higher of about $10\%$) could help reproduce inversion results for the A star. However, this agrees with their second study \citep{2016A&A...596A..73B} where they noted that reducing the efficiency of diffusion allowed the computation of models consistent with the inversion results. This, however, lead their study to inconsistencies in surface composition between the two stars.

We also show the impact of the inclusion of turbulent mixing on the modelling by computing models with a turbulent mixing coefficient fixed at a value of $D_{\textrm{turb}}=7500~cm^2 s^{-1}$. We note that those models tend, for both stars, to be more luminous, hotter and less metallic. The overall agreement with non-seismic data is thus improved. We show in Sec. \ref{Sec:FitHR} the influence of the value of the turbulent mixing coefficient on the optimal results.

\subsubsection{Extension of convective layers}\label{Sec:Ove}
To analyse the effect of the extension of the convective core -- during pre-main sequence -- on the stellar evolution, we include instantaneous overshooting in some of our models. Those are displayed in \mkcolor{Aov} in the figures. We note that the effect on the optimal models is not obvious. Indeed, such models are almost indistinguishable from the reference ones and lie within one another uncertainties. The same goes for the surface compositions retrieved. Including a greater value of the overshooting parameter ($\alpha_{\textrm{ov}}=0.2$ instead of $0.1$) leads to great differences in the behaviours of both stars. Indeed, while the model for 16 Cyg B remains rather similar to that with a lower value of overshooting, thus similar to the one without overshooting, the model for the A component becomes significantly less massive, older, metal poor and with a smaller radius (see Table \ref{Tab:StePar}). The reason for such a difference is that only the A component, because of its slightly greater mass and smaller metallicity, is able to maintain a convective core during the main sequence with such an overshooting parameter value. Therefore, its structure becomes significantly different from that of a model without overshooting which has a radiative core during the main sequence. However, this model maintaining a convective core on the main sequence seems to be a curiosity of the minimisation method. The retro-action of the presence of the convective core ultimately leads to a significant decrease of the optimal mass as well as the metallicity, becoming significantly sub-solar ($\left[\textrm{Fe/H}\right]=-0.39\pm 0.01$). This further indicates the necessity to use non-seismic constraints to obtain proper models. We can therefore safely discard this model. Additionally, we may note that the significantly lower mass of the A model leads to a significantly smaller radius because of the $\Delta$ constraint, which again contradicts observations. Moreover, this pair of model has now a less massive A component than the B. Although, by a simple argument of scaling relations, we obtain a hint that the A component should be heavier than the B. Indeed, from \cite{1995A&A...293...87K} we have that $\frac{M_{\textrm{A}}}{M_{\textrm{B}}} \simeq \left( \frac{\nu_{\textrm{max,A}}}{\nu_{\textrm{max,B}}} \right)^3 \left(\frac{\Delta\nu_A}{\Delta\nu_B}\right)^{-4} \left(\frac{T_{\textrm{eff,A}}}{T_{\textrm{eff,B}}} \right)^{3/2}$. With $\nu_{\textrm{max,A}}=2188 \mu Hz$, $\nu_{\textrm{max,B}}=2561 \mu Hz$ taken from \citet{2017ApJ...835..172L}, and the values of the effective temperatures and large separations presented earlier, we expect the mass of the A component to be larger than the other ($M_{\textrm{A}}\simeq 1.04 M_{\textrm{B}}$). This is what we observe for most of our models.

The inclusion of undershooting below the base of the convective zone has no significant impact on the optimal stellar parameters beside a reduction of the individual uncertainties. Initial compositions and ages for both stars now fall out of each others uncertainties and the models are dismissed as valid candidates to compute binary models in Sect. \ref{Sec:AB}. This statement is clearly illustrated in Fig. \ref{Fig:CompAB} where the initial hydrogen, metal abundances and ages of both stars are plotted against one another. The straight line displays the locus of identical parameters for both stars. In the upper panel, the \mkcolor{Aun} cross, representing the model with undershooting, does not meet the line any more.

\subsubsection{Effects of the atmosphere and mixing length coefficient}\label{Sec:Atm}
We show the influence of the atmosphere on the optimal stellar parameters by using a temperature profile above the photosphere as in \citet{1981ApJS...45..635V} with a specifically calibrated value of $\AMLT=2.02$ (\cshape{VernazzaAMLT} in the figures). We observe that the optimal model is very similar to the reference model. Indeed, both pairs of models lie within the uncertainties of one another. The optimal parameters are given in Table \ref{Tab:Ver}. However, the models become hotter and more luminous. They are thus closer to the observed luminosities and effective temperatures. This indicates that it allows to provide better models in terms of spectroscopic constraints while preserving rather similar stellar parameters compared to the case using the Eddington relation.
We must also point out that the use of a different temperature profile leads to significant changes from the reference models but the calibration of the mixing length parameter compensates for it. To illustrate this,  we compute models with the temperature profile of \citet{1981ApJS...45..635V} while using the reference value of $\AMLT=1.82$ calibrated for an Eddington $T-\tau$ relation. We observe that our models, shown in \mkcolor{Vernazza} in the figures, are very similar to models using a lower value of the mixing length parameter, displayed in \mkcolor{Alpha}. This is especially true for 16 Cygni B for which both models lie within respective uncertainties. For both stars, the computed models are older and lighter.

\begin{figure*}
\begin{minipage}[t]{.49\textwidth}
\centering
\includegraphics[width=0.95\linewidth]{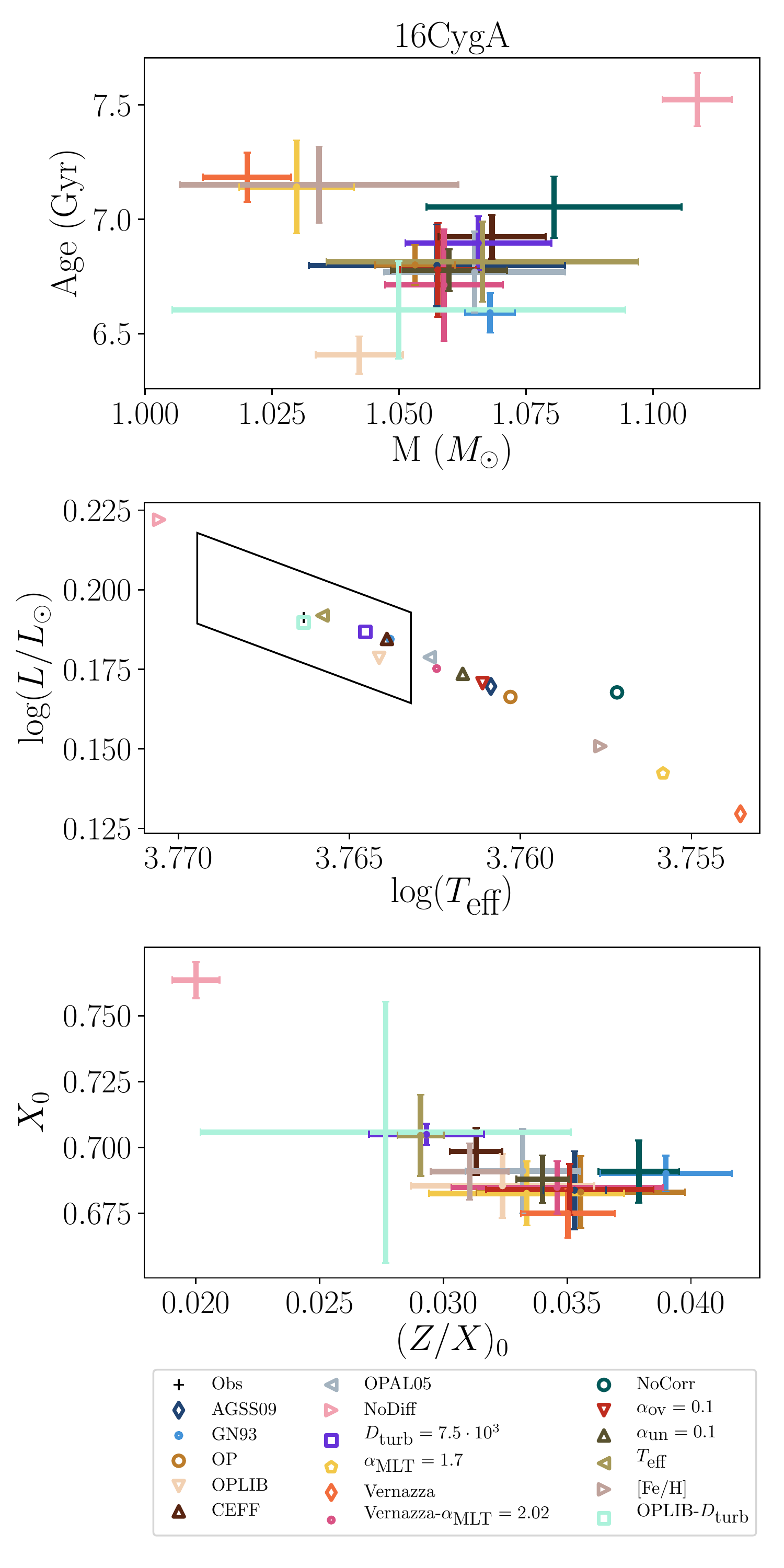}
\caption{Summary of 16 Cyg A best-fit models represented in a Mass - Age diagram (top panel), HR diagram (middle panel) and initial hydrogen abundance versus metal composition diagram. The luminosity and effective temperature constraints from \citet{2013MNRAS.433.1262W} are represented in the HR diagram as a \protect\mkcolor{Obs} box.}\label{Fig:ACal}
\end{minipage}
\hfill
\begin{minipage}[t]{.49\textwidth}
\centering
\includegraphics[width=0.95\linewidth]{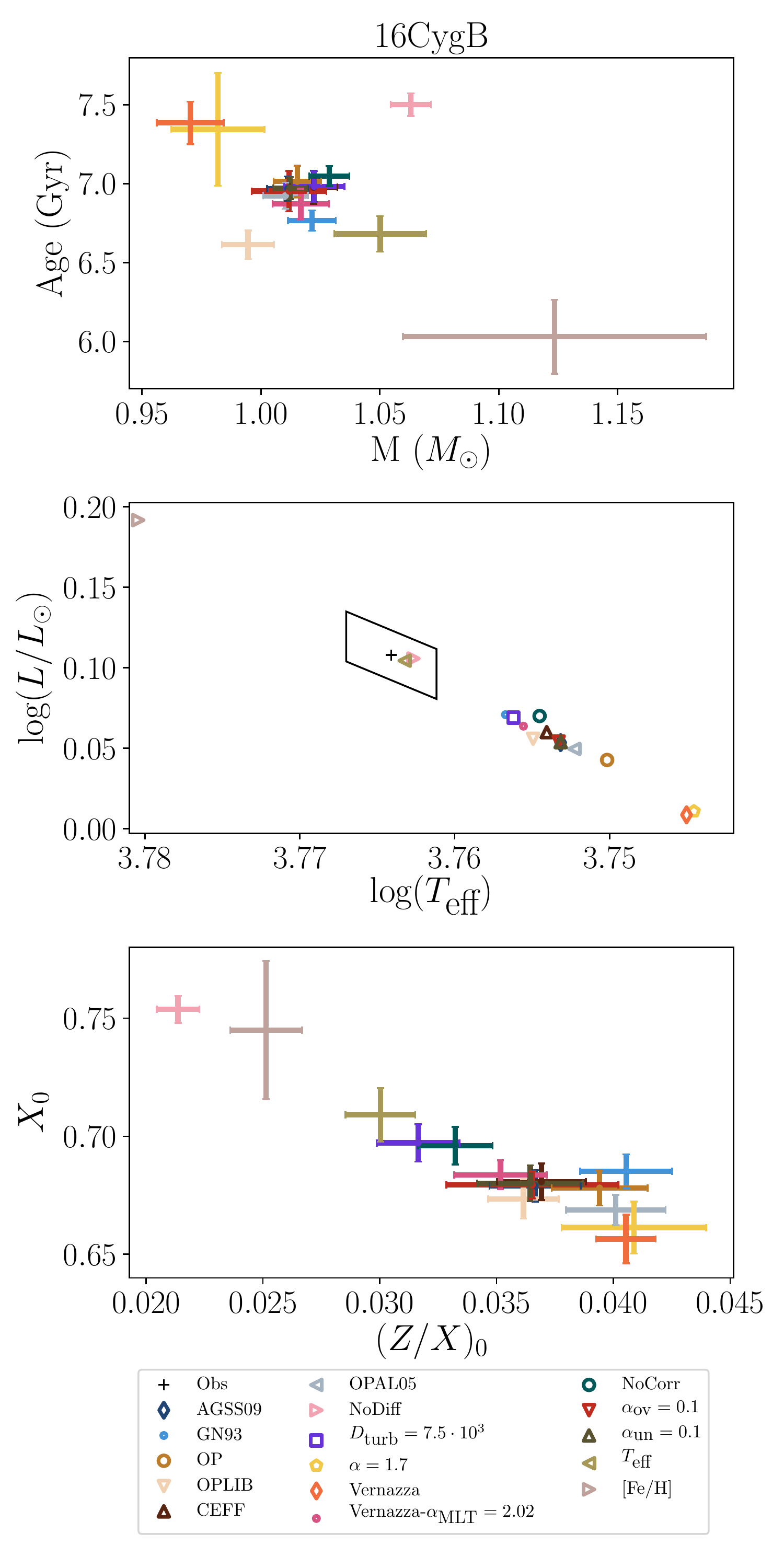}
\caption{Summary of 16 Cyg B best-fit models. The colours are the same as in Fig. \ref{Fig:ACal}}\label{Fig:BCal}
\end{minipage}
\end{figure*}

\begin{figure*}
\begin{minipage}[t]{.49\textwidth}
\centering
\includegraphics[width=0.95\linewidth]{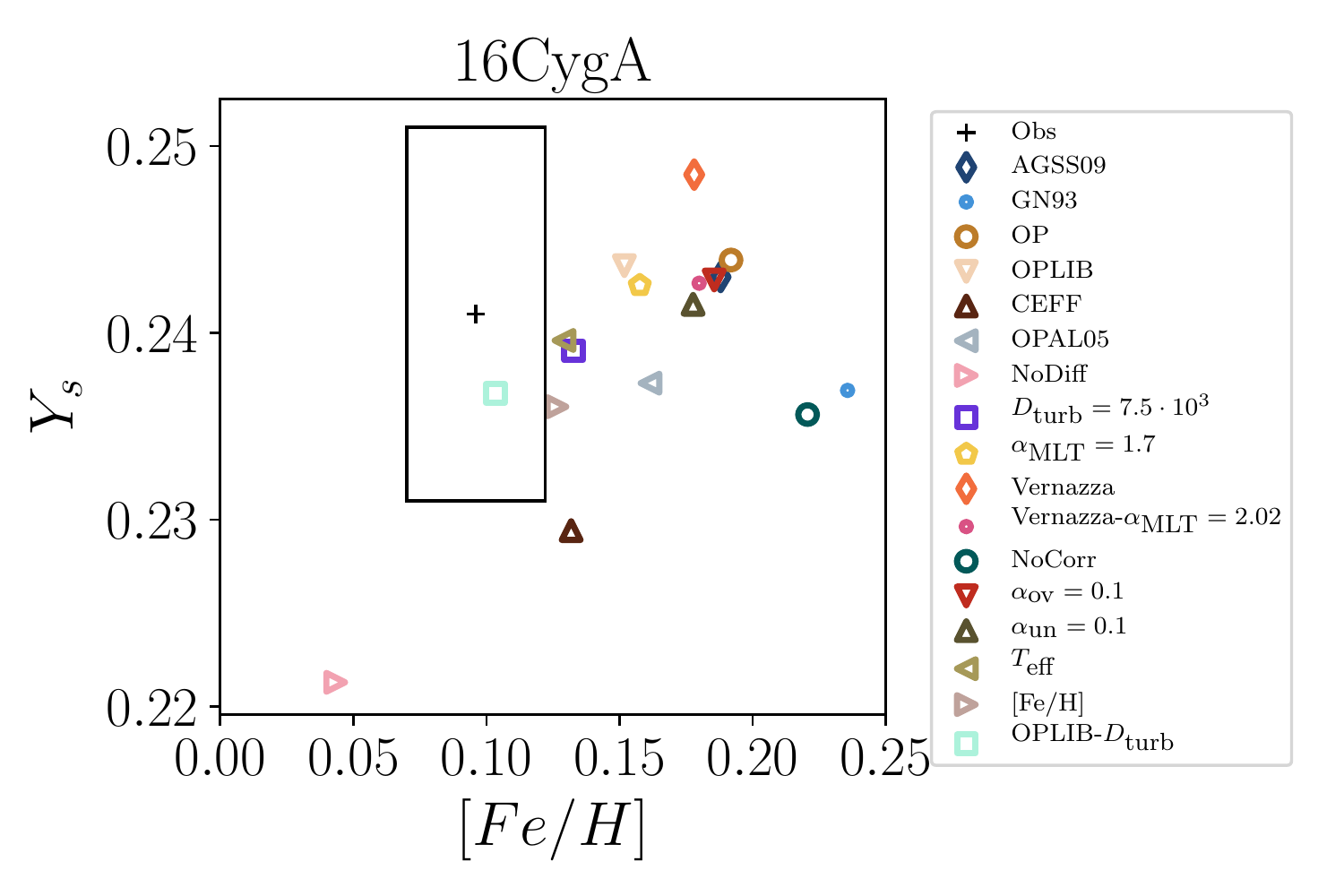}
\caption{Surface helium abundance versus metallicity for 16 Cygni A. The \protect\cshape{Obs} represents the spectroscopic metallicity value computed by \citet{2009A&A...508L..17R} and the surface helium value from \citet{2014ApJ...790..138V} along with the corresponding uncertainties.}\label{Fig:AFeH}
\end{minipage}
\hfill
\begin{minipage}[t]{.49\textwidth}
\centering
\includegraphics[width=0.95\linewidth]{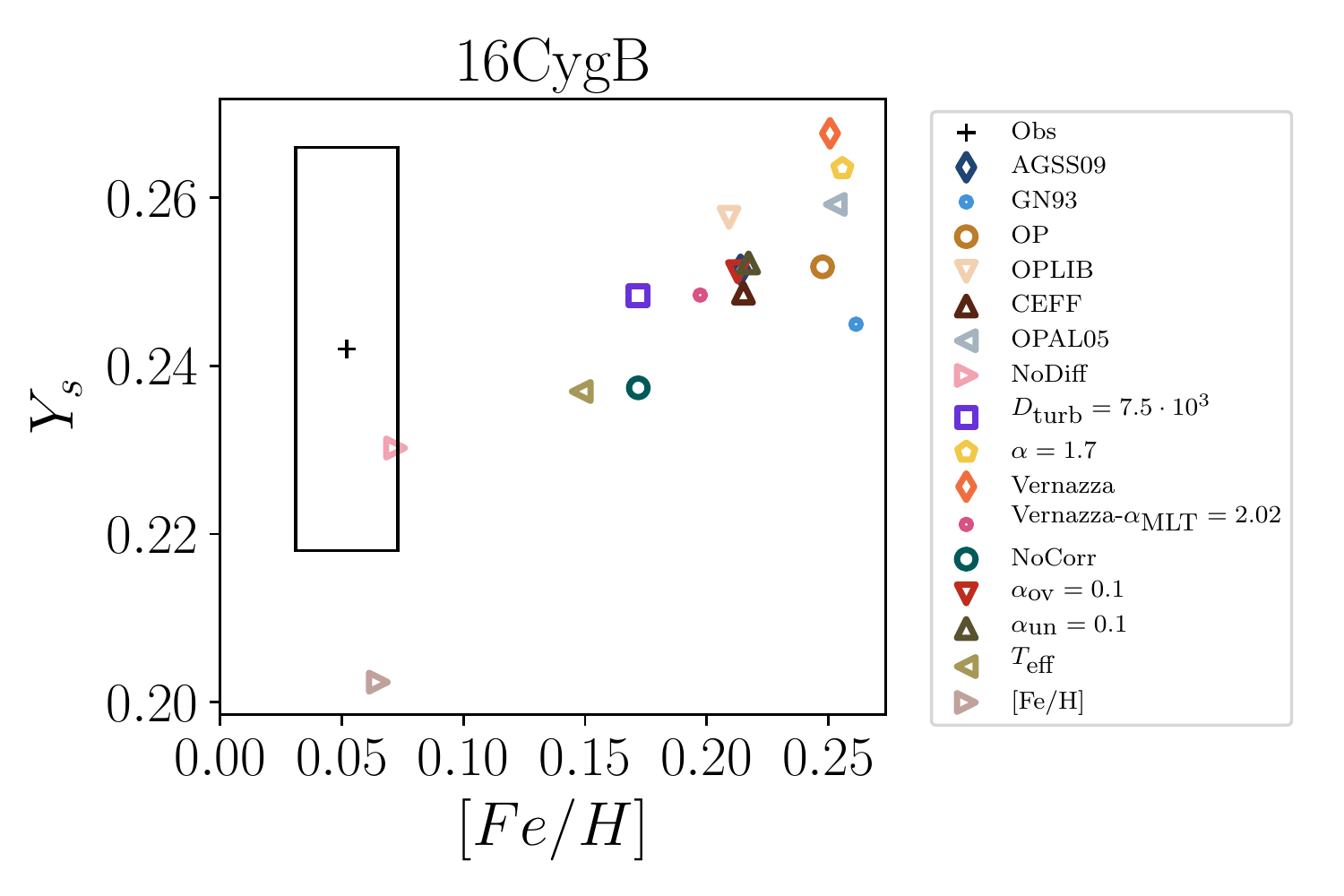}
\caption{Surface helium abundance versus metallicity for 16 Cygni B. The \protect\cshape{Obs} represents the spectroscopic metallicity value computed by \citet{2009A&A...508L..17R} and the surface helium value from \citet{2014ApJ...790..138V} along with the corresponding uncertainties.}\label{Fig:BFeH}
\end{minipage}
\end{figure*}

\subsection{Fitting non-seismic constraints}\label{Sec:NonSei}
To further improve individual models for each star, we may include non-seismic constraints into the minimisation process. We indeed note that, in most cases, they are not accounted for. Therefore, we use those constraints and include additional free parameters to add degrees of freedom. The considered constraints are the effective temperature \citep{2013MNRAS.433.1262W} and the spectroscopic metallicity \citep{2009A&A...508L..17R}. The additional free parameters are the mixing length parameter $\AMLT$ and the turbulent mixing coefficient $D_{\textrm{turb}}$. In what follows, we specify in every case which of those constraints and free parameters are used.

\subsubsection{Accounting for the effective temperature}\label{Sec:FitHR}
From Figs. \ref{Fig:ACal} and \ref{Fig:BCal}, we expect that it is possible to improve the agreement with the observed effective temperature by increasing the value of $\AMLT$. As a consequence of the $\Delta$ constraint, which we showed in Sec. \ref{Sec:ConStePar} to be a proper constraint on the radius, we also expect to produce better model luminosities.

\begin{figure*}
\begin{minipage}[t]{.49\textwidth}
\includegraphics[width=.95\linewidth]{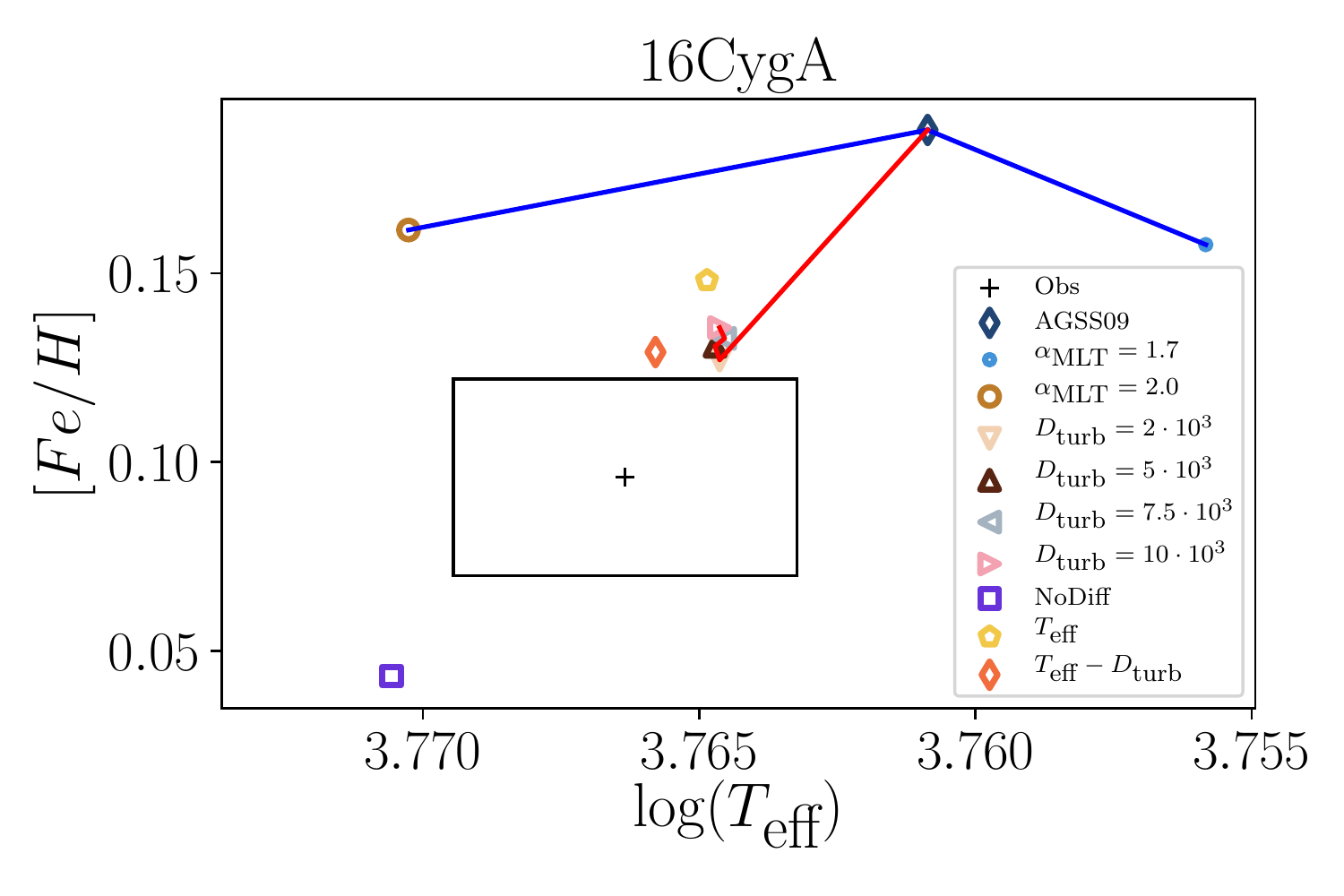}
\caption{Variation of the optimal metallicity and effective temperature for 16 Cyg A with the mixing length parameter and turbulent mixing. Models of different $\AMLT$ values are connected in blue while models with various turbulent mixing are connected by a red line.}
\label{Fig:FeTA}
\end{minipage}
\hfill
\begin{minipage}[t]{.49\textwidth}
\includegraphics[width=.95\linewidth]{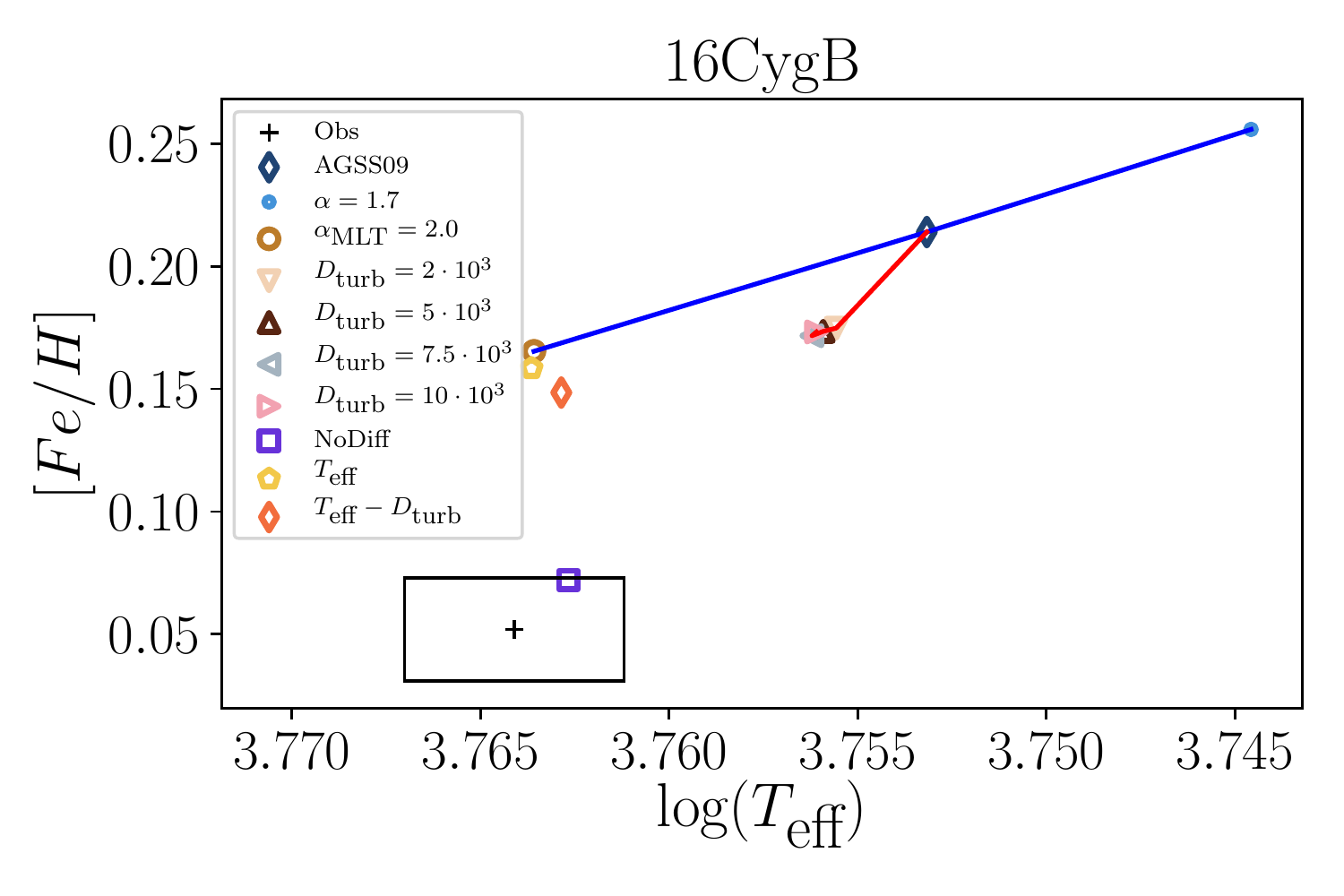}
\caption{Variation of the optimal metallicity and effective temperature for 16 Cyg B with the mixing length parameter and turbulent mixing. Models of different $\AMLT$ values are connected in blue while models with various turbulent mixing are connected by a red line.}
\label{Fig:FeTB}
\end{minipage}
\end{figure*}

To demonstrate that a variation in the mixing length parameter is indeed responsible for the improvement of the model effective temperature, we compute several models with different mixing length parameters. We only include seismic constraints in the fitting procedure to show the influence of $\AMLT$ alone on optimal parameters. This is shown in Figs. \ref{Fig:FeTA} and \ref{Fig:FeTB} for 16 Cyg A and B respectively and the stellar parameters are given in Table \ref{Tab:Alp}. The values considered are $1.7$ [light blue], the solar value of $1.82$ [dark blue], and $2.0$ [brown]. Those models are connected with a blue line to improve visibility. We also display models with several choices for the turbulent mixing coefficient: $D_{\textrm{turb}}=2000~\textrm{cm}^2\textrm{s}^{-1}$ [light pink], $5000~\textrm{cm}^2\textrm{s}^{-1}$ [brown], $7500~\textrm{cm}^2\textrm{s}^{-1}$ [grey], $10000~\textrm{cm}^2\textrm{s}^{-1}$ [dark pink], and no turbulent mixing [dark blue]. These are connected in red and stellar parameters are gathered in Table \ref{Tab:Dif}. We observe that an increase of $\AMLT$ leads to a better agreement with the observed effective temperature for both stars.
Moreover, we note that the inclusion of turbulent mixing improves the agreement in effective temperature and metallicity for both stars. Even so, both stars exhibit too large values of the metallicity compared to the observed ones. We also note that, for 16 Cyg B, turbulent mixing alone is not sufficient for the observed and model $T_{\textrm{eff}}$ to match. In addition, we observe a clear effect of saturation of the turbulent mixing coefficient, which occurs above a threshold value that is already exceeded by the considered values. The results are almost indistinguishable no matter which value is chosen. Therefore, the turbulent mixing and mixing length coefficient both have a impact on the model effective temperature (and thus luminosity) but using the turbulent mixing coefficient as a free parameter would be meaningless. Furthermore, we also display in both figures models which did not include microscopic diffusion of the chemical elements. We observe that those two models differ greatly from the models including turbulent diffusion, both being highly hotter and less metallic. The model for 16 Cygni B even properly reproduces the observed metallicity and effective temperature, although they are not yet part of the constraints.

Finally, we compute two models for each star with a free mixing length coefficient and including $T_{\textrm{eff}}$ as a constraint, with and without turbulent mixing ($D_{\textrm{turb}}=7500~\textrm{cm}^2\textrm{s}^{-1}$). The set of input physics is that of the reference model. Both are displayed in Figs. \ref{Fig:FeTA} and \ref{Fig:FeTB} as orange diamonds labelled `$T_{\textrm{eff}}-D_{\textrm{turb}}$' and as yellow pentagons labelled `$T_{\textrm{eff}}$'. For the A component, we see that including turbulent mixing improves the results. However the agreement with the non-seismic constraints is not improved compared to the model including only turbulent mixing and the opacity table OPLIB with only seismic constraints, that already accounted for the effective temperature, as displayed on Fig. \ref{Fig:ACal}. Conversely, for the B component, the improvement is significant and the effective temperature is now well adjusted. Nevertheless, the inclusion of turbulent mixing does not have a significant impact on the resulting agreement with the non-seismic data. Furthermore, in both cases, the metallicity is still not properly accounted for. The corresponding set of stellar parameters is presented in Tables \ref{Tab:NonSei} and \ref{Tab:NonSeiDturb}.

Analysing our results, we first notice from Table \ref{Tab:NonSei} that the calibrated mixing length parameters are very different. Indeed, looking at Fig.2 from \citet{2015A&A...573A..89M}, who used the same solar reference mixture as we do, we would expect that both adjusted values would remain rather close to the solar calibrated value ($1.82$ in our case) while being smaller as both stars have higher effective temperatures and smaller surface gravities than the Sun. Thus, both differences should translate into a lower mixing length parameter. However, we observe that for 16 Cygni A, and within the error bars, $\alpha_{\textrm{MLT}}$ remains solar while the calibrated value for 16 Cygni B is significantly higher than the solar value. We may need to invoke a special physical process acting on any of the components while being inefficient for the second to produce such a differential effect. Some of the possible scenarios are discussed in Sect. \ref{Sec:DisPhys}.

We note that slightly more massive and less metallic models than previously are favoured. Such an effect stems from the fact that more massive models are hotter and thus in better agreement with the observed effective temperature, while keeping the same density, because of the $\Delta$ constraint. 

Another way to better reproduce the observed position in the HR diagram is to include extra mixing counteracting the diffusion of chemical elements. Indeed, for 16 Cyg B, the model computed without diffusion already accounts for these constraints. Which is striking as those were not yet constraints of the fit. What is more striking is that it also reproduces the spectroscopic metallicity. This strongly suggests that additional mixing processes may be necessary to properly and accurately model this star. For its twin, we note that the inclusion of turbulent mixing, a different opacity table, the Los Alamos one, or the use of a different equation of state, either CEFF or OPAL05, could help us account for the observed luminosity and effective temperature. Therefore, we perform another fit using the OPLIB opacity table, including turbulent mixing with a coefficient of $D_{\textrm{turb}}=2000~\textrm{cm}^2\textrm{s}^{-1}$, and adding the effective temperature to the set of constraints. We do not include the turbulent mixing coefficient into the free parameters as Figs. \ref{Fig:FeTA} and \ref{Fig:FeTB} clearly demonstrate that it saturates. The set of constraints is now composed of our seismic indicators and the effective temperature and the free parameters are the age, mass, and initial composition. We are now able to get a suitable model which, again, accounts for the position in the HR diagram but also for the spectroscopic metallicity, which was not required. This is illustrated as a \cshape{OPLIBDturb} in the figures, with the label `$\textrm{OPLIB}-D_{\textrm{turb}}$'. This shows that modified opacities could help model the 16 Cyg A star as accurately as possible and also reinforces the argument that additional mixing processes might be necessary to model both stars. The values of the several stellar parameters are gathered in Table \ref{Tab:OPLIBDturb}. 

\begin{table}
\centering
\small
\caption{Adjusted stellar parameters including the $T_{eff}$ with the reference set of input physics.}\label{Tab:NonSei}
\begin{tabular}{ccc}
\hline
Quantity & 16CygA & 16CygB \\
\hline
\hline\\[-0.8em]
$M~(M_{\odot})$ & $1.1 \pm 0.1$ & $1.05 \pm 0.02$ \\
$X_0$ & $0.69 \pm 0.04$ & $0.70 \pm 0.01$ \\
$\left(Z/X\right)_0$ & $0.032 \pm 0.002$ & $0.032 \pm 0.002$ \\
$Y_0$ & $0.28 \pm 0.04$ & $0.027 \pm 0.01$ \\
$\left[\textrm{Fe/H}\right]$ & $0.15 \pm 0.05$ & $0.16 \pm 0.03$ \\
$Y_s$ & $0.23 \pm 0.02$ & $0.233 \pm 0.007$ \\
$t~(Gyr)$ & $6.7 \pm 0.4$ & $6.6 \pm 0.1$ \\
$R~(R_{\odot})$ & $1.2 \pm 0.1$ & $1.12 \pm 0.02$ \\
$\alpha_{\textrm{MLT}}$ & $1.9 \pm 0.2$ & $1.99 \pm 0.06$ \\
$\chi^2$ & $1.1$ & $0.2$ \\
\hline
\end{tabular}
\end{table}

\begin{table}
\centering
\small
\caption{Adjusted stellar parameters including the $T_{eff}$ constraint and turbulent mixing with a coefficient of $D_{\textrm{turb}}=7500~\textrm{cm}^2\textrm{s}^{-1}$.}\label{Tab:NonSeiDturb}
\begin{tabular}{ccc}
\hline
Quantity & 16CygA & 16CygB \\
\hline
\hline\\[-0.8em]
$M~(M_{\odot})$ & $1.07 \pm 0.03$ & $1.05 \pm 0.02$ \\
$X_0$ & $0.70 \pm 0.01$ & $0.71 \pm 0.01$ \\
$\left(Z/X\right)_0$ & $0.0291 \pm 0.0009$ & $0.0298 \pm 0.0008$ \\
$Y_0$ & $0.27 \pm 0.02$ & $0.027 \pm 0.01$ \\
$\left[\textrm{Fe/H}\right]$ & $0.13 \pm 0.02$ & $0.15 \pm 0.01$ \\
$Y_s$ & $0.24 \pm 0.01$ & $0.238 \pm 0.009$ \\
$t~(Gyr)$ & $6.8 \pm 0.2$ & $6.7 \pm 0.1$ \\
$R~(R_{\odot})$ & $1.22 \pm 0.03$ & $1.12 \pm 0.03$ \\
$\alpha_{\textrm{MLT}}$ & $1.84 \pm 0.08$ & $1.94 \pm 0.07$ \\
$\chi^2$ & $0.6$ & $0.7$ \\
\hline
\end{tabular}
\end{table}

\subsubsection{Accounting for the metallicity}\label{Sec:FitFeH}
Up to now, the only two models accounting for the spectroscopic metallicity constraint from \citet{2009A&A...508L..17R} are models with a reduced impact of diffusion. Those correspond to the one with the OPLIB opacity table and including turbulent mixing  with a fixed coefficient of $D_{\textrm{turb}}=2000~\textrm{cm}^2\textrm{s}^{-1}$, labelled `$\textrm{OPLIB}-D_{\textrm{turb}}$' in the figures, for 16 Cyg A and the one without diffusion for 16 Cyg B. Both models reproduce the complete set of seismic and non-seismic constraints (that is $\Delta$, $\hat{r}_{01}$, $\hat{r}_{02}$, $A_{\textrm{He}}$, $T_{\textrm{eff}}$, $L$, and $\left[ Fe/H \right]$). However, the spectroscopic metallicity was not yet part of the fitting constraints. Moreover, as Figs. \ref{Fig:AFeH} and \ref{Fig:BFeH} clearly illustrate, most of the computed models do not agree with the spectroscopic metallicities.

As Figs. \ref{Fig:FeTA} and \ref{Fig:FeTB} clearly show, the turbulent mixing coefficient saturates and freeing its value cannot enable us to produce models that reproduce the metallicity. We also note that the impact of the mixing length parameter is mostly focused on the effective temperature. As a consequence, we expect a large variation of this parameter will be necessary to reproduce the metallicity. In order to verify this hypothesis, we compute models with a free mixing length parameter and the metallicity as a constraint. The set of free parameters is made of the age, mass, composition and, mixing length of the star and the constraints are the set of seismic constraints and the metallicity. The results are given in Table \ref{Tab:FeH} and shown in \mkcolor{FeH} with the label $\left[\textrm{Fe/H} \right]$ in Figs. \ref{Fig:ACal} through \ref{Fig:BFeH}. We indeed observe that the necessary variations in $\AMLT$ are incompatible with those to reproduce $T_{\textrm{eff}}$. Indeed, now that the model and observed metallicities agree, the effective temperature values do not. Trying to include both non-seismic constraints, using the mixing length parameter as a free coefficient and either including turbulent mixing or not did not lead to a satisfactory adjustment (i.e. with a reduced $\chi^2$ value inferior to $1$). This clearly shows that we are not able, with the current set of parameters, to reproduce the complete set of seismic and non-seismic constraints without invoking special physical processes. We also note that the model for 16 Cyg B is both too massive and young compared to other studies (e.g. \citealp{2016A&A...596A..73B,2017ApJ...837...47V}). This illustrates that one has to proceed with caution when modelling data as it is possible to provide a model which is representative of these data while having no physical meaning.

\begin{figure}
\centering
\includegraphics[width=0.95\linewidth]{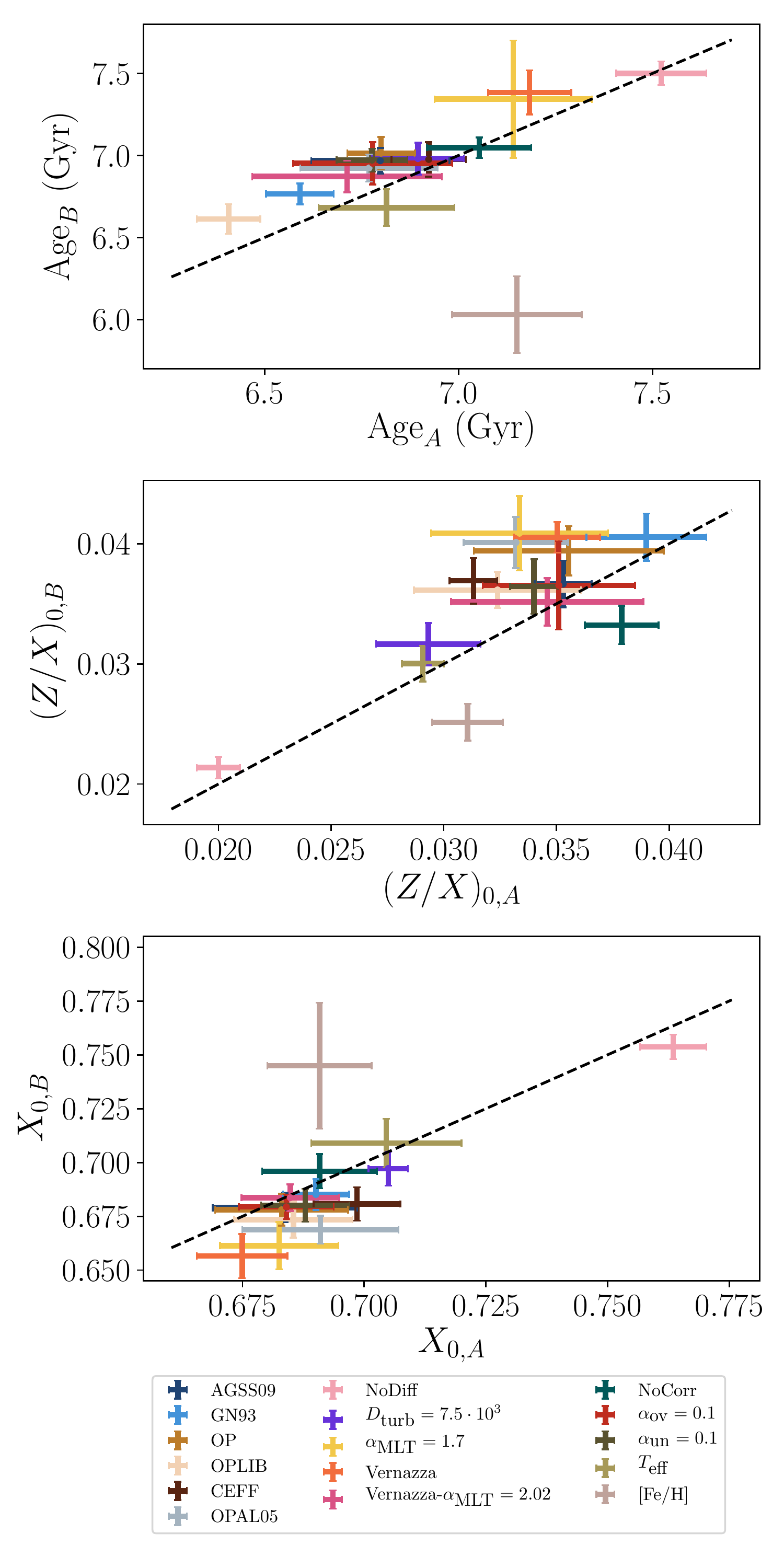}
\caption{Comparison of adjusted stellar parameters for both stars at a given physics. The straight line shows the locus of identical stellar parameters for both stars.}
\label{Fig:CompAB}
\end{figure}

\begin{table}[h]
\centering
\small
\caption{Adjusted stellar parameters including the metallicity constraint.}\label{Tab:FeH}
\begin{tabular}{ccc}
\hline
Quantity & 16CygA & 16CygB \\
\hline
\hline\\[-0.8em]
$M~(M_{\odot})$ & $1.03 \pm 0.03$ & $1.12 \pm 0.06$ \\
$X_0$ & $0.69 \pm 0.01$ & $0.74 \pm 0.03$ \\
$\left(Z/X\right)_0$ & $0.031 \pm 0.003$ & $0.025 \pm 0.001$ \\
$Y_0$ & $0.29 \pm 0.01$ & $0.24 \pm 0.03$ \\
$\left[\textrm{Fe/H}\right]$ & $0.12 \pm 0.04$ & $0.06 \pm 0.04$ \\
$Y_s$ & $0.236 \pm 0.006$ & $0.20 \pm 0.02$ \\
$t~(Gyr)$ & $7.1 \pm 0.2$ & $6.0 \pm 0.2$  \\
$R~(R_{\odot})$ & $1.21 \pm 0.03$ & $1.14 \pm 0.06$ \\
$\AMLT$ & $1.71 \pm 0.04$ & $2.3 \pm 0.2$ \\
$\chi^2$ & $3.2$ & $1.2$ \\
\hline
\end{tabular}
\end{table}

\subsection{Individual best models}\label{Sec:BestMod}In the present section, we summarise and further analyse the two best models we obtain while regarding both stars as separate, that is to say without imposing a common initial composition and age. Those are the only models which simultaneously account for seismic and non seismic constraints and are the ones denoted `$\textrm{OPLIB}-D_{\textrm{turb}}$' for 16 Cyg A and `$\textrm{NoDiff}$' for 16 Cyg B. Table \ref{Tab:BestMod} shows the choice of input physics as well as the set of optimal parameters of those models. As both models do not have the same set of input physics, they may not be regarded as valid candidates to study the system as a whole as is done in Sec. \ref{Sec:AB}. We indeed expect from binary stars with close stellar parameters that their internal physics should overall be identical. The goal of the present section is only to analyse in more depth models which fitted the complete set of considered constraints and to investigate whether those models still may be further improved.

\begin{figure}
\centering
\includegraphics[width=0.95\linewidth]{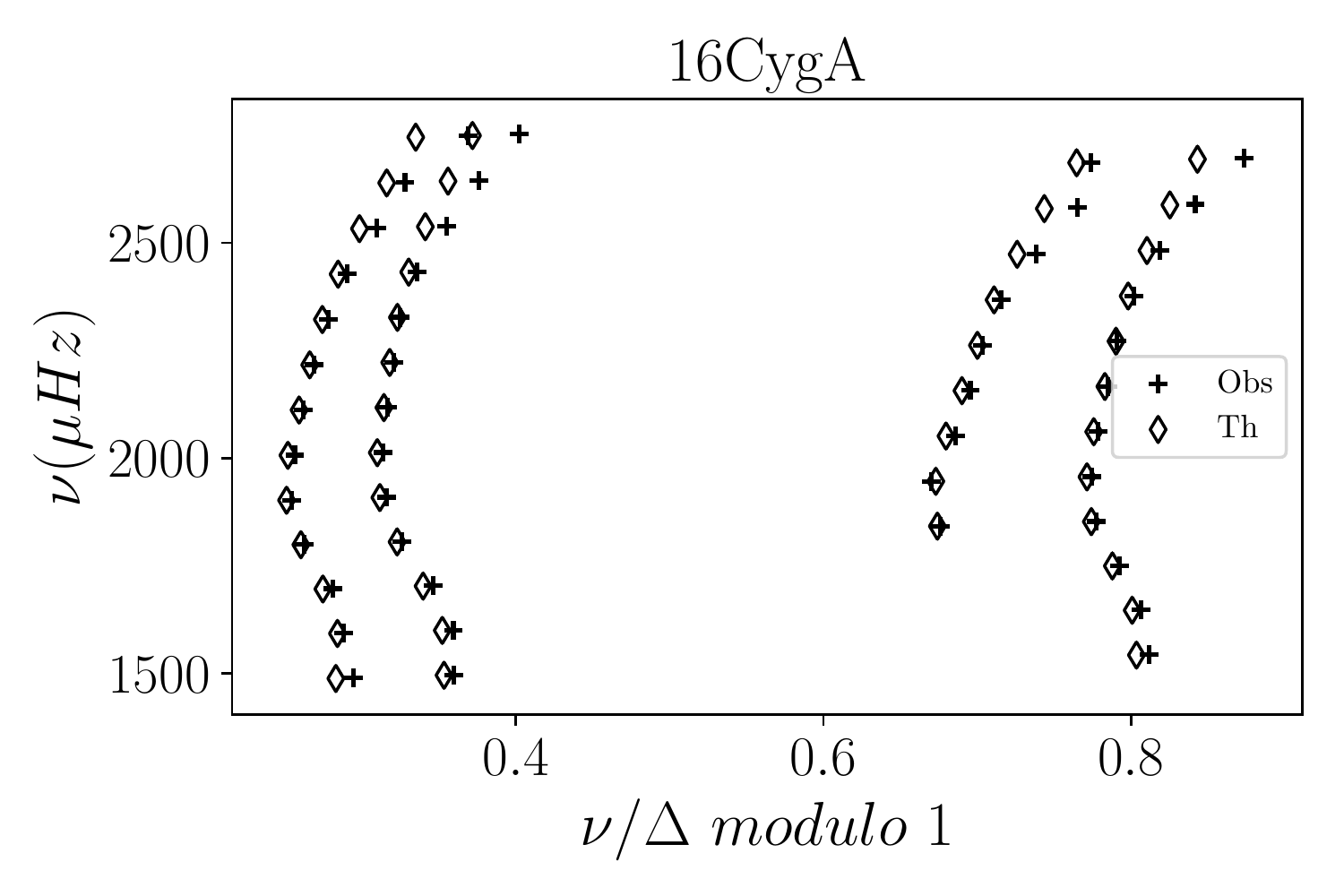}
\caption{\emph{\'Echelle} diagram of 16 Cygni A best-fit model. The crosses are the observed frequencies and the diamonds the theoretical ones.}\label{Fig:EchA}
\end{figure}

\begin{figure}
\centering
\includegraphics[width=0.95\linewidth]{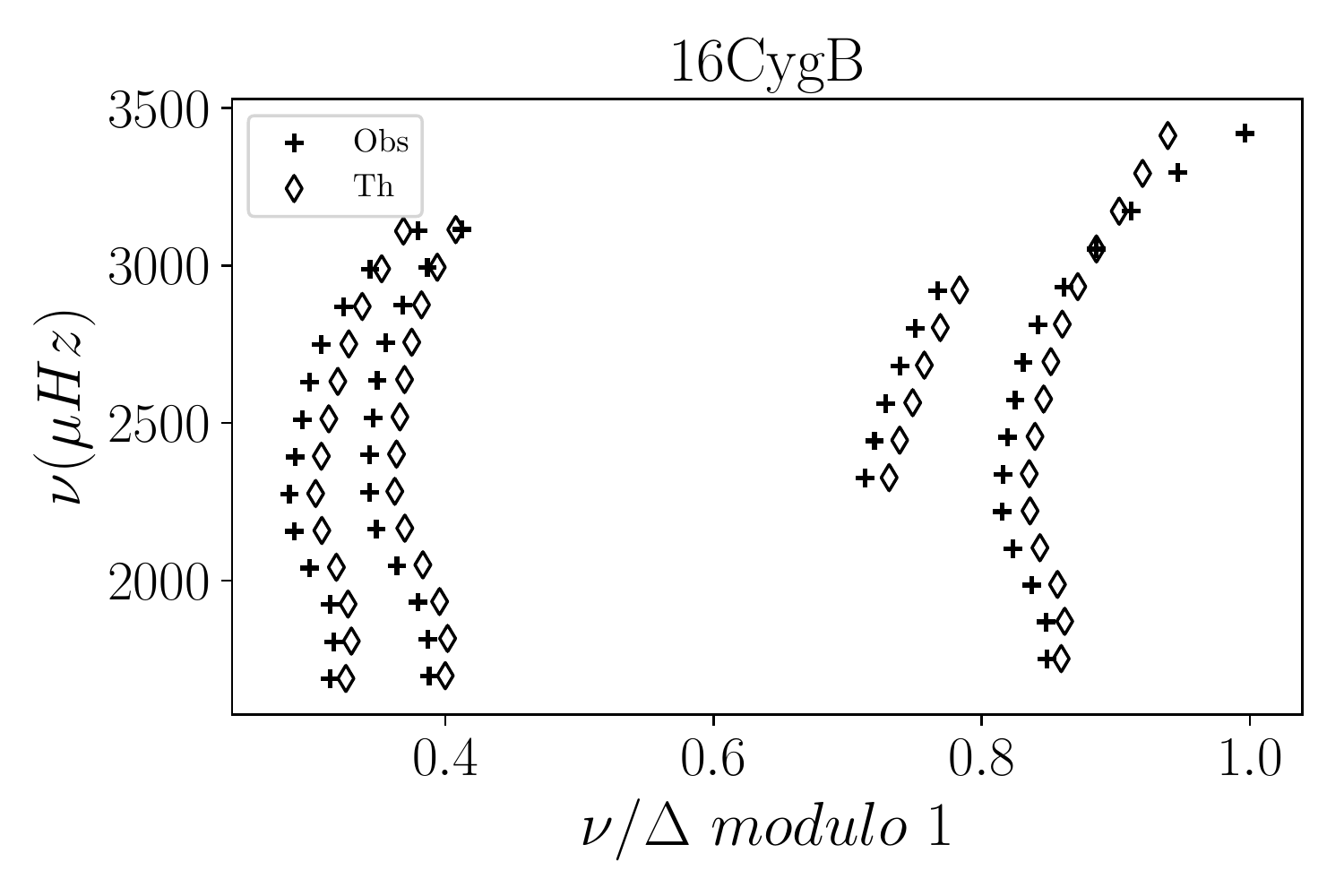}
\caption{\emph{\'Echelle} diagram of 16 Cygni B best-fit model. The crosses are the observed frequencies and the diamonds the theoretical ones.}\label{Fig:EchB}
\end{figure}

We display the \emph{\'echelle} diagrams of each star in Figs. \ref{Fig:EchA} and \ref{Fig:EchB}. We observe that the frequency trend for both stars is well accounted for. However, we note a drift at high frequencies. We expect this effect to mainly result from the surface effects. To illustrate this claim, we display the \emph{\'echelle} diagram of optimal models for both stars computed with seismic indicators which are not corrected for surface effects in Figs. \ref{Fig:EchANoCorr} and \ref{Fig:EchBNoCorr}. We indeed observe that the high frequency drift is reinforced.

\begin{figure*}
\begin{minipage}[t]{.49\textwidth}
\centering
\includegraphics[width=0.95\linewidth]{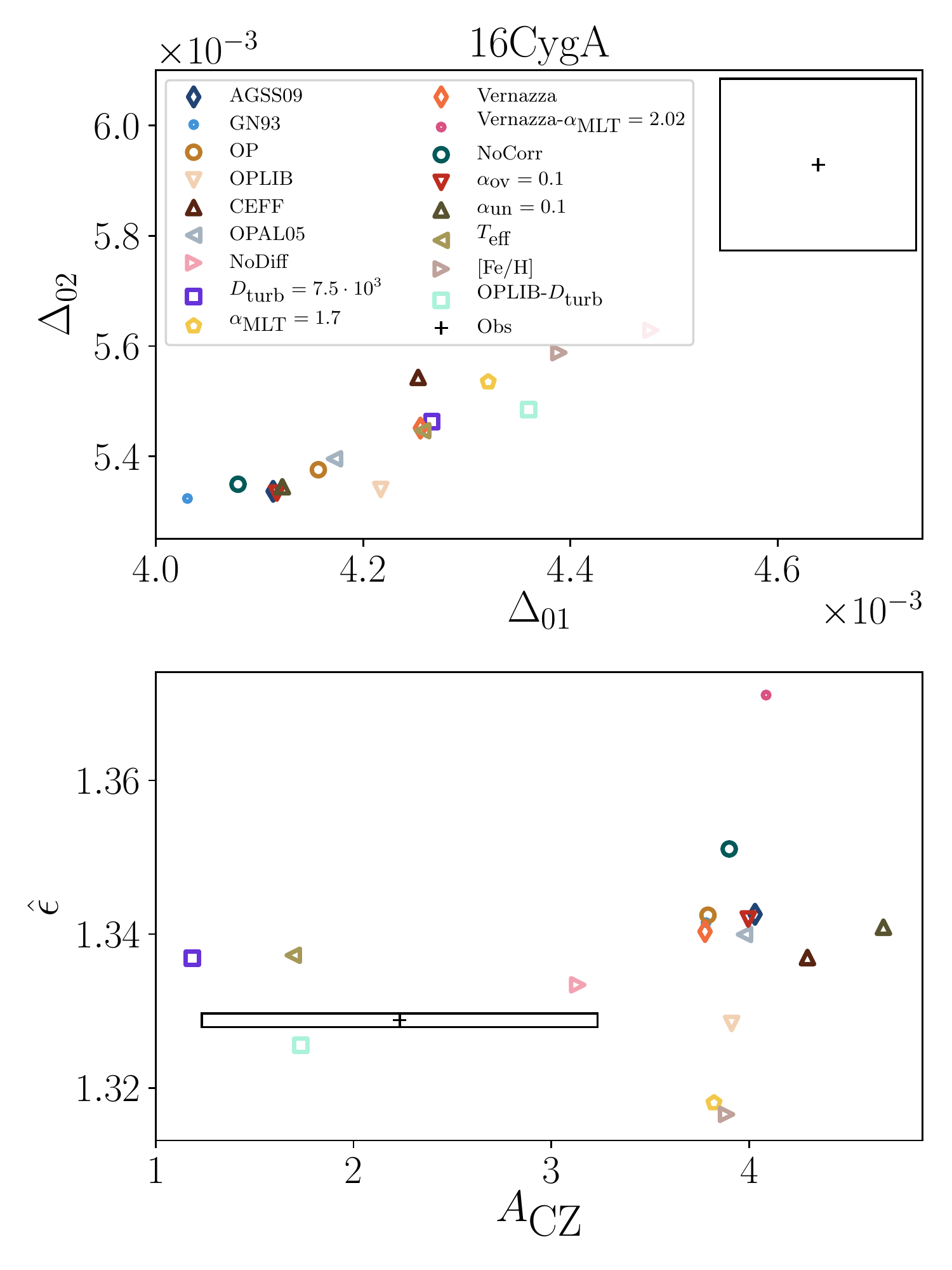}
\caption{Values of complementary seismic indicators for 16 Cyg A. Observed values along with their uncertainties are shown as a\protect\mkcolor{Obs} box.}\label{Fig:IndA}
\end{minipage}
\hfill
\begin{minipage}[t]{.49\textwidth}
\centering
\includegraphics[width=0.95\linewidth]{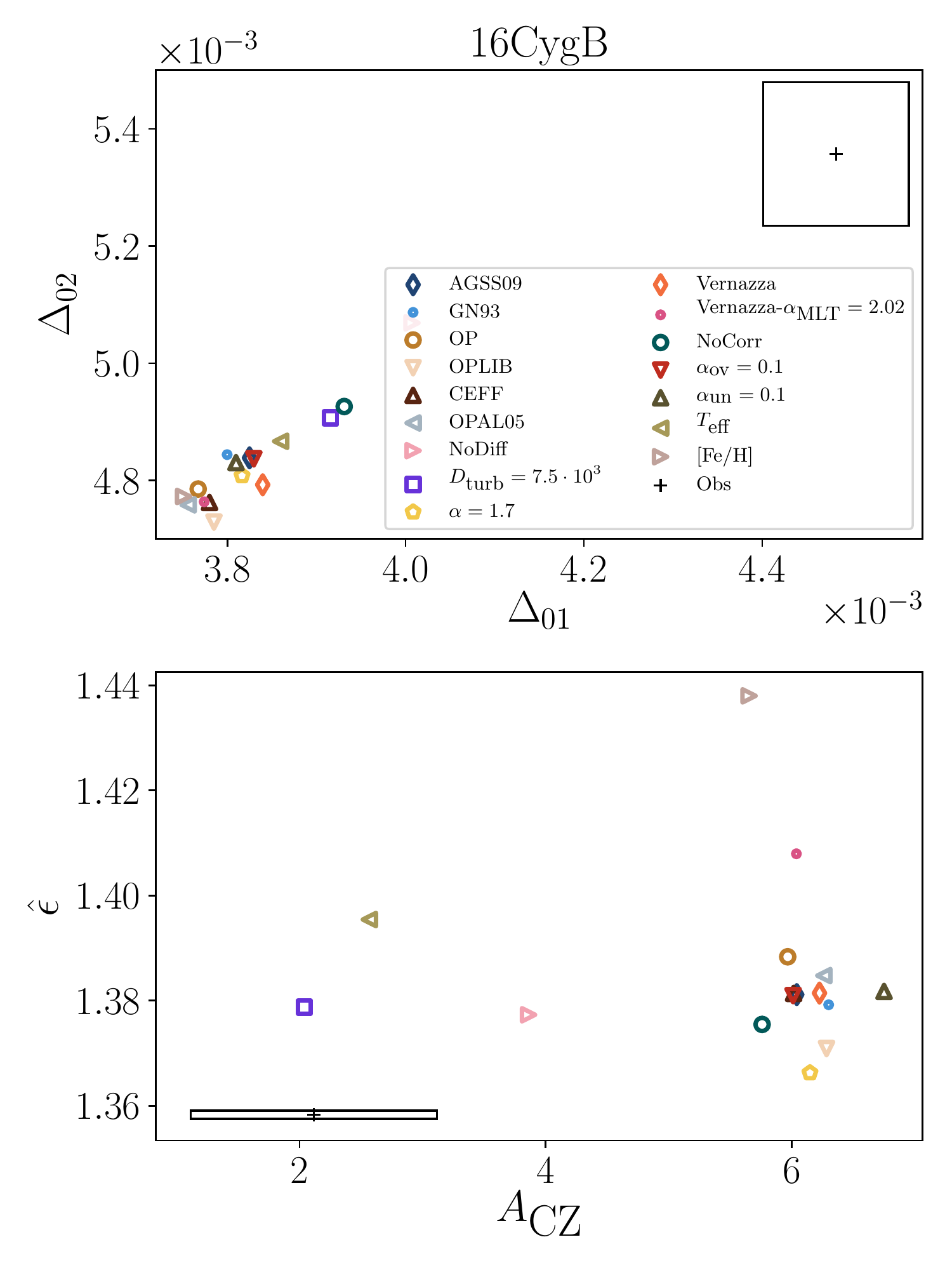}
\caption{Values of complementary seismic indicators for 16 Cyg B. Observed values along with their uncertainties are shown as a \protect\mkcolor{Obs} box.}\label{Fig:IndB}
\end{minipage}
\end{figure*}

We also note that, in the case of 16 Cygni B, the theoretical ridges are shifted with respect to the observed ones. This effect should mainly be due to the $\hat{\epsilon}$ seismic indicator defined in \citet{2019A&A...622A..98F}. It corresponds to an estimator of the constant term in $n$ in the asymptotic expression of frequencies as in \citet{1986HiA.....7..283G} and has been shown to be sensitive to the surface effects. Its value along with several other \who indicators for every computed model are displayed in Figs. \ref{Fig:IndA} and \ref{Fig:IndB} (their definitions and observed values are given in App. \ref{Sec:AddSeiInd}). We observe that the $\hat{\epsilon}$ is not properly accounted for. However, it is worse in the case of 16 Cyg B, which explains why this effect is much more visible.

In App. \ref{Sec:AddSeiInd}, we define other seismic indicators that  were not part of the constraints. Now looking at those indicators displayed in Figs. \ref{Fig:IndA} and \ref{Fig:IndB}, we see that, in most cases, none of them are properly accounted for. In the lower panels, we display the values of the base of the convection zone glitch amplitude and note that only a few models are within the one $\sigma$ uncertainty region. However, one should not be alarmed by this observation as its value is hardly significant in the case of the 16 Cygni system ($A_{\textrm{CZ}}~=~2 \pm 1$ for both stars) and, therefore, bears little information.

We also represent the values for both $\Delta_{01}$ and $\Delta_{02}$ which represent the slopes of the individual frequency ratios $r_{01}$ and $r_{02}$ expressed as a function of the radial order. Again, every model presents a value which is significantly different from the observed ones. Nonetheless, accounting for such data is a complex task and we note that, in the present situation, only the modification of diffusion seems to provide an improvement for both stars.

We do not include the $\hat{\epsilon}$ indicator in the modelling procedure as it has been shown by \citet{2019A&A...622A..98F} (Fig. 4) to be sensitive mostly to the surface effects and highly degenerate in the stellar mass. Moreover, it is tightly correlated to the large separation \citep[see Eq. \ref{Eq:eps} and the asymptotic formulation of the frequencies in ][]{1986HiA.....7..283G}. The $\Delta_{0l}$ indicators are not used as they mostly carry information about central overshooting \citep[][Fig. 3]{2019A&A...622A..98F} which we presumed would not happen as both stars are below the approximative limit of $\sim 1.1M_{\odot}$ and are expected to have a radiative core.

\begin{figure*}
\begin{minipage}[t]{.49\textwidth}
\centering
\includegraphics[width=0.95\linewidth]{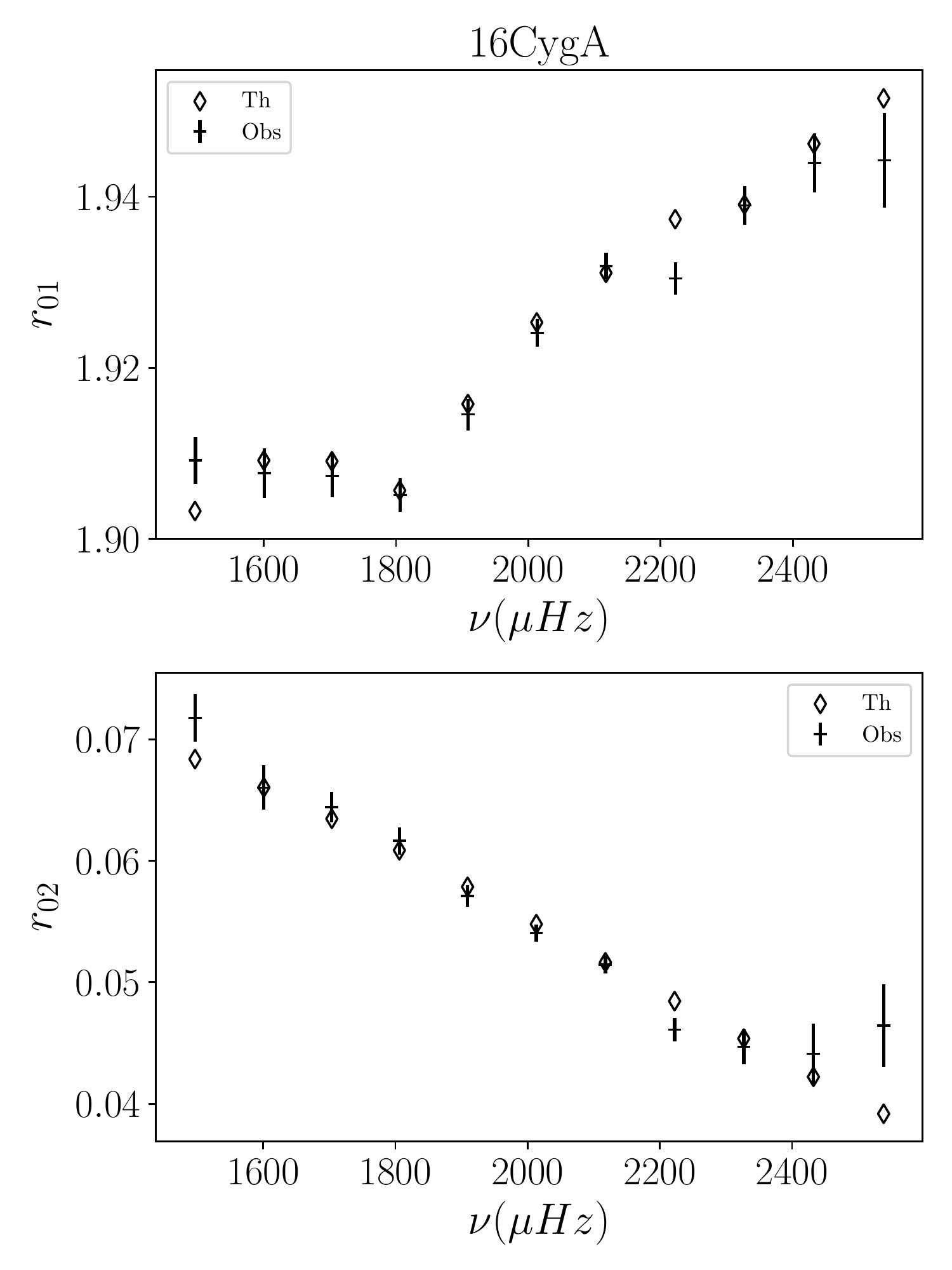}
\caption{Individual ratios for 16 Cygni A. Observed values along with their uncertainties are shown as crosses, best model values are represented by diamonds.}\label{Fig:RatioA}
\end{minipage}
\hfill
\begin{minipage}[t]{.49\textwidth}
\centering
\includegraphics[width=0.95\linewidth]{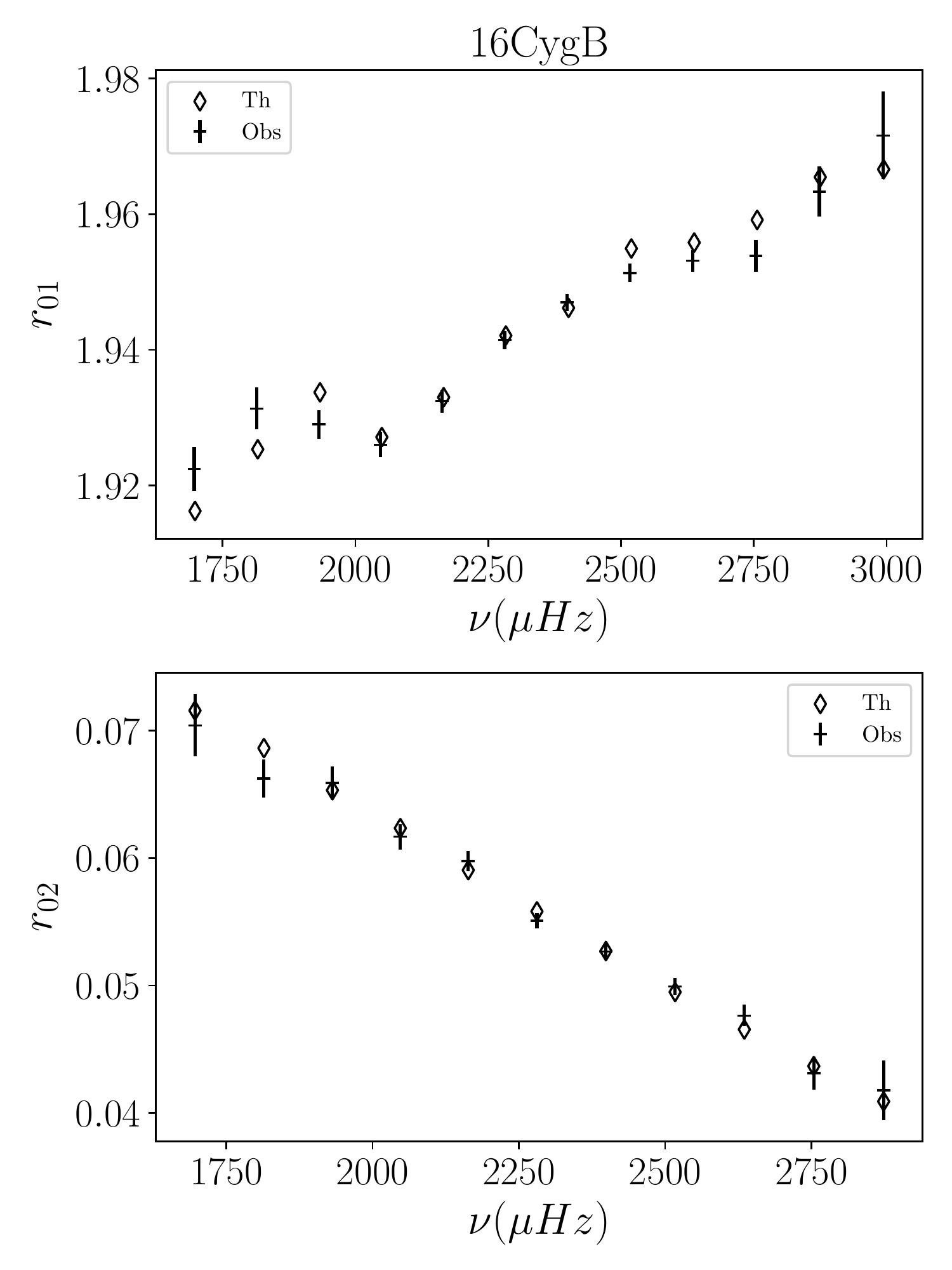}
\caption{Individual ratios for 16 Cygni B. Observed values along with their uncertainties are shown as crosses, best model values are represented by diamonds.}\label{Fig:RatioB}
\end{minipage}
\end{figure*}

Finally, we plot in Figs. \ref{Fig:RatioA} and \ref{Fig:RatioB} the observed individual frequency ratios defined in \cite{2003A&A...411..215R}, which we recall are not used as constraints in our fits, as a function of frequency against the best model ones. We observe that, although the overall agreement is good, the oscillation which is present in the observed ratios is not properly accounted for in the model frequencies. This clearly indicates that some information remains to be exploited to model the system as comprehensively as we can and inversion techniques may be of great help in doing so. However, with the use of only our indicators, instead of individual ratios, the overall trend seems to be well respected in both cases.

\begin{table}
\centering
\small
\caption{Best individual models for each star, accounting for both seismic and non-seismic constraints.}\label{Tab:BestMod}
\begin{tabular}{ccc}
\hline
Quantity & 16CygA & 16CygB \\
\hline
\hline\\[-0.8em]
Solar Ref. & AGSS09 & AGSS09 \\
Opacity & OPLIB & OP \\
Eq. state & Free & Free \\
Atmos & Eddington & Eddington \\
Diffusion & Yes & No \\
Turb. mix. & Yes & No \\
$\alpha_{\textrm{MLT}}$ & $1.82$ & $1.82$ \\
$M~(M_{\odot})$ & $1.07 \pm 0.03$ & $1.063 \pm 0.008$ \\
$X_0$ & $0.70 \pm 0.03$ & $0.754 \pm 0.006$ \\
$\left(Z/X\right)_0$ & $0.029 \pm 0.004$ & $0.0214 \pm 0.009$ \\
$Y_0$ & $0.27 \pm 0.03$ & $0.230 \pm 0.009$\\
$t~(Gyr)$ & $6.8 \pm 0.2$ & $7.50 \pm 0.07$ \\
$D_{\textrm{turb}}$ & $0.2 \cdot 10^4$ & / \\
$\chi^2$ & $0.9$ & $0.6$ \\
\hline
\end{tabular}
\end{table}

\section{The system as a whole}\label{Sec:AB}
In this section, we select the set of individual models which respect the binarity constraint and try to provide models while imposing a common age and initial composition.

\subsection{Accepted models}\label{Sec:Acc}
The models satisfying the binarity constraint are those that have, within one another uncertainties, identical ages and initial composition. Fig. \ref{Fig:CompAB} provides a clear illustration. Only models represented by a cross that meets the line, corresponding to identical stellar parameters in the three panels, are kept. Those models are referred to as the accepted models and are: the reference models (AGSS09, \mkcolor{AGSS09}), those with turbulent mixing ($D_{\textrm{turb}}$, \mkcolor{Dturb}), without diffusion (NoDiff, \mkcolor{NoDiff}), with overshooting ($\alpha_{\textrm{ov}}$, \mkcolor{Aov}), the models with the mixing length coefficient adjusted for the effective temperature  ($T_{\textrm{eff}}$, \mkcolor{TR}), and with a temperature profile above the photosphere as in \citet{1981ApJS...45..635V} with a calibrated $\AMLT$ of $2.02$ (Vernazza $\AMLT$ \mkcolor{VernazzaAMLT}). From those models, we define the range of accepted stellar parameters given in Table \ref{Tab:AccRan}.

Even though models without diffusion are included in the set of accepted models, we emphasise that it does not mean that microscopic diffusion should not be included when modelling the system but that other mixing processes might occur to counteract it. Moreover, those models largely shift the accepted parameter ranges to heavier, older and hydrogen rich -- thus metal and helium poor -- models. This significant difference in composition is a clear illustration of the degeneracy between helium and metal abundances in the helium glitch amplitude.

We previously noted that the only models accounting simultaneously for the complete set of seismic and non-seismic constraints were the model with the OPLIB opacity table and turbulent mixing, for 16 Cygni A, and the one without diffusion for 16 Cygni B. This again hints at the necessity to include non-standard physical processes. However, we must point out that they are incompatible for a joint analysis, as is carried out in the present section, as their ages and compositions are significantly different. Moreover, as we use different opacity tables for both stars, they may not be used simultaneously to analyse the system as a whole.

\begin{table}
\centering
\small
\caption{Accepted stellar range defined as the centroid of the extremum values for each parameter. The uncertainties are the necessary variations to reach those extrema. The accepted models used to define this range are: the reference models, the ones including turbulent mixing, those with overshooting, those without diffusion and those with a temperature profile above the photosphere as in \citet{1981ApJS...45..635V} and with a calibrated $\AMLT$ value.}\label{Tab:AccRan}
\begin{tabular}{ccc}
\hline
Quantity & 16CygA & 16CygB \\
\hline
\hline\\[-0.8em]
 $M~(M_{\odot})$ & $1.08 \pm 0.04$ & $1.03 \pm 0.03$ \\
 $X_0$ & $0.72 \pm 0.05$ & $0.72 \pm 0.04$ \\
 $\left(Z/X\right)_0$ & $0.028 \pm 0.009$ & $0.03 \pm 0.01$ \\
 $Y_0$ & $0.26 \pm 0.05$ & $0.26 \pm 0.05$ \\
 $t~(Gyr)$ & $7.1 \pm 0.5$ & $7.2 \pm 0.4$ \\
\hline
\end{tabular}
\end{table}

\subsection{Binary models}\label{Sec:BinMod}
We now use the individual accepted models to compute models imposing a common age and composition. The adjustment is carried out as in the previous section (that is, with the same set of free parameters and constraints as in Sect. \ref{Sec:ABSep}) only ages and initial compositions are required to be identical, reducing the set of free parameters by three. We use the average values of the initial composition and ages as initial guesses. The set of free parameters is composed of: one value of the age, initial hydrogen and metal fraction for both stars, and a different value of the mass for each star. The set of constraints corresponds to the individual values of the $4$ seismic constraints considered in this paper ($\Delta$, $\hat{r}_{01}$, $\hat{r}_{02}$, and $A_{\textrm{He}}$).

We are not able to provide an exact adjustment of both stars simultaneously (`exact' meaning that the reduced $\chi^2$ value should not exceed a value of $1$). This may result from the fact that the size of the parameters space is reduced by three. Effectively, even though the accepted ages and initial compositions agree within their uncertainties (see Table \ref{Tab:AccRan}), they are not identical and our seismic constraints may be too stringent to allow for an exact simultaneous fit while imposing identical ages and compositions. To illustrate this statement, we have plotted the optimal ages and initial composition of one stellar component against the other for each choice of input physics in Fig. \ref{Fig:CompAB}. We observe in this figure that the three common free parameters almost never simultaneously agree for a given choice of input physics. As further illustration, we may compute the relative difference in the initial metallicity between both stars normalised by the quadratic sum of their uncertainties ($\vert(Z/X)_{0,\textrm{A}}-(Z/X)_{0,\textrm{B}}\vert/\sqrt{\sigma^2\left(Z/X\right)_{0,\textrm{A}}+\sigma^2\left(Z/X\right)_{\textrm{B}}}$). In the most favourable case, we obtain $0.2$, while we get $2.7$ in the least favourable one. This shows that the difference in initial composition for individual models is sometimes significant and may impair the convergence of the Levenberg-Marquardt procedure.

One could argue that our inability to provide an exact adjustment for both stars comes from the reduced number of free parameters. Therefore, we try to include the mixing length parameter into the fitting parameters, allowing it to vary freely and independently for each star. However, this does not improve the results. As a matter of fact, the mixing length parameter value does not significantly vary. We retrieve optimal values of $\alpha_{\textrm{MLT,A}} = 1.8 \pm 0.2$ and $\alpha_{\textrm{MLT,B}} = 1.8 \pm 0.1$ respectively, compared to the fixed solar value of $\AMLT = 1.82$. From \citet{2015A&A...573A..89M} and the solar-twin character of both stars (see also the discussion in Sect. \ref{Sec:FitHR}) one might expect that the mixing length parameter value should remain close to solar.

Although we do not obtain models that exactly reproduce the seismic constraints, in some cases, we may find a reasonable agreement with most of them (but not all, therefore having a reduced $\chi^2$ value greater than $1$). We obtain two sets of convincing results: one for models without diffusion, the other for models with a temperature profile above the photosphere as in \citet{1981ApJS...45..635V} and a corresponding solar calibrated $\alpha_{\textrm{MLT}}=2.02$ value. The optimal model stellar parameters are gathered in Tables \ref{Tab:AB} and \ref{Tab:ABVer}, respectively, and Tables \ref{Tab:IndAB} and \ref{Tab:IndABVer} show the differences between the observed and model seismic indicators normalised to the observed uncertainties. We also show the complete set of 'optimal' model parameters, for each choice of input physics, as well as their relative agreement with the seismic constraints in Tables \ref{Tab:ABAll} and \ref{Tab:IndABAll}. We observe for the models without diffusion in Table \ref{Tab:IndAB} that, out of the $8$ seismic constraints, only the large separation of 16 Cygni A was not properly accounted for. All the other indicators are within the $1\sigma$ uncertainty. For the models with a temperature profile above the photosphere as in \mbox{\citet{1981ApJS...45..635V}}, both the large separation and small frequency ratio between radial and dipolar modes are poorly reproduced.

For the other models, presented in Tables \ref{Tab:ABAll} and \ref{Tab:IndABAll}, we note that it is always the large separation of the A star which is poorly fitted. The other indicators are rather well adjusted but the small separation ratios often fall out, while remaining close, of the $1\sigma$ uncertainty box. This difference in fitting between the several indicators mainly stems from the difference in their relative uncertainties. Indeed, the helium glitch amplitudes have relative uncertainties of about $3\%$ and, by far, are the least stringent constraints. Then come the ratios with relative uncertainties around $0.8\%$. Finally, the $\Delta$ constraint relative uncertainties are of approximately $0.004\%$. Moreover, it was shown by \citet{2019A&A...622A..98F} that the $\hat{r}_{01}$ ratio is mostly sensitive to the evolutionary state of the star, therefore at a given composition and mass, to the stellar age. Then, the large separation is a proxy of the mean stellar density and decreases along the main sequence. As both stars are required to have identical ages and compositions, and with such stringent constraints, we may understand that only one large separation may be fitted at a time and that all the other constraints, from the clear imbalance in the relative uncertainties will adjust to it.
To find a simultaneous agreement for those models that did not reach satisfactory convergence, we either need to relax this assumption, allowing for example different values for the initial composition, or to relax the seismic constraints.

\begin{figure}
\centering
\includegraphics[width=0.95\linewidth]{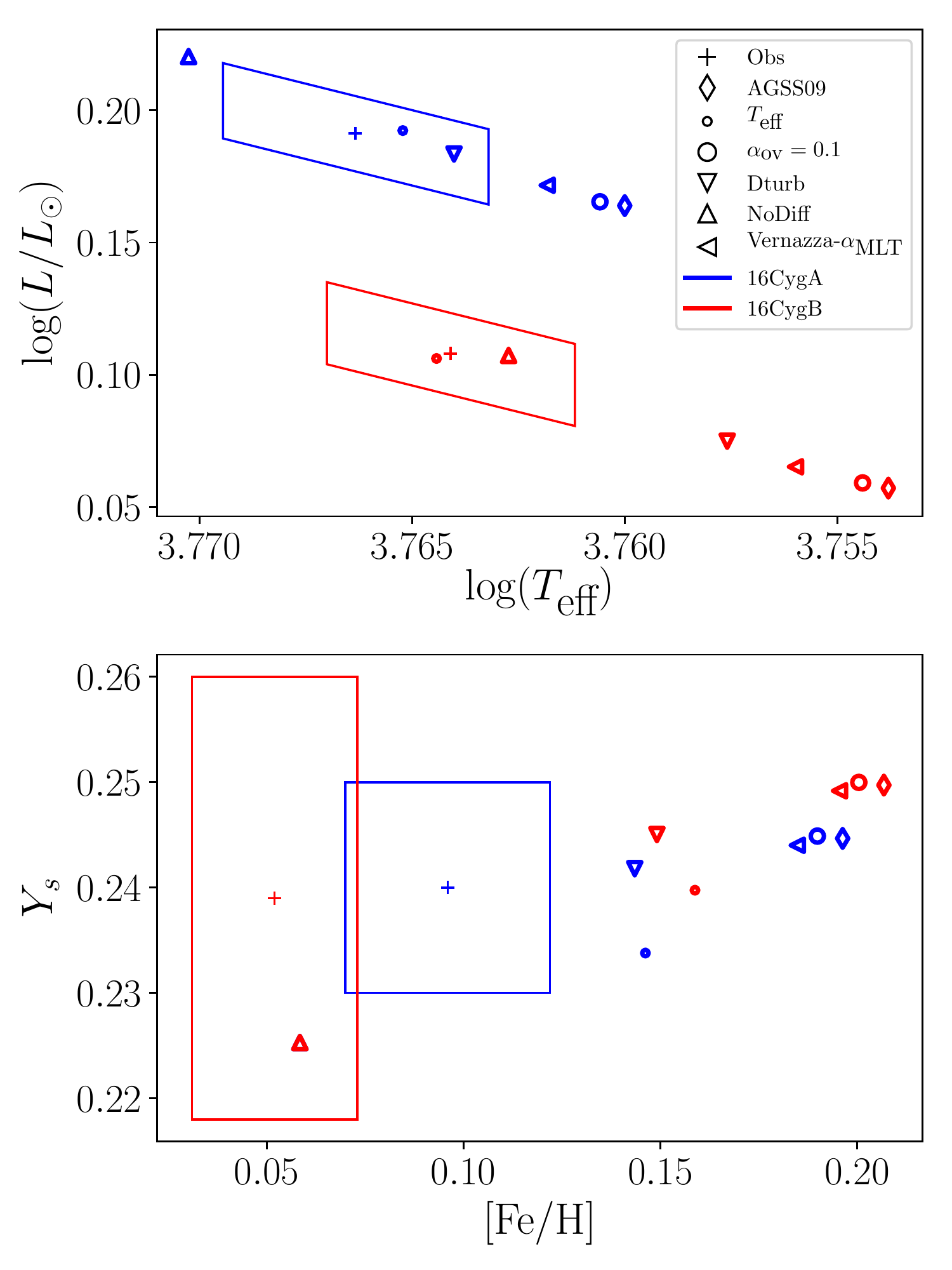}
\caption{Computed values of non-seismic constraints against the observed ones, symbolised by a box, for the system seen as a whole (i.e. with common ages and initial compositions). Each variation in input physics is represented by a different symbol. The colour represents the star, blue for 16 Cyg A and red for 16 Cyg B. The upper panel is a HR diagram. The lower panel shows the surface helium abundance versus the metallicity.}\label{Fig:NonSeiAB}
\end{figure}

We also display in Fig. \ref{Fig:NonSeiAB} the agreement of the models with the non-seismic data for each of the considered variations in input physics, represented by the different symbols. The observed values along with their uncertainties are shown as boxes. We display the results for 16 Cyg A in blue and for 16 Cyg B in red. We note, as in the previous section, that the models for each star are almost constant in radius. We also note that very few models account for the position of the stars in the HR diagram. Actually, those are the models which, individually accounted for these data. It does not come as a surprise as the minimisation aims at finding a compromise between all the seismic constraints for both stars. Then, looking at the lower panel, we note that no model for 16 Cyg A is representative of the surface composition. For 16 Cyg B, only the model without diffusion agrees with these data -- again with no surprise as the individual model already agreed --. Let us add that, as the models without diffusion for both stars must have the same initial composition, their surface compositions are identical and both markers are indistinguishable in the lower panel of the figure.

In a nutshell, we are able to produce two pairs of binary models that are in reasonable agreement with our seismic constraints. However, no model accounts simultaneously for the seismic and non-seismic data of both stars. This may result from a differential effect between both stellar components which could, for example, create differences in compositions as has been discussed by \citet{2019A&A...628A.126M}. This would require the inclusion of non-standard physical processes in the modelling.

\begin{table}
\centering
\small
\caption{Best set of adjusted stellar parameters, models without diffusion, imposing a common age and initial composition for both stars.}\label{Tab:AB}
\begin{tabular}{ccc}
\hline
Quantity & 16CygA & 16CygB \\
\hline
\hline\\[-0.8em]
$M~(M_{\odot})$ & $1.10 \pm 0.01$ & $1.068 \pm 0.004$\\
$\left[\textrm{Fe/H}\right]$ & $0.058 \pm 0.009$ & $0.058 \pm 0.009$ \\
$Y_s$ & $0.225 \pm 0.004 $ & $0.225 \pm 0.004$ \\
$R~(R_{\odot})$ & $1.24 \pm 0.01$ & $1.114 \pm 0.004$ \\
$X_0$ & \multicolumn{2}{c}{$0.759 \pm 0.004$}  \\
$\left(Z/X\right)_0$ & \multicolumn{2}{c}{$0.0207 \pm 0.0004$}  \\
$Y_0$ & \multicolumn{2}{c}{$0.225 \pm 0.004$}  \\
$t~(Gyr)$ & \multicolumn{2}{c}{$7.50 \pm 0.05$}  \\
\hline
\end{tabular}
\end{table}

\begin{table}
\centering
\small
\caption{Best set of adjusted stellar parameters, models with a temperature profile above the photosphere as in \citet{1981ApJS...45..635V} and with $\alpha_{\textrm{MLT}}=2.02$, imposing a common age and initial composition for both stars.}\label{Tab:ABVer}
\begin{tabular}{ccc}
\hline
Quantity & 16CygA & 16CygB \\
\hline
\hline\\[-0.8em]
$M~(M_{\odot})$ & $1.054 \pm 0.006$ & $1.016 \pm 0.006$ \\
$\left[\textrm{Fe/H}\right]$ & $0.18 \pm 0.01$ & $0.19 \pm 0.01$ \\
$Y_s$ & $0.244 \pm 0.003 $ & $0.249 \pm 0.003$ \\
$R~(R_{\odot})$ & $1.216 \pm 0.007$ & $1.105 \pm 0.006$ \\
$\alpha_{MLT}$ & \multicolumn{2}{c}{$2.02$} \\
$X_0$ & \multicolumn{2}{c}{$0.682 \pm 0.005$}  \\
$\left(Z/X\right)_0$ & \multicolumn{2}{c}{$0.0352 \pm 0.0007$}  \\
$Y_0$ & \multicolumn{2}{c}{$0.293 \pm 0.005$}  \\
$t~(Gyr)$ & \multicolumn{2}{c}{$6.82 \pm 0.05$}  \\
\hline
\end{tabular}
\end{table}

\begin{table}
\centering
\small
\caption{Differences between theoretical, model without diffusion, and observed values for the seismic constraints defined as $\delta = \vert I_{\textrm{obs}}-I_{\textrm{th}}\vert$, in the units of the constraint. $\chi^2_{\textrm{red}} = \chi^2/(N-k)$ is the reduced $\chi^2$ value where $N$ is the number of constraints to the fit and $k$ the number of free parameters.}\label{Tab:IndAB}
\begin{tabular}{ccccc}
\hline
Quantity & \multicolumn{2}{c}{16CygA} & \multicolumn{2}{c}{16CygB} \\
 & $\delta$ & $\delta/\sigma$ & $\delta$ & $\delta/\sigma$ \\
\hline
 $\Delta~\left(\mu Hz\right)$ & $0.01$ & $3.0$ & $3 \cdot 10^{-5}$ & $0.007 $ \\
 $A_{\textrm{He}}$ & $0.7$ & $0.7$ & $0.3$ & $0.3$ \\
 $\hat{r}_{01}$ & $10^{-4}$ & $0.4$ & $10^{-4}$ & $0.7$ \\
 $\hat{r}_{02}$ & $10^{-4}$ & $0.4$ & $10^{-4}$ & $0.4$ \\
 $\chi^2_{\textrm{red}}$ & \multicolumn{4}{c}{$3.4$} \\
\hline
\end{tabular}
\end{table}

\begin{table}
\centering
\small
\caption{Same as in Table \ref{Tab:IndAB}, but for models model with a temperature profile above the photosphere as in \citet{1981ApJS...45..635V} and $\alpha_{\textrm{MLT}}=2.02$.}\label{Tab:IndABVer}
\begin{tabular}{ccccc}
\hline
Quantity & \multicolumn{2}{c}{16CygA} & \multicolumn{2}{c}{16CygB} \\
 & $\delta$ & $\delta/\sigma$ & $\delta$ & $\delta/\sigma$ \\
\hline
\hline\\[-0.8em]
 $\Delta~\left(\mu Hz\right)$ & $6 \cdot 10^{-3}$ & $1.3$ & $7 \cdot 10^{-4}$ & $0.1$  \\
 $A_{\textrm{He}}$ & $0.5$ & $0.5$ & $0.6$ & $0.6$ \\
 $\hat{r}_{01}$ & $6 \cdot 10^{-4}$ & $1.9$ & $4 \cdot 10^{-4}$ & $1.1$ \\
 $\hat{r}_{02}$ & $3 \cdot 10^{-4}$ & $1.1$ & $2 \cdot 10^{-4}$ & $0.8$ \\
 $\chi^2_{\textrm{red}}$ & \multicolumn{4}{c}{$2.9$} \\
\hline
\end{tabular}
\end{table}

\subsection{Relaxing the common composition hypothesis}\label{Sec:DifCom}
We noted in the previous section that simultaneously providing models of both stars while imposing them to have identical initial compositions as well as ages is a difficult task. In most cases, this resulted in our inability to build such models. Therefore, we try to model the system requiring only an identical age for both stars. The set of free parameters is thus made of the age, the individual initial hydrogen, and metal abundances and the individual masses. This adds up to $7$ free parameters. The constraints are the seismic indicators used throughout this paper which represent $8$ constraints.

We compute models for the several sets of input physics considered in the previous section. The individual stellar parameters of those models are given in Table \ref{Tab:ABComp}. To quantify whether the improvement of the results is significant given the increased number of parameters, we introduce the Bayesian Information Criterion (BIC) \citep{schwarz1978}. It allows to compare models of different dimensionalities and provides a criterion for making a selection. It has the advantage over the simple $\chi^2$ value to penalise over the number of fitting parameters and may pinpoint overfitting models. Under the assumption that model errors are independent and normally distributed, it takes the form:
\begin{equation}
\textrm{BIC} = \chi^2+k~ln\left(N\right),
\end{equation}\label{Eq:BIC}
with the $\chi^2$ value as defined previously, $k$ the number of free parameters and $N$ the number of constraints.
When comparing different models, the key ingredient is not the BIC value itself but rather the difference between values -- one must however keep in mind that the lower the value, the better --. For the BIC difference to be significant, it has to exceed $2$. BIC differences from $6$ and above will be regarded as strongly significant \citep{doi:10.1080/01621459.1995.10476572}. In this specific case, we compare models from the previous section that had $5$ free parameters and $8$ constraints, the added term to the $\chi^2$ value thus equals $10.4$. In the current section, it is equal to $14.6$. The difference is $4.2$ which the $\chi^2$ improvement has to exceed so that the model may be regarded as improved.

Looking at the models in Table \ref{Tab:ABComp}, we observe that the relative difference in compositions between the two stars is very small (the maximum difference reaches up to $0.07\textrm{dex}$ in $X_0$ and $0.002\textrm{dex}$ in $\left(Z/X\right)_0$). However, that variation alone is able to greatly improve the BIC values, as the comparison of Tables \ref{Tab:IndABAll} and \ref{Tab:IndABComp} shows. Only two models were not significantly improved. The model considering the mixing length parameters calibrated over the effective temperature of the stars could not be improved. Also, the one using the temperature profile of \citet{1981ApJS...45..635V} above the photosphere and a calibrated value of $\AMLT$ is not improved, BIC-wise (the raw $\chi^2$ value did decrease of about $1.7$). The BIC variation is of about $2$ which makes this difference barely relevant. Therefore, in most cases, it is relevant to allow the composition to slightly vary between the two stars. This again points toward the necessity to include non-standard physical processes. We also note that it is often the case that the uncertainties on the individual stellar parameters are degraded. Finally, we note that, in most cases, 16 Cyg B is initially richer in metals and poorer in hydrogen than its twin. This validates the trend we observe in the middle and lower panels of Fig. \ref{Fig:CompAB}. Therefore, we observe that, seismically speaking, the B component of the system is more metallic than the A, which is in opposition with the spectroscopic observations (see Table \ref{Tab:Const}).

\section{Discussion}\label{Sec:Dis}
\subsection{General considerations about the approach}
Compared to other asteroseismic approaches aiming at the modelling of solar-like pulsators, the greatest difference of our approach is that we do not use directly the complete set of individual frequencies or of individual frequency ratios as constraints. Instead, we build seismic indicators, via the \who method, which are as little correlated as possible and relevant to the stellar structure. Moreover, the search for optimal models is carried out by minimising a single cost function comparing simultaneously theoretical seismic and non-seismic constraints to observed data. This has the advantage of avoiding unnecessary correlations.  

Furthermore, the direct use of the helium glitch amplitude in our modelling also makes up the peculiarity of our approach. Indeed, in several other studies, it is not used directly as a constraint to the modelling. For example, \citet{2014ApJ...790..138V,2019MNRAS.483.4678V} calibrate the model helium glitch amplitude with respect to the surface helium abundance in a set of optimal models representative of other constraints (namely individual frequencies, ratios, effective temperature and metallicity) to provide an estimate. In the present case, including the helium glitch amplitude as a constraint to the fit acts as a constraint on the model helium abundance, with some correlation with the metal content as showed in Sect. \ref{Sec:ConStePar}. This means that we do not assume a specific relation between the two quantities and the resulting helium abundance stems from the best model search only. 

We noted that providing models of both stars while requiring a common age and composition proves to be an arduous task. For most choices of micro- and macro-physics, we are not able to produce satisfactory models. However, when completely inhibiting the microscopic diffusion of chemical elements or imposing a temperature relation above the photosphere following the prescription of \citet{1981ApJS...45..635V} while using a specifically calibrated value of the mixing length parameter, we obtained a reasonable agreement, but did not go below the value of $1$ for the reduced $\chi^2$. On the first hand, the fact that including extra mixing produces better results illustrates the need to include non-standard physical processes to properly model such complex data. On the other hand, the improvement of the results while using another temperature profile above the photosphere demonstrates the impact of the surface effects on the seismic indicators. Even though those indicators were defined in such a way to lessen this effect at most. We must also add that non-seismic constraints are not simultaneously accounted for in both stars (see Fig. \ref{Fig:NonSeiAB}). This clearly illustrates the need to include such constraints in the fitting procedure as well as additional physical processes. Some of the possible non-standard physical processes that could be included in the modelling are discussed in Sect. \ref{Sec:DisPhys}. Another way to improve models of the system as a whole is to relax the hypothesis of a common composition. This leads to small differences in composition between both stars (never exceeding $0.07\textrm{dex}$ in $X_0$ and $0.002\textrm{dex}$ in $\left(Z/X\right)_0$). This is however sufficient in many cases to improve the results significantly as the BIC values testify.

\subsection{Impact of the surface effects}\label{Sec:SurEff}
The impact of the surface effects correction of the frequencies on the optimal model is not clear. We computed models that adjusted seismic indicators built over the uncorrected frequencies. Those are displayed in \mkcolor{NoCorr} in the figures throughout the paper. Although it is barely significant we note that those models are heavier and older than models with corrected frequencies.

Furthermore, we noted in Sect. \ref{Sec:BinMod} that the large separation we compute is the most stringent of our indicators with a relative uncertainty of approximately $0.004\%$. However, this specific value of the uncertainty, resulting from the error propagation of the individual frequencies, is unrealistic. To provide a more robust estimate, we can quantify the contribution of the surface effects by computing the difference between corrected and uncorrected values. We obtain $\sigma\left( \Delta_A\right)=0.9~\mu Hz$ and $\sigma\left( \Delta_B\right)=1.0~\mu Hz$, for 16 Cyg A and B respectively, which amounts to $1.1\%$ relative uncertainty. This shows that, even though we build seismic indicators in such way that makes them the less dependent on the surface effects as possible, they are still impacted by the surface conditions.

The impact of the surface effects correction is further illustrated in App. \ref{Sec:NoCorr} where we give the values of the seismic indicators for uncorrected frequencies and display the best-fit models \emph{\'echelle} diagrams of both stars. We also compute models with frequencies corrected according to the prescription of \citet{2014A&A...568A.123B}. The optimal parameters are given in Table \ref{Tab:Ball}. We observe, compared to the reference models, that both stars become older and hydrogen rich. We also note that, while the A component becomes more metallic, the B one is then less metallic. 16 Cygni B also becomes heavier while it is not the case for its twin. Nonetheless, the differences are such that we could include those models in the set of accepted ones.

\subsection{Non-standard physical processes and modelling improvements}\label{Sec:DisPhys}
As was shown by \citet{2019MNRAS.483.4678V} and as we illustrate in Sect. \ref{Sec:Dif}, diffusion is very important in the modelling of solar-like stars. However, the models computed with CLES currently consider only three chemical species, the hydrogen, the helium and the metals. A significant improvement would stem from the consideration of several sub-species in the metals. For example, we may retrieve invaluable information by following the lithium evolution with the star. Indeed, the discussion regarding the effects of potential non-standard processes is tightly linked to the lithium abundance in both stars and its connection with the formation and orbital evolution of planetary systems. \citet{2015A&A...584A.105D} have proposed that, since the B component was orbited by a Jovian planet \citep{1997ApJ...483..457C}, accretion of matter from the planetary disc in the envelope of 16CygB would have triggered fingering convection and thus led to a strong decrease in the lithium abundance. They determined that the accretion of $0.66M_{\oplus}$ would be enough to reproduce the lithium abundance of 16CygB that is 4 times lower than that of 16CygA \citep{1993A&A...274..825F,1997AJ....113.1871K}, despite both stars having very similar rotational and structural properties. It is also interesting to note that, in the broader context of the Li abundance of solar-twins, 16CygA seems to be more Li-rich than similar solar-twins, while 16CygB seems to follow the trend observed with age for these stars \citep{2016A&A...587A.100C}. This could suggest a reverse scenario that should explain the high lithium abundance of 16CygA and in particular the possibility of an increase in lithium abundance related to planet engulfment (e.g. \citealt{2002A&A...386.1039M,2016A&A...587A.100C}). In this context, analysing both scenarios in light of potential traces of such events in seismic indicators may lead to new synergies between asteroseismology and exoplanetology, namely in the analysis of planetary formation and material accretion onto the surface of planet-host stars.

Furthermore, our models do not include rotation, nor rotation-induced mixing. As a matter of fact, \citet{2015MNRAS.446.2959D} and \citet{2019A&A...623A.125B} showed that rotation indeed occurs in both stars taking advantage of the rotational splitting present in stellar oscillation spectra of rotating stars. \citet{2019A&A...623A.125B} even showed that differential rotation occurs in both stars, in a similar way to our Sun. Such a process could  significantly affect the diffusion of chemical elements. This could be argued to be a flaw in our models. However, the additional mixing induced may be approximated by the turbulent mixing of elements as was performed for several models in this study. This prescription consists of an approximation and the models may still be improved. Therefore, we may, in the future, use improved models which account for the rotation. This will be discussed in further papers of the series.

Finally, we noted that choosing a different opacity profile, that of the Los Alamos project, while including turbulent mixing of elements, which counteracts diffusion, allowed to reproduce both seismic and non seismic constraints for 16 Cygni A. This provides clues that we may need to modify the opacity profile of the star to properly account for all the observed constraints. Therefore, inversion techniques could help us to further improve our models. However, the OPLIB opacity table has two different effects on the stellar structure. On the one hand, it modifies slightly the size of the convective envelope. On the other hand, it changes the temperature gradient in the central regions. According to which effects dominates this could be the illustration of the need of non-standard mixing processes as well.

\section{Conclusion}\label{Sec:Con}
With the aim of characterising the 16 Cygni system as thoroughly as possible, we took advantage of the seismic indicators defined via the \who method to provide stringent constraints on the stellar structure and test several choices of micro- and macro-physics. We built those indicators using the frequencies computed over the full length of the \emph{Kepler} data by \citet{2015MNRAS.446.2959D} and corrected for the surface effects according to \citet{2008ApJ...683L.175K}'s power law adjusted by \citet{2015A&A...583A.112S}. The several choices of micro- and macro-physics used in stellar models we tested are: the solar reference mixture, opacity and equation of state tables as well as the inclusion of turbulent mixing or of diffusion of chemical elements, a different choice for the mixing length parameter, the inclusion of overshooting outside of convective regions, a different choice of temperature profile above the photosphere or the effect of the correction of the frequencies for the surface effects. 

Overall, our results agree with previous studies with slight differences according to the choice of physics included in the models. We showed that the use of the \who indicators allows to discriminate between several of those choices. However, we also note that those indicators alone do not suffice to provide a complete adjustment of the stars as, in most cases, the non-seismic constraints (i.e. luminosity, effective temperature, and metallicity) are not satisfied. Therefore, they need to be included in the fitting process to provide the most representative model. 
We show in Sec. \ref{Sec:FitHR} that the mixing length parameter has a clear effect on the modelled effective temperature of the star. Indeed, a value greater than the solar one allows to greatly improve the agreement with the observed value in the case of 16 Cyg B. Moreover, we also study the impact of the inclusion of turbulent mixing and show that it leads to a better fit of the effective temperature and metallicity. However, we observe that the turbulent mixing coefficient saturates and using it as a free parameter of the modelling procedure would be meaningless. Therefore, using a free $\AMLT$ and $T_{\textrm{eff}}$ as a constraint, we are able to produce models in agreement with this constraint. However, the observed metallicities are not reproduced by those models. We also demonstrate in Sec. \ref{Sec:FitFeH} that varying the mixing length parameter while using the metallicity constraint allows to better reproduce its value for both stars but at the cost of the agreement with the observed effective temperatures. This illustrates the necessity to build more complex models in order to reproduce both seismic and non-seismic constraints.

Indeed, we show that, to reproduce the non-seismic constraints, we have to select only specific choices of input physics or even include non-standard physical processes. Indeed, for 16 Cygni A, only a model with a modified opacity profile, from the Los Alamos Opacity Project, and including turbulent mixing of the chemical elements reproduced all constraints. For 16 Cygni B, it was a model that did not include diffusion that was able to account for these constraints. This illustrates in both cases that non-standard physical processes may be necessary to inhibit diffusion and to properly models those stars. Such processes could be the accretion of planetary matter or rotation-induced mixing.

Adjusting both stars simultaneously while imposing a common age and composition proved to be a difficult task. In most cases, with models that were consistent within mutual uncertainties for both stars as initial guesses, we could not obtain a satisfactory adjustment. The large separation of 16 Cyg B, because of the high precision of the $\Delta$ constrain defined in our study, was dominant and the free parameters adapted at best to provide a compromise for the other constraints. This resulted in models that did not exactly fit the large separation of 16 Cygni A while reasonably fitting other constraints (only a few $\sigma$ difference from the required value). Nonetheless, for models without diffusion or with a different temperature profile above the photosphere (that of \citealp{1981ApJS...45..635V}) and a calibrated value of $\AMLT$, we were able to account for the seismic constraints of both stars with a reduced $\chi^2$ of $3.4$ and $2.9$ respectively.
The difficulty to provide satisfactory models of both stars with other choices of input physics indicates that it can be necessary to either relax the common initial composition assumption, the seismic constraints or to invoke special physical processes. For example, we showed in Sect. \ref{Sec:BinMod} that the differences in the initial metallicity between both stellar components of optimal individual models may sometimes be significant. Therefore, we computed models relaxing relaxing the common composition hypothesis in Sec. \ref{Sec:DifCom} and were able to significantly improve the results. We observe that a small difference between the initial compositions of both stars is sufficient.

With the aim of providing a broad sample of reliable models of the system, the extensive analysis of the degeneracies carried out by combining seismic and non-seismic constraints is of prime importance to fully grasp the uncertainties of inverse analysis but also to the extent in which we can constrain physical processes not implemented in standard stellar models linked for example to the effects of accretion of planetary matter, angular momentum transport and their link to both seismic indices and the lithium and beryllium abundances of both stars.

We showed that, even for our models that reproduced both seismic and non-seismic constraints, information remain to be analysed as we observe that other indicators defined in \citet{2019A&A...622A..98F} are not properly represented (i.e. $\hat{\epsilon}$, $\Delta_{01}$, $\Delta_{02}$, and $A_{\textrm{CZ}}$). Further studies could focus on those other constraints.
Finally, as the \who method also proves to provide very stringent seismic constraints, we will, in future studies, undertake the adjustment of the Kepler LEGACY sample \citep{2017ApJ...835..172L} to try and retrieve global trends in solar-like oscillators. This data set contains the best set of solar-like oscillation spectra available to the community to this day as it is composed of 66 solar-like stars which have been continuously observed from space for at least one year. Therefore, we would be able to realise an ensemble study of stellar parameters. For example, we could study the evolution of the amount of central overshooting with the stellar mass.

\begin{acknowledgements}
The authors would like to thank the referee for their careful reading of the paper as well as for their very constructive remarks.\\
M.F. is supported by the FRIA (Fond pour la Recherche en Industrie et Agriculture) - FNRS PhD grant.\\
G.B. acknowledges fundings from the SNF AMBIZIONE grant No 185805 (Seismic inversions and modelling of transport processes in stars).\\
C.Pin\c{c}on is supported by the F.R.S - FNRS  as a Charg\'e de Recherche.\\
P.E. and S.J.A.J.S. have received funding from the European Research Council (ERC) under the European Union's Horizon 2020 research and innovation programme (grant agreement No 833925, project STAREX).\\
C.Pezzotti is sponsored by the Swiss National Science Foundation (project number $200020-172505$).
\end{acknowledgements}

\bibliographystyle{aa}
\bibliography{bibli}

\begin{appendix}
\section{\who decomposition}\label{Sec:WhoDec}
We describe here the basis of functions which are used to represent the oscillation spectrum of a solar-like pulsator.
First, the smooth part of the spectrum is represented by a second-order polynomial of $n$, the radial order. We thus have the following succession of polynomials:
\begin{equation}
p_{lk}\left(n,l'\right) = \delta_{ll'}p_{k}\left(n\right),
\end{equation}
with $p_k\left(n\right) = n^k$, $k=0,1,2$ and $\delta_{ll'}$ the \emph{Kronecker} delta comparing two values of the spherical degree $l$ and $l'$.

Then, the helium glitch is described by the following oscillating functions:
\begin{equation}
\delta\nu_{\textrm{He}} = \sum\limits_{k=-4}^{-5}\left[\cos\left( 4\pi \tau\Delta\nu \tilde{n} \right) + \sin\left( 4\pi \tau\Delta\nu \tilde{n} \right) \right]\tilde{n}^k,
\end{equation}
where $\tau$ is the acoustic depth of the  glitch, $\Delta\nu$ the asymptotic large frequency separation and $\tilde{n}=n+l/2$. $\tilde{n}\Delta\nu$ is actually the first order approximation of $\nu_{l,n}$. The asymptotic large frequency separation is defined as $\Delta\nu = \left(2\int^{R_*}_0 \frac{dr}{c\left( r \right)} \right)^{-1}$ \citep{1980ApJS...43..469T} , with the local radius $r$, the local sound speed $c\left(r\right)$ and $R_*$ the radius of the star at the photosphere.

We must add that the values of $\tau$ and  $\Delta\nu$ are estimated via a model that is representative of the seismic indicators of the smooth part, namely $\Delta$, $\hat{r}_{01}$ and $\hat{r}_{02}$. \citet{2019A&A...622A..98F} showed that the exact value of $\tau \Delta\nu$ has a negligible impact on the amplitude of the glitch. A $10\%$ percent excursion from the optimal value is of negligible impact.

Finally, orthonormalisation of the basis function is carried out via \emph{Gram-Schmidt}'s process. This produces the orthonormal elements over which we may project the frequencies to represent them. We thus retrieve completely independent coefficients.
\FloatBarrier

\section{Additional seismic indicators}\label{Sec:AddSeiInd}
In the present section, we describe supplementary seismic indicators defined in the \who method but that are not used directly in our optimisation for a model representative of the 16 Cygni system. The values of those indicators are given in Table \ref{Tab:AddSeiInd}.

\paragraph{$\hat{\epsilon}$:}
In taking inspiration in the asymptotic formulation of the frequencies \citep{1986HiA.....7..283G}
\begin{equation}
\nu(n,l) \simeq \left(n+\frac{l}{2}+\epsilon\right)\Delta,
\end{equation}
we may construct a vector subspace over which frequencies are represented by the function:
\begin{equation}\label{Eq:eps}
\nu(n,l) = \left(n+\frac{l}{2}+\epsilon\right) \hat{\Delta} = \left(n+\frac{l}{2}\right)\hat{\Delta} + K,
\end{equation}
where $\hat{\Delta}$ and $K$ are free parameters.
By defining an orthonormal basis over this subspace, projecting the frequencies and identifying the several coefficients with the asymptotic expression we may get an expression for $\hat{\epsilon}$.

\paragraph{$A_{\textrm{CZ}}$:}
\begin{equation}
A_\textrm{CZ}~=~\Vert \boldsymbol{\delta\nu_\textrm{CZ}} \Vert,
\end{equation}
where $\boldsymbol{\delta\nu_\textrm{CZ}}$ is the base of the convection zone glitch component.

\paragraph{$\Delta_{0l}$:}
Corresponds to the slope of the individual frequency ratios $r_{0l}$ as a function of the radial order $n$ and is defined as:
\begin{equation}
\Delta_{0l} = \frac{\Delta_{l}}{\Delta_{0}}-1.
\end{equation}

\begin{table}[h]
\centering
\small
\caption{Additional observed seismic indicators. The standard deviations result from the propagation of the uncertainties on the observed frequencies.}\label{Tab:AddSeiInd}
\begin{tabular}{ccc}
\hline
Indicator & 16CygA & 16CygB \\ 
\hline
\hline\\[-0.8em]
$\hat{\epsilon}$ & $1.3288 \pm 0.0009$ & $1.3583 \pm 0.0008$  \\
$A_{\textrm{CZ}}$ & $2 \pm 1$  & $2 \pm 1$ \\
$\Delta_{01}$ & $\left(4.64 \pm 0.09\right)\cdot 10^{-3}$ & $\left(4.48 \pm 0.08\right)\cdot 10^{-3}$ \\
$\Delta_{02}$ & $\left(5.9 \pm 0.2\right)\cdot 10^{-3}$ & $\left(5.4 \pm 0.1\right)\cdot 10^{-3}$ \\
\hline 
\end{tabular}
\end{table}
\FloatBarrier

\section{Impact of high uncertainties modes}\label{Sec:Freq}
\begin{figure}
\centering
\includegraphics[width=0.95\linewidth]{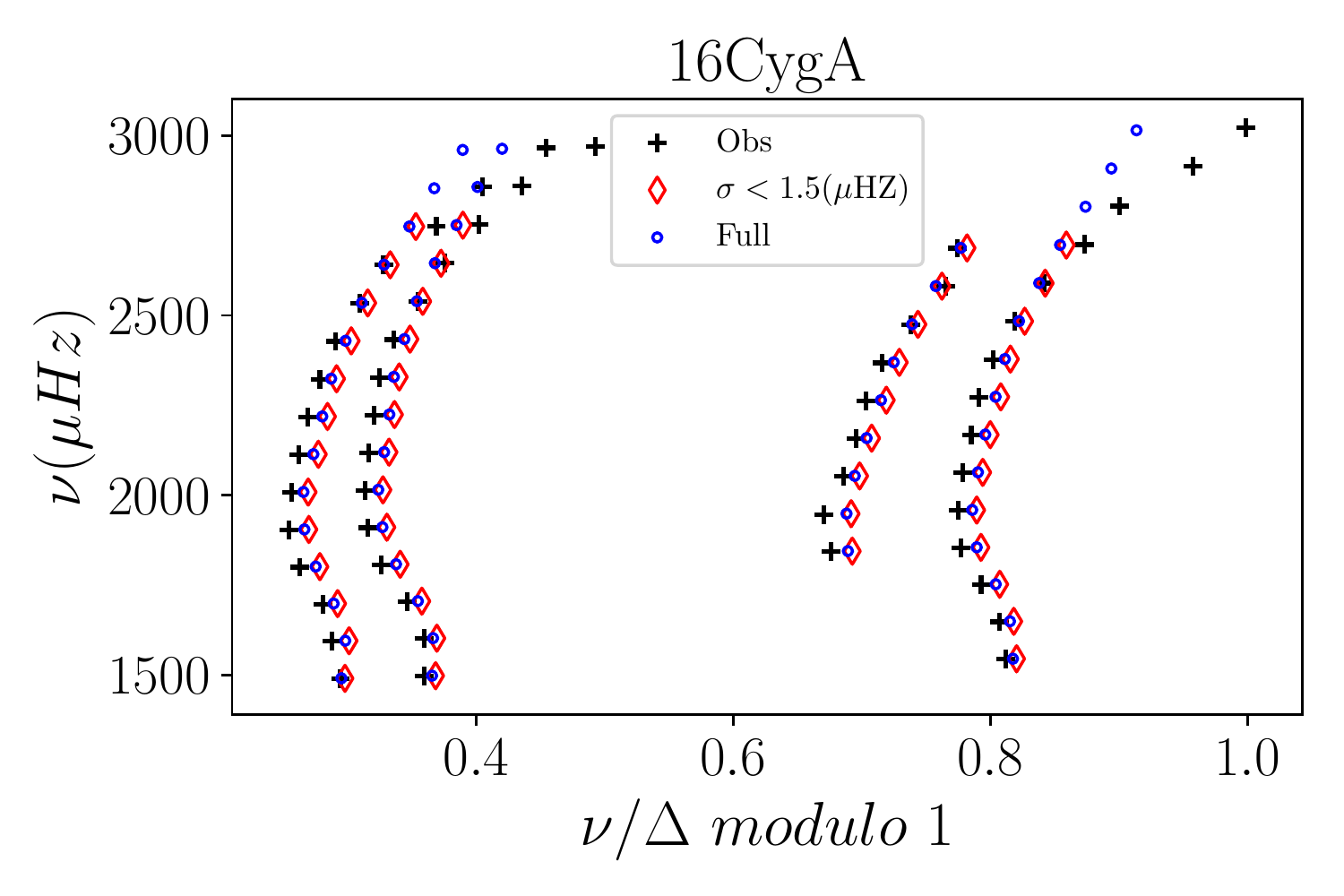}
\caption{\emph{\'Echelle} diagram of 16 Cygni A comparing optimal reference model with full set of frequencies (blue circles) from \citet{2015MNRAS.446.2959D} (black crosses) with the optimal model with a set restricted to frequencies with uncertainties lower than $1.5~\mu Hz$ (red diamonds).}\label{Fig:EchANBoun}
\end{figure}

\begin{figure}
\centering
\includegraphics[width=0.95\linewidth]{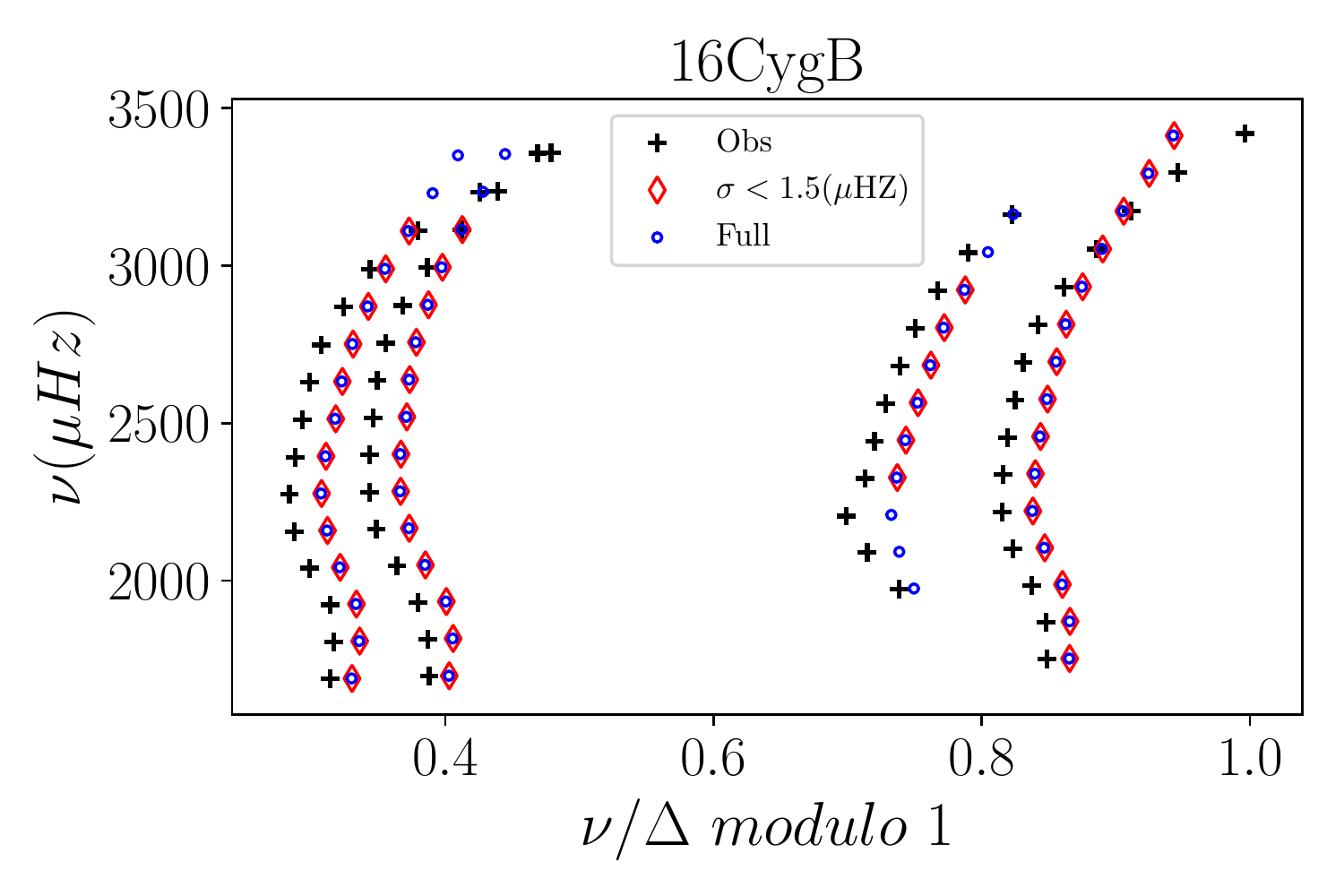}
\caption{\emph{\'Echelle} diagram of 16 Cygni B comparing optimal reference model with full set of frequencies (blue circles) from \citet{2015MNRAS.446.2959D} (black crosses) with the optimal model with a set restricted to frequencies with uncertainties lower than $1.5~\mu Hz$ (red diamonds).}\label{Fig:EchBNBoun}
\end{figure}

From the modes computed by \citet{2015MNRAS.446.2959D} we select those with uncertainties below $1.5~\mu Hz$. Those high uncertainty modes have a limited and negligible impact on the results as our indicators are averaged over the whole spectrum. Furthermore, as the high frequency modes are the ones which are the most affected by the surface effects this may render our results more robust. In Table \ref{Tab:SteParNBound} we show the optimal set of stellar parameters retrieved when using the full set of frequencies. We observe that the results do not vary significantly from the case considering modes with uncertainties below the $1.5~\mu Hz$ threshold, presented in Table \ref{Tab:StePar} under the label AGSS09. Only the uncertainties on the individual parameters are affected which should not be of any concern as the dominant factor remains the choice of input physics. As further validation of this choice, we display in Figs. \ref{Fig:EchANBoun} and \ref{Fig:EchBNBoun} a comparison of the \emph{\'echelle} diagram of optimal models of both stars using the complete set of frequencies (blue circle) with the ones with the reduced set (red diamond) and the observations (black crosses). We observe that the results do not significantly differ, only that the high frequency drift is more visible as more frequencies are displayed.

\begin{table}
\centering
\small
\caption{Stellar parameters retrieved with the reference set of input physics and the complete set of frequencies.}\label{Tab:SteParNBound}
\begin{tabular}{ccc}
\hline
Quantity & 16CygA & 16CygB \\
\hline
\hline\\[-0.8em]
$M~(M_{\odot})$ & $1.06 \pm 0.01$  & $1.011 \pm 0.006$ \\
$X_0$ & $0.684 \pm 0.009$ & $0.679 \pm 0.006$ \\
$\left(Z/X\right)_0$ & $0.035 \pm 0.004$ & $0.037 \pm 0.002$ \\
$Y_0$ & $0.292 \pm 0.009$ & $0.296 \pm 0.006$\\
$\left[\textrm{Fe/H}\right]$ & $0.19 \pm 0.05$ & $0.21 \pm 0.03$ \\
$Y_s$ & $0.243 \pm 0.005$ & $0.251 \pm 0.003$ \\
$R~(R_{\odot})$ & $1.22 \pm 0.01$ & $1.104 \pm 0.006$ \\
$t~(Gyr)$ & $6.8 \pm 0.1$ & $6.97 \pm 0.07$ \\
$\chi^2$ & $1.0$ & $0.7$ \\
\hline
\end{tabular}
\end{table}

\FloatBarrier

\section{Influence of the surface effects}\label{Sec:NoCorr}
\begin{figure}[h]
\centering
\includegraphics[width=0.95\linewidth]{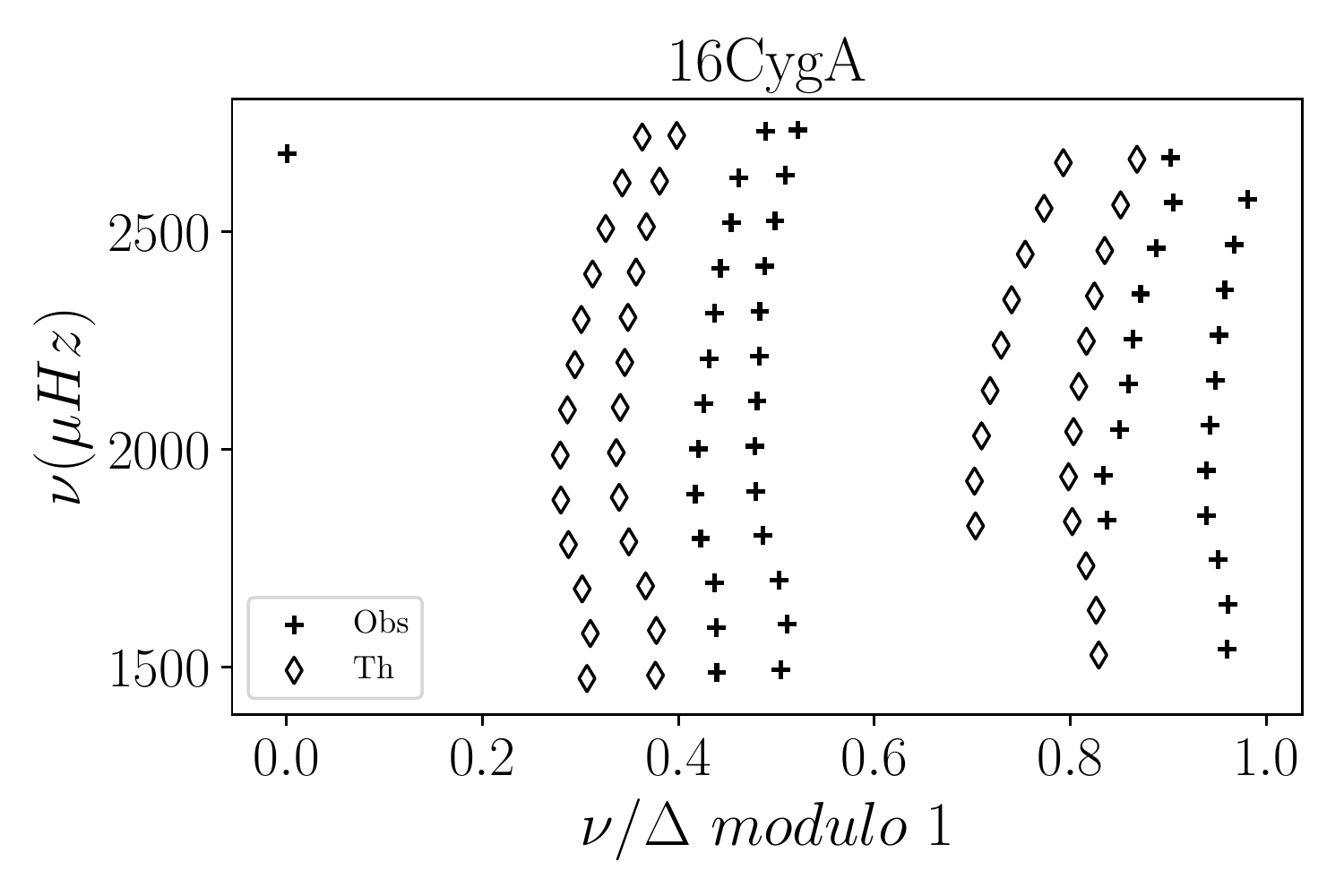}
\caption{\emph{\'Echelle} diagram of 16 Cygni A optimal model calculated with seismic indicators defined over frequencies which are not corrected for the surface effects. The crosses are the observed frequencies and the diamonds the theoretical ones.}\label{Fig:EchANoCorr}
\end{figure}

\begin{figure}[h]
\centering
\includegraphics[width=0.95\linewidth]{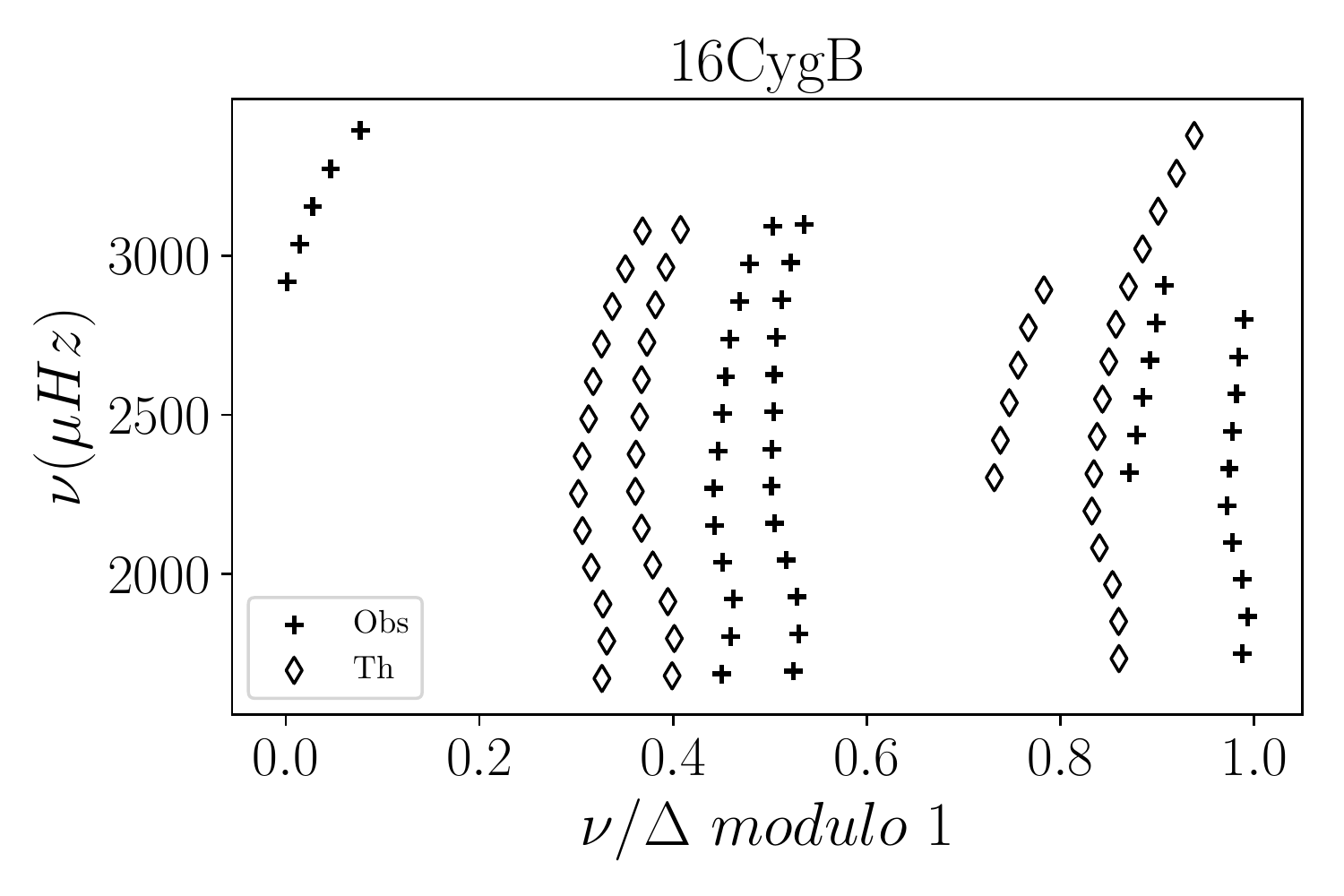}
\caption{\emph{\'Echelle} diagram of 16 Cygni B optimal model calculated with seismic indicators defined over frequencies which are not corrected for the surface effects. The crosses are the observed frequencies and the diamonds the theoretical ones.}\label{Fig:EchBNoCorr}
\end{figure}

We computed both models for frequencies which were not corrected for the surface effects or with theoretical frequencies corrected as in \citet{2014A&A...568A.123B} with the adjusted relation in large separation, effective temperature, surface gravity and opacity of \citet{2018A&A...620A.107M}. Table \ref{Tab:SeiIndNoCorr} displays seismic indicators computed with frequencies which are not corrected for the surface effects. The \emph{\'echelle} diagram for the models computed with these set of indicators are displayed in Figs \ref{Fig:EchANoCorr} and \ref{Fig:EchBNoCorr}. The most striking feature is the large shift between theoretical and observed ridges in both figures which shows that the $\hat{\epsilon}$ indicator is not well accounted for as Fig \ref{Fig:IndA} and \ref{Fig:IndB} show. Moreover, the overall shape of the individual ridges is well represented by the theoretical frequencies.
Table \ref{Tab:Ball} displays the set of optimal parameters for models using frequencies corrected as in \citet{2014A&A...568A.123B}.

\begin{table}[h]
\centering
\small
\caption{Observed seismic indicators with frequencies uncorrected for surface effects. The standard deviations result from the propagation of the uncertainties on the observed frequencies.}\label{Tab:SeiIndNoCorr}
\begin{tabular}{ccc}
\hline
Indicator & 16CygA & 16CygB \\ 
\hline
\hline\\[-0.8em]
$\Delta (\mu Hz)$ & $103.070 \pm 0.005$ & $116.706 \pm 0.004$  \\
$A_{\textrm{He}}$ & $30 \pm 1$  & $33 \pm 1$ \\
$\hat{r}_{01}$ & $\left(3.61 \pm 0.02\right)\cdot 10^{-2}$ & $\left(2.55 \pm 0.02\right)\cdot 10^{-2}$ \\
$\hat{r}_{02}$ & $\left(5.61 \pm 0.03\right)\cdot 10^{-2}$ & $\left(5.41 \pm 0.03\right)\cdot 10^{-2}$ \\
\hline 
\end{tabular}
\end{table}

\begin{table}[h]
\centering
\small
\caption{Adjusted stellar parameters with theoretical frequencies corrected as in \citet{2014A&A...568A.123B} with the adjusted relation in \citet{2018A&A...620A.107M}.}\label{Tab:Ball}
\begin{tabular}{ccc}
\hline
Quantity & 16CygA & 16CygB \\
\hline
\hline\\[-0.8em]
$M~(M_{\odot})$ & $1.07 \pm 0.04$ & $1.027 \pm 0.008$ \\
$X_0$ & $0.687 \pm 0.007$ & $0.692 \pm 0.006$ \\
$\left(Z/X\right)_0$ & $0.039 \pm 0.006$ & $0.034 \pm 0.002$ \\
$Y_0$ & $0.285 \pm 0.009$ & $0.292 \pm 0.006$ \\
$\left[\textrm{Fe/H}\right]$ & $0.23 \pm 0.07$ & $0.19 \pm 0.03$ \\
$Y_s$ & $0.238 \pm 0.005$ & $0.240 \pm 0.04$ \\
$R~(R_{\odot})$ & $1.23 \pm 0.04$ & $1.118 \pm 0.008$ \\
$t~(Gyr)$ & $7.1 \pm 0.2$ & $7.04 \pm 0.08$  \\
$\chi^2$ & $0.8$ & $0.4$ \\
\hline
\end{tabular}
\end{table}

\section{Individual models}\label{Sec:IndMod}
In this section, we summarise the set of input physics used in the reference models and gather the individual stellar parameters as well as the uncertainties propagated during the Levenberg-Marquardt adjustment for each model presented in this paper.  Table \ref{Tab:RefPhy} presents the set of input physics used in the reference model while Table \ref{Tab:PhyVar} summarises the several variations of input physics considered throughout the paper. In the latter, the first column gives the label given to the models considering that specific choice of input physics, the second column is the physical ingredient which is varied upon, the third column is the corresponding value and columns $4$ and $5$ display the $\chi^2$ values obtained for the optimal models of both stars in each case. Finally, Tables \ref{Tab:StePar} through \ref{Tab:IndABComp} give the complete set of stellar parameters obtained for every case considered in the present paper.

\begin{table}[h]
\centering
\caption{Summary of the physical ingredients included in the reference models, denoted AGSS09.}\label{Tab:RefPhy}
\begin{tabular}{c C{6em} c}
\hline\small
Inp. Phys. & Value & Ref. \\ 
\hline
\hline\\[-0.8em]
Solar mixture & AGSS09 & {\cite{2009ARA&A..47..481A}} \\
Eq. of state & Free EOS & {\citet{2003ApJ...588..862C}} \\
Opacity & OPAL & {\citet{1996ApJ...464..943I}} \\
$\AMLT$ & $1.82$ & Solar calibration \\
Overshoot & None & / \\
Diffusion & Yes & {\citet{1994ApJ...421..828T}} \\
Turbulent mix. & None & / \\
Atmosphere & Eddington $T-\tau$ relation & / \\
Surf. eff. corr. & Yes & {\citet{2008ApJ...683L.175K}} \\
\hline
\end{tabular}
\end{table}

\begin{table}[h]
\centering
\caption{Variations in the input physics, corresponding name and reduced $\chi^2$ values.}\label{Tab:PhyVar}
\begin{tabular}{c C{5em} C{6em} c c}
\hline\small
Name & Inp. Phys. & Value & $\chi^2_A$ & $\chi^2_B$ \\ 
\hline
\hline\\[-0.8em]
AGSS09 & Sol. mix. & AGSS09 & $1.0$ & $0.8$ \\
GN93 & Sol. mix. & GN93 & $1.0$ & $1.0$ \\
OP & Opac. & OP & $0.6$ & $0.6$ \\
OPLIB & Opac. & OPLIB & $1.2$ & $1.0$ \\
CEFF & Eq. of state & CEFF & $1.0$ & $0.9$ \\
OPAL05 & Eq. of state & OPAL05 & $0.2$ & $1.1$ \\
No diff. & Diff. & No & $0.9$ & $0.6$ \\
$D_{\textrm{turb}}$ & Turb. mix. & $D_{\textrm{turb}}=7500$ & $0.5$ & $0.9$ \\
$\AMLT=1.7$ & $\AMLT$ & $1.7$ & $1.3$ & $0.3$ \\
Vernazza & Atmos. & {\citet{1981ApJS...45..635V}} & $0.3$ & $0.8$ \\
No surf. corr. & Surf. eff. corr. & No & $0.1$ & $0.0$ \\
$\alpha_{\textrm{ov}}=0.1$ & Overshoot & $0.1$ & $0.8$ & $0.3$ \\
$\alpha_{\textrm{ov}}=0.2$ & Overshoot & $0.2$ & $0.4$ & $0.2$ \\
$\alpha_{\textrm{un}}=0.1$ & Undershoot & $0.1$ & $0.9$ & $0.7$ \\
\hline
\end{tabular}
\end{table}

\begin{table*}[h]
\small
\caption{Summary of the fitted models with only the seismic constraints. The reference model, labelled AGSS09, corresponds to the choice of physics described in Sect. \ref{Sec:Mod} and \ref{Tab:RefPhy}. The individual uncertainties result from the error propagation during the Levenberg-Marquardt adjustment.}\label{Tab:StePar}
\begin{minipage}[t]{.49\textwidth}
\vspace{0em}
\begin{tabular}{*{4}{c}}
\hline
Model & Quantity & 16CygA & 16CygB \\
\hline
\hline\\[-0.8em]
AGSS09 & $M~(M_{\odot})$ & $1.06 \pm 0.02$  & $1.011 \pm 0.009$ \\
& $X_0$ & $0.68 \pm 0.01$ & $0.679 \pm 0.007$ \\
& $\left(Z/X\right)_0$ & $0.035 \pm 0.001$ & $0.037 \pm 0.002$ \\
& $Y_0$ & $0.30 \pm 0.01$ & $0.296 \pm 0.007$\\
& $\left[\textrm{Fe/H}\right]$ & $0.19 \pm 0.01$ & $0.22 \pm 0.03$ \\
& $Y_s$ & $0.243 \pm 0.009$ & $0.251 \pm 0.004$ \\
& $R~(R_{\odot})$ & $1.22 \pm 0.03$ & $1.104 \pm 0.009$ \\
\mkcolor{AGSS09} & $t~(Gyr)$ & $6.8 \pm 0.2$ & $6.97 \pm 0.08$ \\
\hline
GN93 & $M~(M_{\odot})$ & $1.068 \pm 0.005$ & $1.02 \pm 0.01$ \\
& $X_0$ & $0.690 \pm 0.007$ & $0.685 \pm 0.007$ \\
& $\left(Z/X\right)_0$ & $0.039 \pm 0.003$ & $0.040 \pm 0.002$ \\
& $Y_0$ & $0.283 \pm 0.008$ & $0.288 \pm 0.007$\\
& $\left[\textrm{Fe/H}\right]$ & $0.23 \pm 0.03$ & $0.26 \pm 0.03$ \\
& $Y_s$ & $0.237 \pm 0.004$ & $0.245 \pm 0.005$ \\
& $R~(R_{\odot})$ & $1.222 \pm 0.006$ & $1.11 \pm 0.01$ \\
\mkcolor{GN93} & $t~(Gyr)$ & $6.59 \pm 0.09$ & $6.76 \pm 0.06$ \\
\hline
OP & $M~(M_{\odot})$ & $1.053 \pm 0.008$ & $1.01 \pm 0.01$ \\
& $X_0$ & $0.68 \pm 0.01$ & $0.678 \pm 0.007$ \\
& $\left(Z/X\right)_0$ & $0.035 \pm 0.004$ & $0.039 \pm 0.002$ \\
& $Y_0$ & $0.30 \pm 0.01$ & $0.296 \pm 0.007$\\
& $\left[\textrm{Fe/H}\right]$ & $0.19 \pm 0.06$ & $0.25 \pm 0.03$ \\
& $Y_s$ & $0.244 \pm 0.008$ & $0.252 \pm 0.005$ \\
& $R~(R_{\odot})$ & $1.217 \pm 0.009$ & $1.11 \pm 0.01$ \\
\mkcolor{OP} & $t~(Gyr)$ & $6.80 \pm 0.09$ & $7.0 \pm 0.1$ \\
\hline
OPLIB & $M~(M_{\odot})$ & $1.042 \pm 0.009$ & $0.99 \pm 0.01$ \\
& $X_0$ & $0.68 \pm 0.01$ & $0.673 \pm 0.008$ \\
& $\left(Z/X\right)_0$ & $0.032 \pm 0.004$ & $0.036 \pm 0.001$ \\
& $Y_0$ & $0.30 \pm 0.01$ & $0.303 \pm 0.008$\\
& $\left[\textrm{Fe/H}\right]$ & $0.16 \pm 0.05$ & $0.20 \pm 0.01$ \\
& $Y_s$ & $0.244 \pm 0.007$ & $0.258 \pm 0.005$ \\
& $R~(R_{\odot})$ & $1.21 \pm 0.01$ & $1.10 \pm 0.01$ \\
\mkcolor{OPLIB} & $t~(Gyr)$ & $6.41 \pm 0.08$ & $6.61 \pm 0.09$ \\
\hline
CEFF & $M~(M_{\odot})$ & $1.07 \pm 0.01$ & $1.02 \pm 0.01$ \\
& $X_0$ & $0.698 \pm 0.009$ & $0.681 \pm 0.008$ \\
& $\left(Z/X\right)_0$ & $0.031 \pm 0.001$ & $0.037 \pm 0.002$ \\
& $Y_0$ & $0.280 \pm 0.009$ & $0.294 \pm 0.008$\\
& $\left[\textrm{Fe/H}\right]$ & $0.12 \pm 0.02$ & $0.22 \pm 0.03$ \\
& $Y_s$ & $0.229 \pm 0.005$ & $0.249 \pm 0.005$ \\
& $R~(R_{\odot})$ & $1.22 \pm 0.01$ & $1.11 \pm 0.01$ \\
\mkcolor{CEFF} & $t~(Gyr)$ & $6.9 \pm 0.1$ & $7.0 \pm 0.1$ \\
\hline
OPAL05 & $M~(M_{\odot})$ & $1.06 \pm 0.02$ & $1.010 \pm 0.009$ \\
& $X_0$ & $0.69 \pm 0.02$ & $0.669 \pm 0.006$ \\
& $\left(Z/X\right)_0$ & $0.033 \pm 0.002$ & $0.040 \pm 0.002$ \\
& $Y_0$ & $0.29 \pm 0.02$ & $0.304 \pm 0.004$\\
& $\left[\textrm{Fe/H}\right]$ & $0.16 \pm 0.03$ & $0.25 \pm 0.03$ \\
& $Y_s$ & $0.237 \pm 0.009$ & $0.259 \pm 0.004$ \\
& $R~(R_{\odot})$ & $1.22 \pm 0.02$ & $1.10 \pm 0.01$ \\
\mkcolor{OPAL05} & $t~(Gyr)$ & $6.8 \pm 0.2$ & $6.92 \pm 0.08$ \\
\hline
No diff. & $M~(M_{\odot})$ & $1.109 \pm 0.007$ & $1.063 \pm 0.008$ \\
& $X_0$ & $0.763 \pm 0.007$ & $0.754 \pm 0.006$ \\
& $\left(Z/X\right)_0$ & $0.020 \pm 0.001$ & $0.0214 \pm 0.0009$ \\
& $Y_0$ & $0.22 \pm 0.01$ & $0.230 \pm 0.009$\\
& $\left[\textrm{Fe/H}\right]$ & $0.04 \pm 0.02$ & $0.07 \pm 0.02$ \\
& $Y_s$ & $0.221 \pm 0.007$ & $0.230 \pm 0.006$ \\
& $R~(R_{\odot})$ & $1.237 \pm 0.008$ & $1.123 \pm 0.009$ \\
\mkcolor{NoDiff} & $t~(Gyr)$ & $7.5 \pm 0.1$ & $7.50 \pm 0.07$ \\
\hline
\end{tabular}
\end{minipage}
\hfill
\begin{minipage}[t]{.49\textwidth}
\vspace{0em}
\begin{tabular}{*{4}{c}}
\hline
Model & Quantity & 16CygA & 16CygB \\
\hline
\hline\\[-0.8em]
$D_{\textrm{turb}}$ & $M~(M_{\odot})$ & $1.07 \pm 0.01$ & $1.02 \pm 0.01$ \\
& $X_0$ & $0.705 \pm 0.004$ & $0.697 \pm 0.008$ \\
& $\left(Z/X\right)_0$ & $0.029 \pm 0.002$ & $0.032 \pm 0.002$ \\
& $Y_0$ & $0.274 \pm 0.004$ & $0.281 \pm 0.008$\\
& $\left[\textrm{Fe/H}\right]$ & $0.14 \pm 0.03$ & $0.17 \pm 0.03$ \\
& $Y_s$ & $0.239 \pm 0.003$ & $0.248 \pm 0.006$ \\
& $R~(R_{\odot})$ & $1.22 \pm 0.02$ & $1.11 \pm 0.01$ \\
\mkcolor{Dturb} & $t~(Gyr)$ & $6.9 \pm 0.1$ & $7.0 \pm 0.1$ \\
\hline
$\AMLT=1.7$ & $M~(M_{\odot})$ & $1.03 \pm 0.01$ & $0.98 \pm 0.02$ \\
& $X_0$ & $0.68 \pm 0.01$ & $0.66 \pm 0.01$ \\
& $\left(Z/X\right)_0$ & $0.033 \pm 0.004$ & $0.041 \pm 0.003$ \\
& $Y_0$ & $0.30 \pm 0.01$ & $0.31 \pm 0.01$\\
& $\left[\textrm{Fe/H}\right]$ & $0.16 \pm 0.07$ & $0.26 \pm 0.04$ \\
& $Y_s$ & $0.242 \pm 0.007 $ & $0.263 \pm 0.006$ \\
& $R~(R_{\odot})$ & $1.21 \pm 0.01$ & $1.09 \pm 0.02$ \\
\mkcolor{Alpha} & $t~(Gyr)$ & $7.1 \pm 0.2$ & $7.3 \pm 0.4$ \\
\hline
Vernazza & $M~(M_{\odot})$ & $1.020 \pm 0.009$ & $0.97 \pm 0.01$ \\
& $X_0$ & $0.675 \pm 0.009$ & $0.66 \pm 0.01$ \\
& $\left(Z/X\right)_0$ & $0.035 \pm 0.002$ & $0.040 \pm 0.001$ \\
& $Y_0$ & $0.30 \pm 0.01$ & $0.32 \pm 0.01$ \\
& $\left[\textrm{Fe/H}\right]$ & $0.17 \pm 0.03$ & $0.25 \pm 0.01$ \\
& $Y_s$ & $0.248 \pm 0.005$ & $0.268 \pm 0.006$ \\
& $R~(R_{\odot})$ & $1.20 \pm 0.01$ & $1.09 \pm 0.02$  \\
& $t~(Gyr)$ & $7.2 \pm 0.1$ & $7.4 \pm 0.1$ \\
\hline
No surf. & $M~(M_{\odot})$ & $1.08 \pm 0.02$ & $1.028 \pm 0.009$ \\
corr. & $X_0$ & $0.69 \pm 0.01$ & $0.696 \pm 0.008$ \\
& $\left(Z/X\right)_0$ & $0.038 \pm 0.002$ & $0.033 \pm 0.002$ \\
& $Y_0$ & $0.28 \pm 0.01$ & $0.281 \pm 0.008$\\
& $\left[\textrm{Fe/H}\right]$ & $0.22 \pm 0.03$ & $0.17 \pm 0.03$ \\
& $Y_s$ & $0.236 \pm 0.007$ & $0.237 \pm 0.005$ \\
& $R~(R_{\odot})$ & $1.24 \pm 0.03$ & $1.12 \pm 0.01$ \\
\mkcolor{NoCorr} & $t~(Gyr)$ & $7.0 \pm 0.1$ & $7.05 \pm 0.06$ \\
\hline
$\alpha_{\textrm{ov}}=0.1$ & $M~(M_{\odot})$ & $1.058 \pm 0.007$ & $1.01 \pm 0.02$ \\
& $X_0$ & $0.68 \pm 0.01$ & $0.679 \pm 0.006$ \\
& $\left(Z/X\right)_0$ & $0.035 \pm 0.003$ & $0.036 \pm 0.004$ \\
& $Y_0$ & $0.30 \pm 0.01$ & $0.297 \pm 0.007$\\
& $\left[\textrm{Fe/H}\right]$ & $0.19 \pm 0.05$ & $0.20 \pm 0.04$ \\
& $Y_s$ & $0.243 \pm 0.006$ & $0.251 \pm 0.004$ \\
& $R~(R_{\odot})$ & $1.218 \pm 0.09$ & $1.10 \pm 0.02$ \\
\mkcolor{Aov} & $t~(Gyr)$ & $6.8 \pm 0.2$ & $6.9 \pm 0.1$ \\
\hline
$\alpha_{\textrm{ov}}=0.2$ & $M~(M_{\odot})$ & $0.920 \pm 0.009$ & $1.016 \pm 0.009$ \\
& $X_0$ & $0.700 \pm 0.006$ & $0.682 \pm 0.007$ \\
& $\left(Z/X\right)_0$ & $0.0106 \pm 0.0002$ & $0.036 \pm 0.002$ \\
& $Y_0$ & $0.293 \pm 0.006$ & $0.293 \pm 0.007$ \\
& $\left[\textrm{Fe/H}\right]$ & $-0.39 \pm 0.01$ & $0.21 \pm 0.02$ \\
& $Y_s$ & $0.217 \pm 0.002$ & $0.249 \pm 0.004$ \\
& $R~(R_{\odot})$ & $1.16 \pm 0.01$ & $1.11 \pm 0.01$ \\
& $t~(Gyr)$ & $7.16 \pm 0.06$ & $6.97 \pm 0.09$ \\
\hline
$\alpha_{\textrm{un}}=0.1$ & $M~(M_{\odot})$ & $1.06 \pm 0.01$ & $1.012 \pm 0.007$ \\
& $X_0$ & $0.688 \pm 0.009$ & $0.680 \pm 0.008$ \\
& $\left(Z/X\right)_0$ & $0.034 \pm 0.001$ & $0.036 \pm 0.002$ \\
& $Y_0$ & $0.289 \pm 0.009$ & $0.295 \pm 0.008$\\
& $\left[\textrm{Fe/H}\right]$ & $0.17 \pm 0.02$ & $0.22 \pm 0.03$ \\
& $Y_s$ & $0.241 \pm 0.005$ & $0.252 \pm 0.005$ \\
& $R~(R_{\odot})$ & $1.22 \pm 0.01$ & $1.105 \pm 0.008$ \\
\mkcolor{Aun} & $t~(Gyr)$ & $6.78 \pm 0.09$ & $6.97 \pm 0.07$ \\
\hline
\end{tabular}
\end{minipage}
\end{table*}

\begin{table}[h]
\centering
\small
\caption{Adjusted stellar parameters using the temperature profile from \citet{1981ApJS...45..635V} with a solar calibrated value of $\AMLT=2.02$.}\label{Tab:Ver}
\begin{tabular}{ccc}
\hline
Quantity & 16CygA & 16CygB \\
\hline
\hline\\[-0.8em]
$M~(M_{\odot})$ & $1.06 \pm 0.01$ & $1.02 \pm 0.01$ \\
$X_0$ & $0.68 \pm 0.01$ & $0.684 \pm 0.006$ \\
$\left(Z/X\right)_0$ & $0.035 \pm 0.004$ & $0.035 \pm 0.002$ \\
$Y_0$ & $0.29 \pm 0.01$ & $0.292 \pm 0.006$ \\
$\left[\textrm{Fe/H}\right]$ & $0.20 \pm 0.03$ & $0.18 \pm 0.06$ \\
$Y_s$ & $0.243 \pm 0.006$ & $0.248 \pm 0.04$ \\
$R~(R_{\odot})$ & $1.22 \pm 0.01$ & $1.10 \pm 0.01$ \\
$t~(Gyr)$ & $6.7 \pm 0.2$ & $6.9 \pm 0.1$  \\
$\AMLT$ & \multicolumn{2}{c}{$2.02$} \\
\hline
\end{tabular}
\end{table}

\begin{table}[h]
\centering
\small
\caption{Results of the modelling considering only seismic constraints with different values of the mixing length coefficient.}\label{Tab:Alp}
\begin{tabular}{*{4}{c}}
\hline
$\AMLT$ & Quantity & 16CygA & 16CygB \\
\hline
\hline\\[-0.8em]
$1.7$ & $M~(M_{\odot})$ & $1.03 \pm 0.01$ & $0.98 \pm 0.02$ \\
& $X_0$ & $0.68 \pm 0.01$ & $0.66 \pm 0.01$ \\
& $\left(Z/X\right)_0$ & $0.033 \pm 0.004$ & $0.041 \pm 0.003$ \\
& $Y_0$ & $0.30 \pm 0.01$ & $0.31 \pm 0.01$\\
& $\left[\textrm{Fe/H}\right]$ & $0.16 \pm 0.07$ & $0.26 \pm 0.04$ \\
& $Y_s$ & $0.242 \pm 0.007 $ & $0.263 \pm 0.006$ \\
& $R~(R_{\odot})$ & $1.21 \pm 0.01$ & $1.09 \pm 0.02$ \\
& $t~(Gyr)$ & $7.1 \pm 0.2$ & $7.3 \pm 0.4$ \\
\hline
$1.82$ & $M~(M_{\odot})$ & $1.06 \pm 0.02$  & $1.011 \pm 0.009$ \\
& $X_0$ & $0.68 \pm 0.01$ & $0.679 \pm 0.007$ \\
& $\left(Z/X\right)_0$ & $0.035 \pm 0.001$ & $0.037 \pm 0.002$ \\
& $Y_0$ & $0.30 \pm 0.01$ & $0.296 \pm 0.007$\\
& $\left[\textrm{Fe/H}\right]$ & $0.19 \pm 0.01$ & $0.22 \pm 0.03$ \\
& $Y_s$ & $0.243 \pm 0.009$ & $0.251 \pm 0.004$ \\
& $R~(R_{\odot})$ & $1.22 \pm 0.03$ & $1.104 \pm 0.009$ \\
& $t~(Gyr)$ & $6.8 \pm 0.2$ & $6.97 \pm 0.08$ \\
\hline
$2.0$ & $M~(M_{\odot})$ & $1.108 \pm 0.009$ & $1.058 \pm 0.009$ \\
& $X_0$ & $0.705 \pm 0.007$ & $0.705 \pm 0.06$ \\
& $\left(Z/X\right)_0$ & $0.033 \pm 0.002$ & $0.032 \pm 0.002$ \\
& $Y_0$ & $0.272 \pm 0.007$ & $0.272 \pm 0.06$\\
& $\left[\textrm{Fe/H}\right]$ & $0.16 \pm 0.03$ & $0.16 \pm 0.04$ \\
& $Y_s$ & $0.227 \pm 0.004 $ & $0.232 \pm 0.004$ \\
& $R~(R_{\odot})$ & $1.23 \pm 0.01$ & $1.12 \pm 0.01$ \\
& $t~(Gyr)$ & $6.4 \pm 0.1$ & $6.6 \pm 0.1$ \\
\hline
\end{tabular}
\end{table}

\begin{table}[h]
\centering
\small
\caption{Results of the modelling considering only seismic constraints and including turbulent mixing with different values of the turbulent mixing coefficient.}\label{Tab:Dif}
\begin{tabular}{*{4}{c}}
\hline
$D_{\textrm{turb}}\left(\textrm{cm}^2\textrm{s}^{-1} \right)$ & Quantity & 16CygA & 16CygB \\
\hline
\hline\\[-0.8em]
$2000$ & $M~(M_{\odot})$ & $1.07 \pm 0.01$ & $1.024 \pm 0.008$ \\
& $X_0$ & $0.70 \pm 0.01$ & $0.697 \pm 0.007$ \\
& $\left(Z/X\right)_0$ & $0.029 \pm 0.001$ & $0.032 \pm 0.002$ \\
& $Y_0$ & $0.27 \pm 0.01$ & $0.281 \pm 0.007$\\
& $\left[\textrm{Fe/H}\right]$ & $0.12 \pm 0.02$ & $0.17 \pm 0.02$ \\
& $Y_s$ & $0.236 \pm 0.007 $ & $0.246 \pm 0.005$ \\
& $R~(R_{\odot})$ & $1.22 \pm 0.01$ & $1.109 \pm 0.009$ \\
& $t~(Gyr)$ & $6.9 \pm 0.1$ & $6.99 \pm 0.07$ \\
\hline
$5000$ & $M~(M_{\odot})$ & $1.1 \pm 0.1$ & $1.023 \pm 0.007$ \\
& $X_0$ & $0.7 \pm 0.1$ & $0.697 \pm 0.006$ \\
& $\left(Z/X\right)_0$ & $0.03 \pm 0.01$ & $0.032 \pm 0.001$ \\
& $Y_0$ & $0.3 \pm 0.1$ & $0.280 \pm 0.006$\\
& $\left[\textrm{Fe/H}\right]$ & $0.1 \pm 0.2$ & $0.17 \pm 0.02$ \\
& $Y_s$ & $0.24 \pm 0.07 $ & $0.247 \pm 0.004$ \\
& $R~(R_{\odot})$ & $1.2 \pm 0.1$ & $1.109 \pm 0.007$ \\
& $t~(Gyr)$ & $6.8 \pm 0.8$ & $6.99 \pm 0.06$ \\
\hline
$7500$ & $M~(M_{\odot})$ & $1.07 \pm 0.01$ & $1.02 \pm 0.01$ \\
& $X_0$ & $0.705 \pm 0.004$ & $0.697 \pm 0.008$ \\
& $\left(Z/X\right)_0$ & $0.029 \pm 0.002$ & $0.032 \pm 0.002$ \\
& $Y_0$ & $0.274 \pm 0.004$ & $0.281 \pm 0.008$\\
& $\left[\textrm{Fe/H}\right]$ & $0.14 \pm 0.03$ & $0.17 \pm 0.03$ \\
& $Y_s$ & $0.239 \pm 0.003$ & $0.248 \pm 0.006$ \\
& $R~(R_{\odot})$ & $1.22 \pm 0.02$ & $1.11 \pm 0.01$ \\
& $t~(Gyr)$ & $6.9 \pm 0.1$ & $7.0 \pm 0.1$ \\
\hline
$10000$ & $M~(M_{\odot})$ & $1.03 \pm 0.01$ & $0.98 \pm 0.02$ \\
& $X_0$ & $0.70 \pm 0.01$ & $0.66 \pm 0.01$ \\
& $\left(Z/X\right)_0$ & $0.029 \pm 0.001$ & $0.032 \pm 0.001$ \\
& $Y_0$ & $0.27 \pm 0.01$ & $0.280 \pm 0.006$\\
& $\left[\textrm{Fe/H}\right]$ & $0.14 \pm 0.02$ & $0.17 \pm 0.02$ \\
& $Y_s$ & $0.240 \pm 0.007 $ & $0.248 \pm 0.004$ \\
& $R~(R_{\odot})$ & $1.22 \pm 0.01$ & $1.109 \pm 0.007$ \\
& $t~(Gyr)$ & $6.9 \pm 0.1$ & $6.99 \pm 0.06$ \\
\hline
\end{tabular}
\end{table}

\begin{table}[h]
\centering
\small
\caption{Adjusted stellar parameters including the effective temperature constraint with the OPLIB opacity table and with a fixed turbulent mixing coefficient.}\label{Tab:OPLIBDturb}
\begin{tabular}{cc}
\hline
Quantity & 16CygA \\
\hline
\hline\\[-0.8em]
$M~(M_{\odot})$ & $1.05 \pm 0.07$  \\
$X_0$ & $0.70 \pm 0.08$  \\
$\left(Z/X\right)_0$ & $0.028 \pm 0.001$  \\
$Y_0$ & $0.27 \pm 0.08$  \\
$\left[\textrm{Fe/H}\right]$ & $0.1 \pm 0.1$ \\
$Y_s$ & $0.24 \pm 0.03$ \\
$R~(R_{\odot})$ & $1.21 \pm 0.05$ \\
$t~(Gyr)$ & $6.6 \pm 0.4$  \\
$D_{\textrm{turb}}$ & $0.2 \cdot 10^4$ \\
\hline
\end{tabular}
\end{table}

\begin{table}[h]
\centering
\small
\caption{Adjusted stellar parameters imposing a common age and initial composition for both stars for the different variations in physics.}\label{Tab:ABAll}
\begin{tabular}{*{4}{c}}
\hline
Model & Quantity & 16CygA & 16CygB \\
\hline
\hline\\[-0.8em]
AGSS09 & $M~(M_{\odot})$ & $1.05 \pm 0.01$ & $1.015 \pm 0.006$ \\
& $\left[\textrm{Fe/H}\right]$ & $0.20 \pm 0.01$ & $0.21 \pm 0.01$ \\
& $Y_s$ & $0.245 \pm 0.003 $ & $0.250 \pm 0.003$ \\
& $R~(R_{\odot})$ & $1.21 \pm 0.01$ & $1.106 \pm 0.007$ \\
& $X_0$ & \multicolumn{2}{c}{$0.681 \pm 0.006$}  \\
& $\left(Z/X\right)_0$ & \multicolumn{2}{c}{$0.0360 \pm 0.0009$}  \\
& $Y_0$ & \multicolumn{2}{c}{$0.294 \pm 0.006$}  \\
& $t~(Gyr)$ & \multicolumn{2}{c}{$6.87 \pm 0.04$}  \\
\hline
$D_{\textrm{turb}}$ & $M~(M_{\odot})$ & $1.063 \pm 0.006$ & $1.024 \pm 0.005$ \\
& $\left[\textrm{Fe/H}\right]$ & $0.144 \pm 0.009$ & $0.15 \pm 0.01$ \\
& $Y_s$ & $0.242 \pm 0.003 $ & $0.245 \pm 0.004$ \\
& $R~(R_{\odot})$ & $1.220 \pm 0.007$ & $1.109 \pm 0.006$ \\
& $X_0$ & \multicolumn{2}{c}{$0.702 \pm 0.005$}  \\
& $\left(Z/X\right)_0$ & \multicolumn{2}{c}{$0.0300 \pm 0.0006$}  \\
& $Y_0$ & \multicolumn{2}{c}{$0.277 \pm 0.005$}  \\
& $t~(Gyr)$ & \multicolumn{2}{c}{$0.692 \pm 0.006$}  \\
\hline
No Diff. & $M~(M_{\odot})$ & $1.10 \pm 0.01$ & $1.068 \pm 0.004$\\
& $\left[\textrm{Fe/H}\right]$ & $0.058 \pm 0.009$ & $0.058 \pm 0.009$ \\
& $Y_s$ & $0.225 \pm 0.004 $ & $0.225 \pm 0.004$ \\
& $R~(R_{\odot})$ & $1.24 \pm 0.01$ & $1.114 \pm 0.004$ \\
& $X_0$ & \multicolumn{2}{c}{$0.759 \pm 0.004$}  \\
& $\left(Z/X\right)_0$ & \multicolumn{2}{c}{$0.0207 \pm 0.0004$}  \\
& $Y_0$ & \multicolumn{2}{c}{$0.225 \pm 0.004$}  \\
& $t~(Gyr)$ & \multicolumn{2}{c}{$7.50 \pm 0.05$}  \\
\hline
$\alpha_{\textrm{ov}}=0.1$ & $M~(M_{\odot})$ & $1.052 \pm 0.007$ & $1.013 \pm 0.006$ \\
& $\left[\textrm{Fe/H}\right]$ & $0.19 \pm 0.01$ & $0.20 \pm 0.01$ \\
& $Y_s$ & $0.245 \pm 0.003 $ & $0.250 \pm 0.003$ \\
& $R~(R_{\odot})$ & $1.214 \pm 0.008$ & $1.105 \pm 0.006$ \\
& $X_0$ & \multicolumn{2}{c}{$0.681 \pm 0.005$}  \\
& $\left(Z/X\right)_0$ & \multicolumn{2}{c}{$0.0355 \pm 0.0007$}  \\
& $Y_0$ & \multicolumn{2}{c}{$0.294 \pm 0.005$}  \\
& $t~(Gyr)$ & \multicolumn{2}{c}{$6.85 \pm 0.04$}  \\
\hline
$\AMLT$ & $M~(M_{\odot})$ & $1.072 \pm 0.004$ & $1.039 \pm 0.003$ \\
from $T_{\textrm{eff}}$ & $\left[\textrm{Fe/H}\right]$ & $0.145 \pm 0.007$ & $0.159 \pm 0.007$ \\
fit & $Y_s$ & $0.234 \pm 0.002 $ & $0.240 \pm 0.002$ \\
& $R~(R_{\odot})$ & $1.226 \pm 0.006$ & $1.114 \pm 0.004$ \\
& $\AMLT$ & $1.85$ & $1.99$ \\
& $X_0$ & \multicolumn{2}{c}{$0.696 \pm 0.003$}  \\
& $\left(Z/X\right)_0$ & \multicolumn{2}{c}{$0.0320 \pm 0.0004$}  \\
& $Y_0$ & \multicolumn{2}{c}{$0.282 \pm 0.003$}  \\
& $t~(Gyr)$ & \multicolumn{2}{c}{$6.63 \pm 0.05$}  \\
\hline
Vernazza & $M~(M_{\odot})$ & $1.054 \pm 0.006$ & $1.016 \pm 0.006$ \\
calibrated & $\left[\textrm{Fe/H}\right]$ & $0.18 \pm 0.01$ & $0.19 \pm 0.01$ \\
$\AMLT$ & $Y_s$ & $0.244 \pm 0.003 $ & $0.249 \pm 0.003$ \\
& $R~(R_{\odot})$ & $1.216 \pm 0.007$ & $1.105 \pm 0.006$ \\
& $\AMLT$ & \multicolumn{2}{c}{$2.02$} \\
& $X_0$ & \multicolumn{2}{c}{$0.682 \pm 0.005$}  \\
& $\left(Z/X\right)_0$ & \multicolumn{2}{c}{$0.0352 \pm 0.0007$}  \\
& $Y_0$ & \multicolumn{2}{c}{$0.293 \pm 0.005$}  \\
& $t~(Gyr)$ & \multicolumn{2}{c}{$6.82 \pm 0.05$}  \\
\hline
\end{tabular}
\end{table}

\begin{table}[h]
\centering
\small
\caption{Differences between theoretical and observed values for the seismic constraints, for the several variations in physics, defined as $\delta = \vert I_{\textrm{obs}}-I_{\textrm{th}}\vert$, in the units of the constraint. $\chi^2_{\textrm{red}} = \chi^2/(N-k)$ is the reduced $\chi^2$ value where $N$ is the number of constraints to the fit and $k$ the number of free parameters. The BIC value is defined in Sec. \ref{Sec:DifCom}.}\label{Tab:IndABAll}
\begin{tabular}{*{6}{c}}
\hline
Model & Quantity & \multicolumn{2}{c}{16CygA} & \multicolumn{2}{c}{16CygB} \\
& & $\delta$ & $\delta/\sigma$ & $\delta$ & $\delta/\sigma$ \\
\hline
\hline\\[-0.8em]
AGSS09 & $\Delta~\left(\mu Hz\right)$ & $0.2$ & $47.4$ & $10^{-3}$ & $0.2$ \\
& $A_{\textrm{He}}$ & $0.7$ & $0.7$ & $0.7$ & $0.7$ \\
& $\hat{r}_{01}$ & $5 \cdot 10^{-5}$ & $0.2$ & $10^{-4}$ & $0.8$ \\
& $\hat{r}_{02}$ & $6 \cdot 10^{-4}$ & $2.0$ & $4 \cdot 10^{-4}$ & $1.4$ \\
& $\chi^2_{\textrm{red}}$ & \multicolumn{4}{c}{$751.9$} \\
& BIC & \multicolumn{4}{c}{$2266.2$} \\
\hline
$D_{\textrm{turb}}$ & $\Delta~\left(\mu Hz\right)$ & $0.06$ & $13.6$ & $7 \cdot 10^{-3}$ & $1.6$ \\
& $A_{\textrm{He}}$ & $0.4$ & $0.4$ & $0.2$ & $0.2$ \\
& $\hat{r}_{01}$ & $2 \cdot 10^{-5}$ & $0.1$ & $3 \cdot 10^{-4}$ & $1.3$ \\
& $\hat{r}_{02}$ & $4 \cdot 10^{-4}$ & $1.2$ & $3 \cdot 10^{-4}$ & $1.1$ \\
& $\chi^2_{\textrm{red}}$ & \multicolumn{4}{c}{$63.3$} \\
& BIC & \multicolumn{4}{c}{$200.4$} \\
\hline
No Diff. & $\Delta~\left(\mu Hz\right)$ & $0.01$ & $3.0$ & $3 \cdot 10^{-5}$ & $0.007 $ \\
& $A_{\textrm{He}}$ & $0.7$ & $0.7$ & $0.3$ & $0.3$ \\
& $\hat{r}_{01}$ & $10^{-4}$ & $0.4$ & $10^{-4}$ & $0.7$ \\
& $\hat{r}_{02}$ & $10^{-4}$ & $0.4$ & $10^{-4}$ & $0.4$ \\
& $\chi^2_{\textrm{red}}$ & \multicolumn{4}{c}{$3.4$} \\
& BIC & \multicolumn{4}{c}{$20.7$} \\
\hline
$\alpha_{\textrm{ov}}=0.1$ & $\Delta~\left(\mu Hz\right)$ & $0.3$ & $59.7$ & $1 \cdot 10^{4}$ & $0.03$ \\
& $A_{\textrm{He}}$ & $0.8$ & $0.8$ & $0.6$ & $0.6$ \\
& $\hat{r}_{01}$ & $3 \cdot 10^{-5}$ & $0.1$ & $2 \cdot 10^{-4}$ & $0.9$ \\
& $\hat{r}_{02}$ & $5 \cdot 10^{-4}$ & $1.6$ & $4 \cdot 10^{-4}$ & $1.6$ \\
& $\chi^2_{\textrm{red}}$ & \multicolumn{4}{c}{$1188.8$} \\
& BIC & \multicolumn{4}{c}{$3576.7$} \\
\hline
$\AMLT$ & $\Delta~\left(\mu Hz\right)$ & $0.2$ & $52.8$ & $10^{-3}$ & $0.2$ \\
from $T_{\textrm{eff}}$ & $A_{\textrm{He}}$ & $0.2$ & $0.2$ & $0.5$ & $0.5$ \\
fit & $\hat{r}_{01}$ & $2 \cdot 10^{-4}$ & $1.0$ & $4 \cdot 10^{-4}$ & $2.0$ \\
& $\hat{r}_{02}$ & $4 \cdot 10^{-4}$ & $1.3$ & $6 \cdot 10^{-4}$ & $2.2$ \\
& $\chi^2_{\textrm{red}}$ & \multicolumn{4}{c}{$928.4$} \\
& BIC & \multicolumn{4}{c}{$2795.6$} \\
\hline
Vernazza & $\Delta~\left(\mu Hz\right)$ & $0.01$ & $3.0$ & $7 \cdot 10^{-4}$ & $0.1$ \\
calibrated & $A_{\textrm{He}}$ & $0.7$ & $0.7$ & $0.3$ & $0.3$ \\
$\AMLT$ & $\hat{r}_{01}$ & $10^{-4}$ & $0.4$ & $10^{-4}$ & $0.7$ \\
& $\hat{r}_{02}$ & $10^{-4}$ & $0.4$ & $10^{-4}$ & $0.4$ \\
& $\chi^2_{\textrm{red}}$ & \multicolumn{4}{c}{$2.9$} \\
& BIC & \multicolumn{4}{c}{$19.2$} \\
\hline
\end{tabular}
\end{table}

\begin{table}[h]
\centering
\small
\caption{Same as \ref{Tab:ABAll}, but without imposing a common composition for the two stars.}\label{Tab:ABComp}
\begin{tabular}{*{4}{c}}
\hline
Model & Quantity & 16CygA & 16CygB \\
\hline
\hline\\[-0.8em]
AGSS09 & $M~(M_{\odot})$ & $1.05 \pm 0.04$ & $1.017 \pm 0.007$ \\
& $\left[\textrm{Fe/H}\right]$ & $0.20 \pm 0.05$ & $0.23 \pm 0.02$ \\
& $Y_s$ & $0.25 \pm 0.01 $ & $0.251 \pm 0.004$ \\
& $R~(R_{\odot})$ & $1.21 \pm 0.04$ & $1.106 \pm 0.008$ \\
& $X_0$ & $0.68 \pm 0.02$ & $0.679 \pm 0.006$ \\
& $\left(Z/X\right)_0$ & $0.037 \pm 0.003$ & $0.038 \pm 0.002$  \\
& $Y_0$ & $0.29 \pm 0.02$ & $0.295 \pm 0.006$ \\
& $t~(Gyr)$ & \multicolumn{2}{c}{$6.86 \pm 0.06$}  \\
\hline
$D_{\textrm{turb}}$ & $M~(M_{\odot})$ & $1.1 \pm 0.1$ & $1.02 \pm 0.05$ \\
& $\left[\textrm{Fe/H}\right]$ & $0.1 \pm 0.1$ & $0.2 \pm 0.1$ \\
& $Y_s$ & $0.24 \pm 0.04 $ & $0.25 \pm 0.03$ \\
& $R~(R_{\odot})$ & $1.2 \pm 0.1$ & $1.11 \pm 0.05$ \\
& $X_0$ & $0.70 \pm 0.06$ & $0.70 \pm 0.05$ \\
& $\left(Z/X\right)_0$ & $0.029 \pm 0.007$ & $0.031 \pm 0.009$  \\
& $Y_0$ & $0.27 \pm 0.06$ & $0.28 \pm 0.04$ \\
& $t~(Gyr)$ & \multicolumn{2}{c}{$6.9 \pm 0.1$}  \\
\hline
No Diff. & $M~(M_{\odot})$ & $1.11 \pm 0.02$ & $1.064 \pm 0.006$\\
& $\left[\textrm{Fe/H}\right]$ & $0.04 \pm 0.02$ & $0.06 \pm 0.02$ \\
& $Y_s$ & $0.221 \pm 0.008 $ & $0.228 \pm 0.005$ \\
& $R~(R_{\odot})$ & $1.23 \pm 0.02$ & $1.123 \pm 0.007$ \\
& $X_0$ & $0.763 \pm 0.008$ & $0.756 \pm 0.005$ \\
& $\left(Z/X\right)_0$ & $0.0200 \pm 0.0008$ & $0.021 \pm 0.001$ \\
& $Y_0$ & $0.221 \pm 0.008$ & $0.228 \pm 0.005$ \\
& $t~(Gyr)$ & \multicolumn{2}{c}{$7.51 \pm 0.07$}  \\
\hline
$\alpha_{\textrm{ov}}=0.1$ & $M~(M_{\odot})$ & $1.05 \pm 0.02$ & $1.010 \pm 0.009$ \\
& $\left[\textrm{Fe/H}\right]$ & $0.19 \pm 0.03$ & $0.19 \pm 0.02$ \\
& $Y_s$ & $0.245 \pm 0.007$ & $0.250 \pm 0.003$ \\
& $R~(R_{\odot})$ & $1.21 \pm 0.02$ & $1.10 \pm 0.01$ \\
& $X_0$ & $0.68 \pm 0.01$ & $0.681 \pm 0.006$  \\
& $\left(Z/X\right)_0$ & $0.035 \pm 0.002$ & $0.035 \pm 0.002$  \\
& $Y_0$ & $0.29 \pm 0.01$ & $0.295 \pm 0.006$  \\
& $t~(Gyr)$ & \multicolumn{2}{c}{$6.86 \pm 0.05$}  \\
\hline
$\AMLT$ & $M~(M_{\odot})$ & $1.072 \pm 0.004$ & $1.039 \pm 0.003$ \\
from $T_{\textrm{eff}}$ & $\left[\textrm{Fe/H}\right]$ & $0.145 \pm 0.007$ & $0.159 \pm 0.007$ \\
fit & $Y_s$ & $0.234 \pm 0.002 $ & $0.240 \pm 0.002$ \\
& $R~(R_{\odot})$ & $1.226 \pm 0.006$ & $1.114 \pm 0.004$ \\
& $\AMLT$ & $1.85$ & $1.99$ \\
& $X_0$ & $0.696 \pm 0.003$ & $0.696 \pm 0.003$  \\
& $\left(Z/X\right)_0$ & $0.0320 \pm 0.0004$ & $0.0320 \pm 0.0004$ \\
& $Y_0$ & $0.282 \pm 0.003$ & $0.282 \pm 0.003$  \\
& $t~(Gyr)$ & \multicolumn{2}{c}{$6.63 \pm 0.05$}  \\
\hline
Vernazza & $M~(M_{\odot})$ & $1.05 \pm 0.02$ & $1.017 \pm 0.008$ \\
calibrated & $\left[\textrm{Fe/H}\right]$ & $0.18 \pm 0.02$ & $0.20 \pm 0.02$ \\
$\AMLT$ & $Y_s$ & $0.244 \pm 0.007 $ & $0.249 \pm 0.004$ \\
& $R~(R_{\odot})$ & $1.21 \pm 0.02$ & $1.105 \pm 0.009$ \\
& $\AMLT$ & \multicolumn{2}{c}{$2.02$} \\
& $X_0$ & $0.68 \pm 0.01$ & $0.683 \pm 0.007$ \\
& $\left(Z/X\right)_0$ & $0.035 \pm 0.002$ & $0.035 \pm 0.001$ \\
& $Y_0$ & $0.29 \pm 0.01$ & $0.293 \pm 0.007$ \\
& $t~(Gyr)$ & \multicolumn{2}{c}{$6.82 \pm 0.06$}  \\
\hline
\end{tabular}
\end{table}

\begin{table}[h]
\centering
\small
\caption{Same as \ref{Tab:IndABAll}, but without imposing a common composition for the two stars.}\label{Tab:IndABComp}
\begin{tabular}{*{6}{c}}
\hline
Model & Quantity & \multicolumn{2}{c}{16CygA} & \multicolumn{2}{c}{16CygB} \\
& & $\delta$ & $\delta/\sigma$ & $\delta$ & $\delta/\sigma$ \\
\hline
\hline\\[-0.8em]
AGSS09 & $\Delta~\left(\mu Hz\right)$ & $0.1$ & $31.7$ & $6 \cdot 10^{-3}$ & $1.3$ \\
& $A_{\textrm{He}}$ & $0.9$ & $0.9$ & $1.5$ & $1.5$ \\
& $\hat{r}_{01}$ & $4 \cdot 10^{-5}$ & $0.1$ & $5\cdot 10^{-4}$ & $0.2$ \\
& $\hat{r}_{02}$ & $6 \cdot 10^{-4}$ & $1.9$ & $4 \cdot 5 \cdot 10^{-4}$ & $1.7$ \\
& $\chi^2_{\textrm{red}}$ & \multicolumn{4}{c}{$1020.9$} \\
& BIC & \multicolumn{4}{c}{$1035.5$} \\
\hline
$D_{\textrm{turb}}$ & $\Delta~\left(\mu Hz\right)$ & $3 \cdot 10^{-3}$ & $0.2$ & $2 \cdot 10^{-5}$ & $0.05$ \\
& $A_{\textrm{He}}$ & $0.3$ & $0.3$ & $0.6$ & $0.6$ \\
& $\hat{r}_{01}$ & $3 \cdot 10^{-5}$ & $0.1$ & $3 \cdot 10^{-4}$ & $1.4$ \\
& $\hat{r}_{02}$ & $2 \cdot 10^{-4}$ & $0.6$ & $3 \cdot 10^{-4}$ & $1.2$ \\
& $\chi^2_{\textrm{red}}$ & \multicolumn{4}{c}{$4.4$} \\
& BIC & \multicolumn{4}{c}{$19.0$} \\
\hline
No Diff. & $\Delta~\left(\mu Hz\right)$ & $3\cdot 10^{-3}$ & $0.6$ & $0.2\cdot 10^{-3}$ & $0.05$ \\
& $A_{\textrm{He}}$ & $0.002$ & $0.002$ & $0.02$ & $0.02$ \\
& $\hat{r}_{01}$ & $10^{-4}$ & $0.5$ & $0.4\cdot 10^{-4}$ & $0.2$ \\
& $\hat{r}_{02}$ & $0.2\cdot 10^{-4}$ & $0.07$ & $0.5 \cdot 10^{-4}$ & $0.2$ \\
& $\chi^2_{\textrm{red}}$ & \multicolumn{4}{c}{$0.8$} \\
& BIC & \multicolumn{4}{c}{$15.3$} \\
\hline
$\alpha_{\textrm{ov}}=0.1$ & $\Delta~\left(\mu Hz\right)$ & $0.2$ & $49.1$ & $3\cdot 10^{-4}$ & $0.007$ \\
& $A_{\textrm{He}}$ & $0.7$ & $0.7$ & $0.2$ & $0.2$ \\
& $\hat{r}_{01}$ & $3 \cdot 6 \cdot 10^{-5}$ & $0.2$ & $2 \cdot 10^{-4}$ & $0.9$ \\
& $\hat{r}_{02}$ & $6 \cdot 10^{-4}$ & $1.7$ & $3 \cdot 10^{-4}$ & $1.2$ \\
& $\chi^2_{\textrm{red}}$ & \multicolumn{4}{c}{$2426.7$} \\
& BIC & \multicolumn{4}{c}{$2441.3$} \\
\hline
$\AMLT$ & $\Delta~\left(\mu Hz\right)$ & $0.2$ & $52.8$ & $10^{-3}$ & $0.2$ \\
from $T_{\textrm{eff}}$ & $A_{\textrm{He}}$ & $0.2$ & $0.2$ & $0.5$ & $0.5$ \\
fit & $\hat{r}_{01}$ & $2 \cdot 10^{-4}$ & $1.0$ & $4 \cdot 10^{-4}$ & $2.0$ \\
& $\hat{r}_{02}$ & $4 \cdot 10^{-4}$ & $1.3$ & $6 \cdot 10^{-4}$ & $2.2$ \\
& $\chi^2_{\textrm{red}}$ & \multicolumn{4}{c}{$928.4$} \\
& BIC & \multicolumn{4}{c}{$2795.6$} \\
\hline
Vernazza & $\Delta~\left(\mu Hz\right)$ & $2 \cdot 10^{-4}$ & $0.05$ & $5 \cdot 10^{-6}$ & $0.001$ \\
calibrated & $A_{\textrm{He}}$ & $0.5$ & $0.5$ & $0.7$ & $0.7$ \\
$\AMLT$ & $\hat{r}_{01}$ & $6 \cdot 10^{-4}$ & $1.9$ & $10^{-4}$ & $0.9$ \\
& $\hat{r}_{02}$ & $3 \cdot 10^{-4}$ & $1.1$ & $2 \cdot 10^{-4}$ & $0.8$ \\
& $\chi^2_{\textrm{red}}$ & \multicolumn{4}{c}{$7.1$} \\
& BIC & \multicolumn{4}{c}{$21.7$} \\
\hline
\end{tabular}
\end{table}
\FloatBarrier

\end{appendix}

\end{document}